\documentclass[a4paper]{article}
\usepackage[margin=2.5cm]{geometry}

\usepackage[utf8]{inputenc}
\usepackage{graphicx}
\usepackage{algorithmic, algorithm}
\usepackage{caption}
\usepackage{amsmath}
\usepackage{amssymb}
\usepackage{enumitem}
\usepackage{dsfont}
\usepackage{booktabs}
\usepackage{float}
\usepackage{subcaption}
\usepackage[numbers]{natbib}

\usepackage{color}
\usepackage{placeins}
\usepackage{setspace}
\usepackage{hyperref}

\usepackage{verbatim}
\usepackage{mathtools}
\usepackage{bbm}
\usepackage{mathrsfs}

\DeclarePairedDelimiter{\ceil}{\lceil}{\rceil}

\definecolor{Code}{rgb}{0,0,0}
\definecolor{Decorators}{rgb}{0.5,0.5,0.5}
\definecolor{Numbers}{rgb}{0.5,0,0}
\definecolor{MatchingBrackets}{rgb}{0.25,0.5,0.5}
\definecolor{Keywords}{rgb}{0,0,1}
\definecolor{self}{rgb}{0,0,0}
\definecolor{Strings}{rgb}{0,0.63,0}
\definecolor{Comments}{rgb}{0,0.63,1}
\definecolor{Backquotes}{rgb}{0,0,0}
\definecolor{Classname}{rgb}{0,0,0}
\definecolor{FunctionName}{rgb}{0,0,0}
\definecolor{Operators}{rgb}{0,0,0}
\definecolor{Background}{rgb}{0.98,0.98,0.98}

\providecommand{\keywords}[1]{\textbf{\textit{Keywords---}} #1}

\makeatletter
\newcommand{\pushright}[1]{\ifmeasuring@#1\else\omit\hfill$\displaystyle#1$\fi\ignorespaces}
\newcommand{\pushleft}[1]{\ifmeasuring@#1\else\omit$\displaystyle#1$\hfill\fi\ignorespaces}
\makeatother

\title{Queue-reactive Hawkes models for the order flow}
\date{}
\author{
Peng Wu$^1$, Marcello Rambaldi $^1$, Jean-François Muzy$^2$, Emmanuel Bacry$^1$ \\
\small  
Université Paris-Dauphine and CEREMADE CNRS-UMR 7534\\
\small
SPE UMR 6134, CNRS, Université de Corse, 20250 Corte, France
}

\begin{document}

\maketitle

\begin{abstract}
In this work we introduce two variants of multivariate Hawkes models with an explicit dependency on various queue sizes aimed at modeling the stochastic time evolution of a limit order book. The models we propose thus integrate the influence of both the current book state and the past order flow.
The first variant considers the flow of order arrivals at a specific price level
as independent from the other one and describes this flow by adding a Hawkes component to the arrival rates provided by the continuous time Markov "Queue Reactive" model of Huang et al. \cite{rsb1}. Empirical calibration using Level-I order book data from Eurex future assets (Bund and DAX) show that the Hawkes term dramatically improves the pure "Queue-Reactive" model not only for the description of the order flow properties (as e.g. the statistics of inter-event times) but also with respect to the shape of the queue distributions. The second variant we introduce describes the joint dynamics of all events occurring at best bid and ask sides of some order book during a trading day. This model can be considered as a queue dependent extension of the multivariate Hawkes order-book model of Bacry et al. \cite{thibault1}. We provide an explicit way to calibrate this model either with a Maximum-Likelihood method or with a Least-Square approach. Empirical estimation from Bund and DAX level-I order book data allow us to recover the main features of Hawkes interactions 
uncovered in Bacry et al. in \cite{thibault1} but also to unveil their joint dependence on bid and ask queue sizes. We notably find that while the market order or mid-price changes rates can mainly be functions 
on the volume imbalance this is not the case for the arrival rate of limit or cancel orders.
Our findings also allows us to 
clearly bring to light various features that distinguish small and large tick assets.
\end{abstract}

\newcommand{\refpr}{p_{\mbox{\tiny ref}}}
\newcommand{\midpr}{p_{\mbox{\tiny mid}}}
\newcommand{\plev}{Q}

\newcommand{\ixa}{\ell}
\newcommand{\ixb}{m}
\newcommand{\E}[1]{\mathbb{E}\left[#1\right]}
\newcommand{\quantile}{qt}

\newcommand{\cred}[1]{{\color{red} #1}}

\providecommand{\keywords}[1]
{
	\small	
	\textbf{\textit{Keywords---}} #1
}
\keywords{Limit order book, market micro structure, Hawkes process, high frequency data, jump Markov process, ergodic properties, market simulator}

\section{Introduction}
Building faithful models for the Limit Order Book (LOB) is a longstanding issue on which many efforts have been invested in the quantitative finance community. A rich literature of theoretical and empirical studies of limit order books has emerged in the last decade (see, e.g., \cite{Gould:2013hg} and \cite{Abergel2016} for a recent review). Modeling the LOB is a challenging task due to its intricate dependence structure. Indeed, the configuration of the limit order book is determined by the arrival of multiple types of orders: limit, cancel and market orders in the simplest setting, and the way these orders arrive on the market is non-trivial. For example, it is well known that order arrival inter-event times present strong and persistent autocorrelation (see e.g. \cite{Chakraborti:2011em}), implying that past order flow influences the current state of the book. At the same time, anecdotal as well as empirical evidence (\cite{Lehalle2018}) suggests that market participants look at the state of the order book in order to take their trading decisions.\\
\noindent
Models for the LOB can be roughly divided into two main classes. On one side are the models developed by the economics community where the focus is on the behavior of rational agents that act strategically to optimize their utility function (see e.g. \cite{parlour2008limit}). 
Another stream of literature, beginning notably with \cite{smith2003}, focus instead on the overall statistical properties of LOBs and assumes a certain simplified dynamics for the order flow in order to build mathematically tractable models that can reproduce at least partially some of these observed properties. This work contributes to the latter and builds on previous works in this field.\\
\noindent
As stated above, in the pioneering work \cite{smith2003}, the order book is seen as a purely stochastic system - a so called \emph{zero intelligence} model - where orders arrive randomly, and where the interest is making testable predictions based on measurable inputs.
\cite{cont2010} is one of the first paper to clearly frame the problem of LOB modeling in the context of queuing theory and Markov chains. By leveraging the properties of Markov chains, the authors are able to derive several conditional probabilities such as the likelihood of a mid-price move or the probability of a limit order execution before a price change. 
The authors of \cite{Abergel2010} keep the same assumption of Poisson-driven independent queues and prove, using the theory of infinitesimal generators and Lyapunov stability criteria, the importance of the cancellation structure to ensure the stability of the LOB distribution and also show that under their model the price process converges to a Wiener process.
Although the hypothesis made by these models are in disagreement with empirical facts, they  present the advantage of being very tractable and to allow the derivation of many useful quantities analytically. In \cite{abergel1}, the authors drop the assumption of uncorrelated order flow and introduce some memory effect by choosing to model the rate of limit and market order arrival, $\lambda^L$ and $\lambda^M$ by a Hawkes process (\cite{hawkes1971})
\begin{equation}
\label{eq:shm}
\lambda^\ixa(t) = \mu^\ixa + \sum_{\ixb} \int \phi^{\ixa \ixb}(t-s) dN^\ixb_s \; .
\end{equation}
By setting the kernel function $\phi$ to an exponential form $\phi(t)=\alpha e^{-\beta t}$ the process $(\lambda, N)$ has the Markov property and thus the authors are able to use a similar machinery to \cite{Abergel2010} in order to study the limiting behavior of their model.
In \cite{thibault1} and \cite{rambaldi2017role}, the authors also use multivariate Hawkes processes to analyze the order flow interaction at the fist level of the order book.
Their model is calibrated without any assumption on the Hawkes kernel shapes using a non-parametric 
method.

\noindent
In \cite{rsb1}, the authors focus instead on the influence of the current state of the LOB on trading decisions. They propose a simple Markov model where the order flow arrival intensity depends only on the currently state of LOB through the available volume. They established conditions under which their model possess ergodic properties, making it possible to reproduce the empirical LOB queue size distributions as the invariant distribution of a Markov process. More recently, \cite{Lu2018} extended the model of \cite{rsb1} by allowing the order book dynamics to depend also on the type of the order that led to a complete depletion of a level (i.e. a market or cancel order) and also by taking into account the order size. \cite{Lu2018} thus depart slightly form the pure Markovian framework. They then discuss optimal market making strategies in the context of the model and also assess their performance on real data. 

\noindent
In this paper, we aim at contributing to this stream of literature by building on the work of \cite{rsb1} on one side and on the one of \cite{abergel1} and \cite{thibault1} on the other side by presenting a model where both dependence on past order flow and dependence on the state of the LOB are present. More precisely, we propose to consider stochastic LOB models that are multivariate Hawkes models whose parameters explicitly depend on the queue sizes of the order book. 
For a given set of event types $\ixa$ that can occur in the considered LOB model, 
if $\vec{q}(t)$ represents the state of the book queues at some given time $t$, such a model 
could be simply written as the following generalization of the standard multivariate Hawkes model (Eq. \eqref{eq:shm}):
\begin{equation}
\label{eq:general_qrh}
\lambda^\ixa(t) = \mu^\ixa\left(\vec{q}(t)\right) + \sum_{\ixb} \int \phi^{\ixa \ixb}\left(t-s,\vec{q}(t)\right) dN^\ixb_s \; .
\end{equation}
where we have accounted for the possible explicit dependence on the queue sizes of both the exogenous intensities $\mu^\ixa$ and interaction kernels $\phi^{\ixa \ixb}$.
We call this class of models {\em Queue Reactive Hawkes} (QRH) models. Let us notice that the issue of considering both self-excitation effects and dependence on some given state within a single model has also been considered very recently (during the completion of the present work) by Morariu-Patrichi and Pakkanen \cite{Morariu2018}. These authors proposed a general framework called ``state dependent Hawkes process" where the Hawkes kernels $\phi$ are functions of some state process $X(t)$ that can take a finite number of values and that switches from one state to another one when an event of the Hawkes process occurs and according to a transition rule that depends on the type of this event. The specific application of this framework to LOB modeling proposed in \cite{Morariu2018} mainly consists in considering either the volume imbalance or the spread as the state variable. Let us mention another very recent and related work in the paper of Daw and Pender \cite{DawPender2018} that defined and studied a Markov process constructed a pair of inter-dependent processes $(N_t,Q_t)$, where $N_t$ is a counting process and $Q_t$ a queuing process.

\noindent
In this paper our purpose is twofold: We first investigate in which respect adding Hawkes self- and cross-
excitation properties to the ``Queue Reactive" model of Huang et al. \cite{rsb1}, may improve this model. 
To achieve this goal, we consider a simple version, referred to as QRH-I, of the general model described by \eqref{eq:general_qrh} where the Hawkes kernels do not depend on the queue sizes and the various queues are considered as independent. In a second part, we consider a queue state dependent version of level-I LOB Hawkes model
introduced by Bacry et al. \cite{thibault1} where we account for bid and ask queue interactions
and we suppose that both exogenous intensity and interaction kernel matrix share the same multiplicative queue 
dependencies. We call this model the QRH-II order book model.
By calibrating these models using high frequency data from Eurex future markets,
we show that both models achieve a better fit of the data than their pure queue reactive 
or Hawkes restrictions. Let us emphasize that, like the QR model of \cite{rsb1}, the QRH-I model can be considered as a model for both the order flow and the queue state whereas, within the QRH-II model, our main concern will be to improve the order flow description, in particular we will consider the queue as an exogenous imput.

\noindent
The rest of paper is organized as follows. In Section \ref{sec:part1} we elaborate on the QRH-I model, i.e. on
a single-queue model that consists in adding an order flow dependence as provided by a multivariate Hawkes process to 
the Queue Reactive (QR) model introduced in \cite{rsb1}. We show how such a model can be calibrated using a maximum likelihood approach and prove that, very much like the QR model, under some reasonable assumption, the queue size
admits an invariant distribution. We then compare the likelihood of QRH-I model with both a standard Hawkes model with no state-dependence and with the model of \cite{rsb1} on real data form the Eurex exchange. Comparisons with empirical 
data show that QRH-I model represents an important improvement of the QR model not only with respect to the inter-event
time statistics buy also regarding the predicted shape of the equilibrium queue size distribution.
In Section~\ref{sec:part2}, we start instead from the point of view of level-I book model of \cite{thibault1}, that is a multivariate Hawkes model for all events occurring at the first level of the order book. The QRH-II model is defined by considering a multiplicative dependence of Hawkes kernel matrix and exogenous intensities 
on both the best bid and best ask queue states. We show that such a state dependence is a very meaningful 
assumption. The empirical results provided by the model calibration using Eurex future data are then discussed.
Concluding remarks and prospects for future research are provided in Section~\ref{sec:conclusions}.
Technical results like the proof of the ergodicity of QRH-I, the model calibration issues by Maximum Likelihood or
Least-Square approaches are given in the Appendices. 
\section{A Queue Reactive Hawkes model for a fixed-price best limit}\label{part1}
\label{sec:part1}

\subsection{Adding memory to the Queue Reactive model of Huang et al.}
As mentioned in the introduction, Huang, Lehalle and Rosenbaum present, in \cite{rsb1}, a model where the order flow arrival at a given price level is modeled as an inhomogeneous Poisson process with an 
intensity that depends only on the current state of the order book 
and in particular on the queue sizes. They name this property {\em Queue Reactive} (QR). 
Let us briefly recall here the main lines of the QR approach which will be shared by our model.

\noindent
The order book is seen as a $2K$-dimensional vector, where $K$ represents the number of available levels on each side, the prices living on a grid whose resolution is the tick size $tick$, i.e., the unit of price variations. The bid and ask sides are separated by a reference price $\refpr$ which is equal to the midprice $\midpr$ (i.e., the mean value of best bid and best ask prices) if the spread is an odd multiple of the tick size and equal to $\midpr \pm \frac{tick}{2}$, whichever is closer to the previous $\refpr$, if the spread is an even multiple of the tick size. The price levels on the bid side are denoted as $\{\plev_{-i}\}_{i=1\dots K}$, those at the ask side as $\{\plev_{i}\}_{i=1\dots K}$ 
while the quantities available at these levels by $q_{\pm i}$. The queue sizes are modified by the arrival of limit, market and cancel orders. For the sake of simplicity, all orders are assumed of unitary volume, so a limit order adds one unit to the queue, while a market or a cancel order subtract one unit.
We will denote by $\lambda^L_i$, $\lambda^M_i$ and $\lambda_i^C$
the arrival intensities of respectively limit, market and cancel orders on the queue $i$. 
In the simplest version of the QR model of Huang et al., all the queue sizes are independent one from each other and, for some given queue $i$, the intensity $\lambda^L_i$, $\lambda^M_i$ and $\lambda_i^C$
depends only on the queue size $q_i$, namely:
\begin{equation}
\label{eq:qr_model}
\begin{split}
\lambda_i^L (t) &= \mu_i^L(q_i(t^-))\\
\lambda_i^C (t) &= \mu_i^C(q_i(t^-))\\
\lambda_i^M (t) &= \mu_i^M({q_i(t^-)})
\end{split}
\end{equation}
where the functions $\{\mu_{i}^\ixa(q)\}_{i,\ixa}$ are the parameters of the model. They correspond to the rates of a birth-death Markov process and can easily be estimated via maximum likelihood which, in this case, amounts to the computation of conditional empirical means of intensities defined in \cite{rsb1}. As the labeling of a price level is relative to the reference price, when $\refpr{}$ changes, the level labels also change. Hence, the estimation is performed on intervals where $\refpr{}$ is constant and each period is regarded as an independent realization of the process. As shown in \cite{rsb1}, the QR model
and its extension accounting for the queue interactions is an ergodic continuous time jump 
Markov process provided the limit order rate is bounded for large queue sizes, and the rate at which orders are removed
is larger than the rate which increases the queues. In that respect, the QR model represents 
a simple and parsimonious Markov model that allows one to account for the state dependent nature of the book dynamics and that is able to describe the (stationary) distribution of the queue sizes. 

\noindent
Our goal is to consider an extension of Huang \emph{et al.} QR model with independent queues that is able to account for both queue reactive and for the memory effects in the order flow. This can be done by combining the dependence on the current state of LOB with the one on past order flow events as given by a multivariate Hawkes process within a QRH model. Note that, as discussed in the introduction (Eq. \eqref{eq:general_qrh}), in principle both Hawkes kernels and the baseline intensities could depend on the LOB state and indeed we will explore such a possibility in the next section. Here however we are interested in the simplest modification of the model of Huang \emph{et al.} that accounts for the dependence on the past order flow. In that respect our model introduces state-dependence in the multivariate Hawkes framework by allowing exclusively 
exogenous intensities to depend on the queue size.
Furthermore, since our database mainly concerns the best bid and ask, we consider only the positions $Q_{\pm 1}$.  By bid-ask symmetry (supported by empirical observations), both processes can be assumed to have the same law and  we will henceforth drop the subscript $i$.
Accordingly, $N^{L}_t$, $N^C_t$, $N^M_t$, $\lambda^{L}(t)$, $\lambda^C(t)$ and $\lambda^M(t)$ will denote the counting processes and their associated intensities defined by the arrivals of respectively limit, cancel and market orders at best bid (or alternatively at best ask). We refer to this variant of the QRH model 
as the QRH-I model since it mainly concerns a single LOB queue (at best bid or best ask).
For $\ell, m \in \{L,M,C \}$ (for respectively limit, market and cancel orders), the QRH-I model thus defines $\lambda^\ixa(t)$ as:
\begin{equation}\label{eq:qrh_model}
\lambda^\ixa(t) = \mu^\ixa(q(t^-)) + \sum_{\ixb} \int_0^{t} \phi^{\ixa \ixb}(t-s) dN_s^\ixb   \;.
\end{equation}
where the queue size $q(t)$ is simply $q(t) = q(0)+N^{L}_t-N^M_t-N^C_t$.
Since market and cancel orders can only be sent when the queue is non-empty, in the case when $q(t) =0$, the previous expression should be replaced by $\lambda^\ixa(t) = 0$ for $\ixa = M,C$.
The baseline intensities $\{\mu^\ixa(q)\}$ depend on the LOB state $q$ while the Hawkes kernels $\phi^{\ixa \ixb}(t)$ account for the effect of past orders of type $\ixb$ occurrence on the current intensity $\lambda^\ixa(t)$.
In full rigor, to complete the model definition, one should specify the law of the the initial 
queue size $q(0)$. Since, has shown below, we will consider a situation when the queue process is a component of an ergodic vector Markov process, the choice of this law is not pertinent an we simply choose $q(0)=0$.

\noindent
Let us note that the intensity function of the QR model (with independent queues) and the one of a standard Hawkes model could both be treated as a special case of the intensity function of QRH-I model.
One recovers the QR model when the Hawkes kernels are zero and a standard multivariate Hawkes model for constant baseline intensities.
We proceed as in \cite{rsb1} and we assume that the model \eqref{eq:qrh_model} holds in periods when the reference price is constant, and furthermore that such periods can be considered as independent realizations. Note that by doing so we reset the Hawkes memory every time there is a change in the reference price. We will discuss this point below when analyzing the empirical results and we will drop this assumption in the model we present in Section~\ref{sec:part2}. 

\noindent
The model parameters can be estimated using the maximum likelihood method. 
The log-likelihood $L$ of a $D$-dimensional point process where the components do not share any parameters has the following general form (see \cite{Daley}, page 21)
\begin{equation}\label{llk_fnc}
L(\theta) = \sum_{\ixa{}=1}^D L_\ixa{}(\theta), 
\quad \text{with} \quad
L_\ixa(\theta) =  \int_0^T \log \lambda^\ixa(t; \theta|\mathscr{F}_t) dN^\ixa_t  - \int_0^T \lambda^\ixa(t; \theta|\mathscr{F}_t) dt\
\end{equation}
where $\theta$ denotes the parameter set and $\lambda^\ixa$ is the intensity function of the $\ixa$-th component.

\noindent
To use the method in practice, a parametric form must be specified for the interaction kernels $\phi^{\ixa \ixb}$ in Eq. \eqref{eq:qrh_model}. A standard choice is to consider that $\phi^{\ixa \ixb}$ can be written as sum of exponential kernels: 
\begin{equation}
\label{eq:hawk_kernels}
\phi^{\ixa\ixb}(t) = \sum_{u=1}^U \alpha^{\ixa \ixb}_u \beta_u e^{-\beta_u(t - s)}
\end{equation}
where $\alpha^{\ixa \ixb}_u$ are parameters of the model and $\beta_u$, $U$ are suitably chosen hyper-parameters. This choice also presents the important advantage that the resulting log-likelihood is a convex function of the model parameters. To faciliate the notation, we use $\vec{\mu}$ and $\vec{\alpha}$ to represent all $\mu^\ixa$ and $\alpha^{\ixa \ixb}_u$. With such parametrization, $\theta =(\vec{\mu}, \vec{\alpha})$.The log-likelihood of the QRH-I model \eqref{eq:qrh_model} thus reads:
\begin{equation}
\begin{aligned}
L(\theta) &= 
\sum_{\ixa=1}^D \sum_{k=1}^{N^\ixa} \log \Big( \mu^\ixa(q(t_k)) + \sum_{ \ixb=1}^D \sum_{u=1}^U \alpha^{\ixa \ixb}_u \beta_u \int_0^{t} e^{-\beta_u(t - s)} dN_s^ \ixb \Big)\\
&- \sum_{\ixa=1}^D \int_0^{T} \Big( \mu^\ixa(q(s)) + \sum_{ \ixb=1}^D \sum_{u=1}^U \alpha^{\ixa \ixb}_u \beta_u \int_0^{s} e^{-\beta_u(s - v)} dN_v^ \ixb \Big) ds \\
\end{aligned}
\end{equation}
As we show in Appendix \ref{sec:app2}, the specific choice of a sum of exponential functions (Eq. \eqref{eq:hawk_kernels}) allows for a computationally efficient calculation of the log-likelihood and of its gradient.

\noindent
Another important advantage of the parametrization \eqref{eq:hawk_kernels} is that it allows us
to work within the framework of Markov processes.
Let us note $I=\{C,M\}$ the set of order types that decrease the queue size and $J=\{L\}$ the set of order types that increase the queue size. 
The queue size $q(t)$ thus simply corresponds to:
\begin{equation}
\label{eq:def_qt}
q(t) = \sum_{\ixb \in J} N^\ixb_t - \sum_{\ixa \in I} N^\ixa_t \; .
\end{equation}
If one defines
\begin{equation}
o_{\ixa \ixb u}(t) = \int_0^{t} \alpha^{\ixa \ixb}_u e^{-\beta_u(t - s)} dN_s^\ixb \;,
\end{equation}
and denote by $\vec{o}(t)$ the vector obtained by a vertical stacking of $o_{\ixa \ixb u}(t)$
($u \in \{1,\ldots,U\}$, $\ixa,\ixb \in \{L,M,C\})$,
then $\binom{q(t)}{\overrightarrow{o}(t)}$ is vector Markov process.
This property can be proved exactly along the same lines as in Proposition 2.2 of \cite{Jedidi2013}. Moreover,
if one assumes that 
$$\sum_{\ixa \in I}\mu^\ixa(q) \geq c_\ixa q \; \mbox{and} \; \mu^\ixb(q) \leq c^* ,\forall \ixb \in J \;,$$ we show, in Appendix
\ref{sec:app1}, with the help of Lyapunov functions approach, that the process $\binom{q(t)}{\overrightarrow{o}(t)}$ is  V-uniformly ergodic which notably means that $q(t)$ admits an invariant distribution and that this equilibrium is reached exponentially fast.

\subsection{Calibration results}

\paragraph{Data}
In this study we use tick by tick level L1 data of Bund future and DAX index future traded on the Eurex electronic future market. The data span the period from October 1st 2013 to September 30th 2014.
The dataset consists in snapshots of the first level of the order book, each with a timestamp indicating the record time with microsecond precision, that provide prices and outstanding quantities. Every time a trade occurs a specific line is added to the dataset, thus allowing to precisely determine the type of order (i.e. limit order, cancellation, or market order\footnote{With a slight abuse of language, in this paper we use the term ``market order'' to denote any order that immediately gives rise to a trade, regardless to whether or not it has a limit price.}) that lead to a change in the LOB. 
The Eurex future market is open from 8 a.m. to 10 p.m., Frankfurt time, however thorough this paper we only consider the time slot from 9 a.m. to 9 p.m. in order to capture the most active period. 
In Table \ref{tab:stats}, we report some descriptive statistics of our datasets. We note that, from the micro-structural point of view, the Bund future can be considered as a large tick asset, with and average spread very close to one, while the DAX has a considerably smaller perceived tick size compared to the Bund. This difference at the market microstructure level will reflect also in the queue dynamics and in the result of our model. Indeed, the queues on the Bund are often large as the midprice stays constant for relatively long periods of time while bids/offers accumulate at the best quotes, while on the other hand on the DAX the midprice changes more frequently thus resulting in slimmer queues at the best quotes. Further details on the datasets as well as a more detailed description of the inter-event time distribution can be found in \cite{rambaldi2017role}. 

\noindent
Let us emphasize that since in our setting the order book is seen as a collection of independent queues, we impose \emph{a priori} a strict bid-ask symmetry and all the results presented in Sec. 2 correspond to averaging estimations obtained on the bid and ask sides. 

\begin{table}[tb]
\centering
\begin{tabular}{lccccccc}
\toprule
&  \# $L$ &  \# $C$ &  \# $M$ &  Avg. spread & Med. spread &  AES & Med. inter-event time\\
\midrule
Bund & $5.41\times10^{7}$& $4.67\times10^{7}$& $6.29\times10^{6}$& 1.012 &1.0 & 6.34&$4.89\times10^{-4}$ \\
DAX &  $5.46\times10^{6}$ & $5.62\times10^{6}$& $6.68\times10^{5}$& 1.591& 2.0& 1.30 &$1.73\times10^{-3}$ \\
\bottomrule
\end{tabular}
\caption{Descriptive statistics of our dataset. Average number of limit, cancel and market order at the best quotes per day. Average and median spread (measured in ticks) and average order sizes expressed in contracts. }\label{tab:stats}
\end{table}

\begin{figure}[H]
	\centering
	\includegraphics[width=0.48\textwidth]{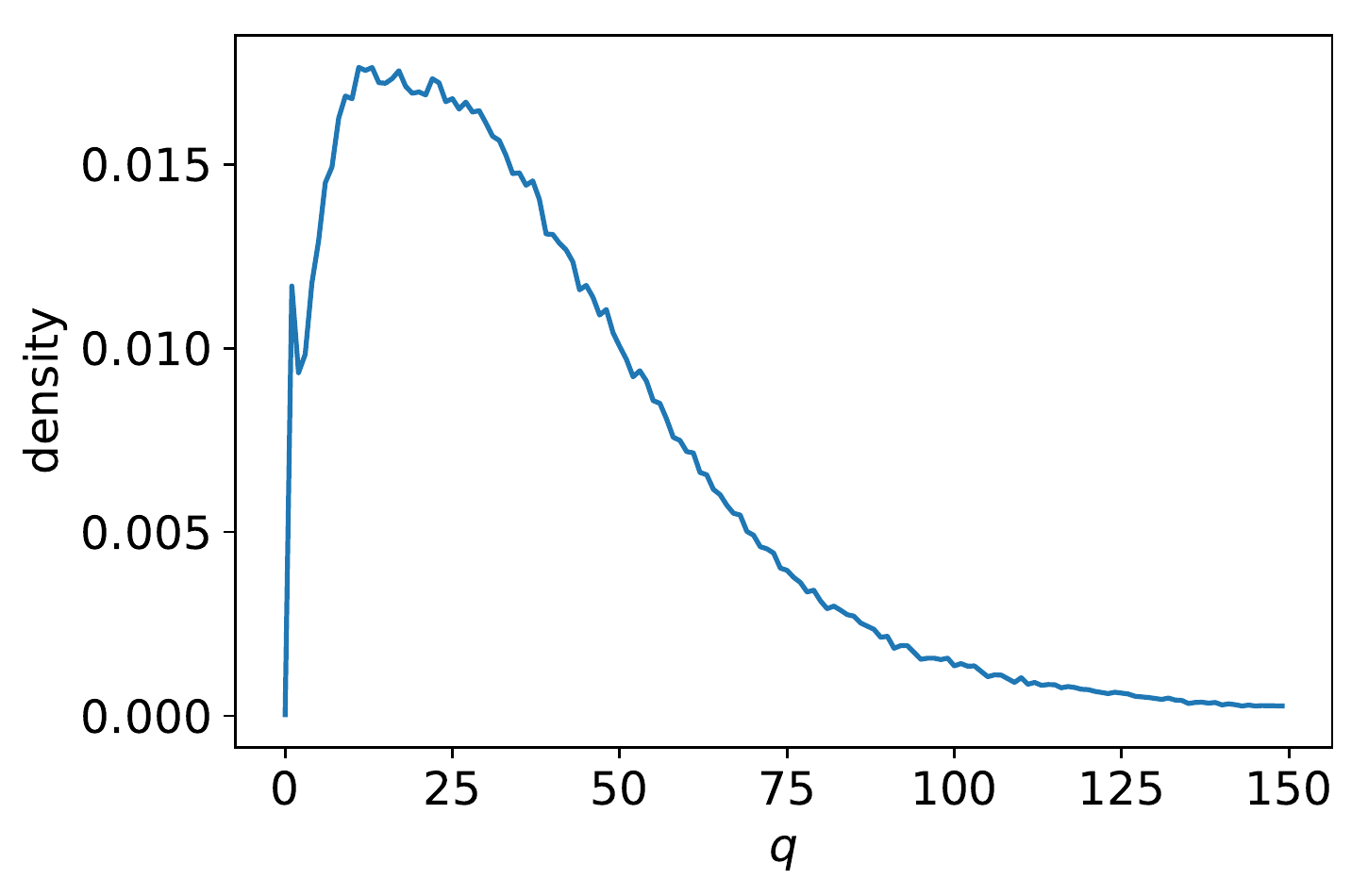}
	\includegraphics[width=0.48\textwidth]{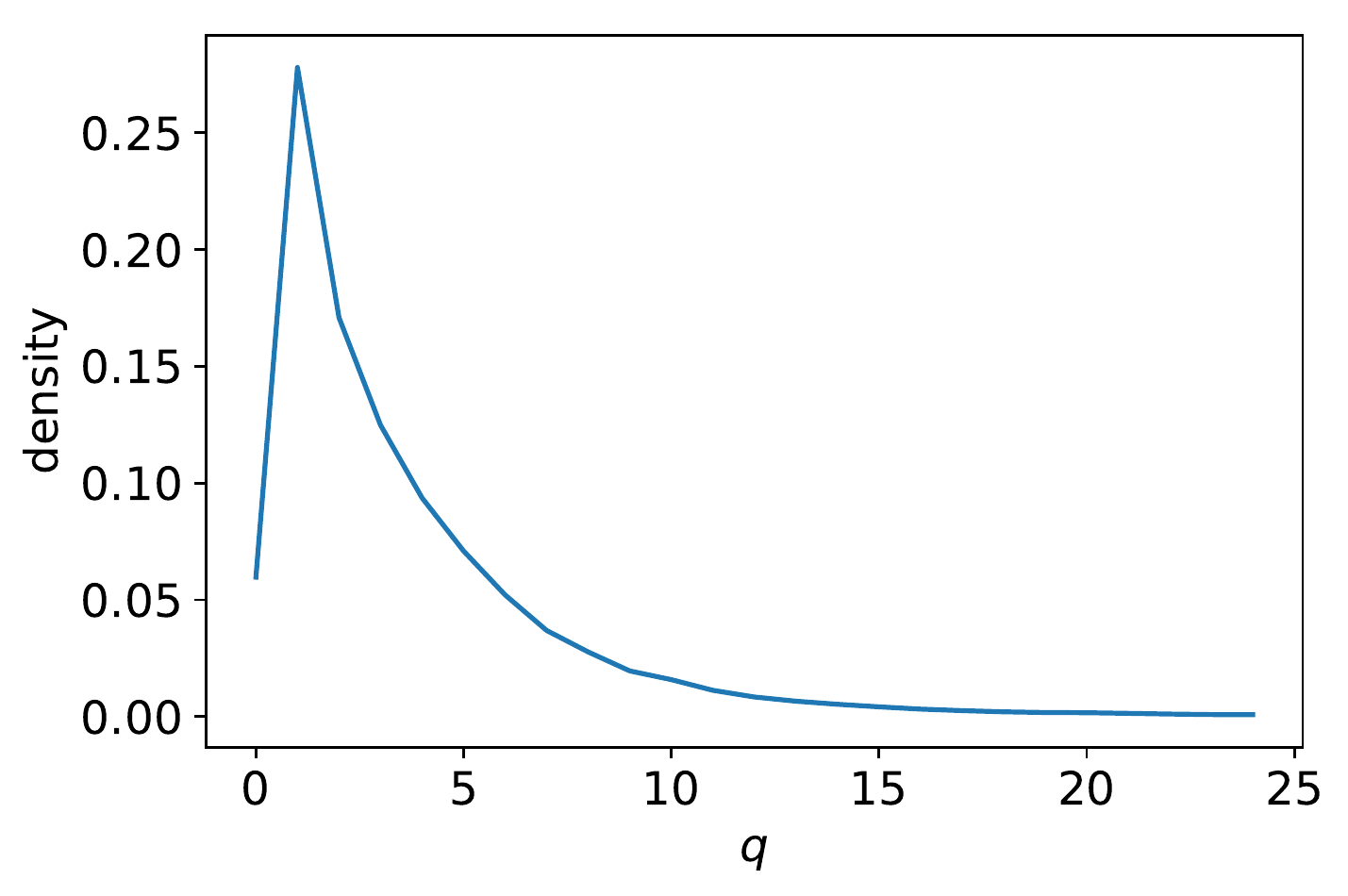}
	\caption{Empirical distribution of the queue states (measured in unit of AES) as defined in \eqref{def_state}. Left: Bund future. Right: DAX future.}
	\label{fig:density_states}
\end{figure}

\paragraph{Estimation and goodness-of-fit analysis}
In order to estimate the parameters of our model, for each day in our sample we first compute the reference price $\refpr{}$ as specified at the beginning of this section.
Then we determine the queue sizes, as in \cite{rsb1}, we assume that orders sizes for all events is a constant corresponding to the average event size (AES), defined as the average volume of all types of orders arriving at $Q_{\pm 1}$. 
We therefore measure the queue $q$ in units of AES as 
\begin{equation}\label{def_state}
q(t) = \ceil[\big]{\frac{v(t)}{\text{AES}}} 
\end{equation}
where $\ceil[\big]{}$ is the ceiling function and $v(t)$ is the volume available at time $t$ in the queue. In Figure \ref{fig:density_states} we show the empirical distribution of the so-defined $q(t)$ for Bund (left panel) and DAX (right panel). These distributions are obtained by sampling the book state every 30s over the whole time period.
Notice that the state $q(t)$ is set to $0$ if and only if $v(t) = 0$. 
We observe that the Bund future presents a broader and smoother distribution as compared to the one of the DAX,
which is on the contrary more concentrated on small queue sizes. This is a direct consequence of the different perceived tick sizes as we observed above. 

\noindent
Once $\refpr{}$, $Q_{\pm 1}$, and $q_{\pm 1}$ are determined, we divide each day in periods where $\refpr{}$ is constant. Then, each period is considered as an independent sample and we determine the parameters of our model by numerically optimizing the joint log-likelihood over all the so-obtained independent samples. In total, we have $1,207,099$ periods for the Bund and $417,581$ for the DAX.
Notice that statistical estimation purpose, in the results reported below we considered only samples that contain at least a total of 20 events and disregarded the other ones.
 Notice that if $s_k$ stands for the length of the realization $k$ and $k_t$ the index of the realization located around time $t$, 
the quantity 
$$\tau_m = \frac{\E{s_k^2}}{\E{s_k}}$$ 
represents the average length of the realization $k_t$
if one chooses $t$ at random (i.e. with a uniform probability). 
This quantity is pertinent when performing averages over a fixed grid of times $\{t_j\}_j$.
For the Bund we have $\tau_m \simeq 100$ $s$ while for the DAX we estimated $\tau_m \simeq 16$ $s$.

\noindent
As we pointed out above, in order to estimate our model we need to fix the number of exponential decays $U$ as well as the values of the decays $\beta$ themselves. We take $U=3$ and we set $\beta_1=60s^{-1}$, $\beta_2=1500s^{-1}$, $\beta_3=5500s^{-1}$ for the Bund and $\beta_1=40s^{-1}$, $\beta_2=2100s^{-1}$, $\beta_3=5200s^{-1}$ for the DAX. We experimented with several combinations of $U$ and $\beta$ and we found these ones to represent a good compromise between the total number of parameters to estimate (the number of parameters $\alpha$ grows linearly with $U$) and the model goodness of fit as measured by (penalized) log-likelihood.

\noindent
The maximum likelihood allows us to perform a quantitative comparison of the QR and QRH-I models in terms of goodness of fit, which is one of the central results of this paper. For completeness we also consider a standard Hawkes model, i.e. with no dependence on the queue state.  In Table~\ref{tab:likelihood} we report the log-likelihood values for the three models as well as the Akaike Information Criterion (AIC) score
\begin{equation}
\text{AIC} = 2k - 2L
\end{equation}
and the Schwartz information criterion (BIC) score
\begin{equation}
\text{BIC} = k\log N - 2L
\end{equation}
where $k$ is the number of parameters, $L$ is the log-likelihood and $N$ is the total sample size (number of events in our case). These scores allow one to compare nested models by their likelihood while taking into account the different number of parameters (lower score is better). By looking at the values reported in Table~\ref{tab:likelihood}, we observe that the QRH-I model has better scores in terms of AIC and BIC for both assets.
We can also use the likelihood ratio test in order to compare the models. Indeed the QRH-I model reduces to the QR model when all the $\alpha$ are set tot zero. Likewise, the QRH-I model reduces to a standard Hawkes model when, $\forall \ixa$, $\mu^\ixa(q)= \mu^\ixa$, i.e the dependence on the queue state is dropped. We report the test statistics 
\begin{equation}
\text{LR} = 2(L(\hat{\theta}_1) - L({\hat{\theta}_0}))
\end{equation}
where $\hat{\theta}_1$ and $\hat{\theta}_0$ are the maximum likelihood estimates for the null and for the alternative model respectively,
and $p$-values for the likelihood ratio test in Table~\ref{tab:lr}. We note that both the QR and the Hawkes model are rejected with a very high degree of significance when compared to the QRH-I model. 

\begin{table}
	\centering
	\begin{tabular}{lcccc}
		\toprule
		&\multicolumn{4}{c}{Bund}\\
		\midrule
		&$L$ & AIC & BIC & \# parameters\\
		\midrule
		QR & $2.046\times 10^7$ & {$-4.093\times 10^7$}& {$-4.092\times 10^7$} & 450\\
		QRH & {$2.055\times 10^8$} & {$-4.110\times 10^8$} & {$-4.110\times 10^8$} & 477\\
		\midrule
		&\multicolumn{4}{c}{DAX}\\
		\midrule
		&$L$ & AIC & BIC & \# parameters\\
		\midrule
		QR & {$7.268\times 10^5$}  & {$-1.453\times 10^6$}& {$-1.452\times 10^6$} & 75\\
		QRH & {$9.506\times 10^6$} & {$-1.901\times 10^7$} & {$-1.901\times 10^7$} & 102\\
		\bottomrule
	\end{tabular}
	\caption{Log-likelihood, AIC, and BIC values for the three considered models for Bund and DAX data.}
	\label{tab:likelihood}
\end{table}

\begin{table}
	\centering
	\begin{tabular}{lccc}
		\toprule
		\multicolumn{4}{c}{Bund}\\
		\midrule
		& Difference of log-likelihood & df & $p$-value \\
		\midrule
		$H_0 =$ QR, $H_1 = $ QRH  & $3.7\cdot 10^8$ & 27 & $< 10^{-16}$\\
		\midrule
		\multicolumn{4}{c}{DAX}\\
		\midrule
		& Difference of log-likelihood & df & $p$-value \\
		\midrule
		$H_0 =$ QR, $H_1 = $ QRH  & $1.8\cdot 10^7$ & 27 & $< 10^{-16}$\\
		\bottomrule
	\end{tabular}
	\caption{Likelihood ratio test statistic and $p$-values for the case where the null hypothesis is the QR model and for the case where the null hypothesis is a standard Hawkes model. The ``degree of freedom" (``df") value indicates  the difference in the number of parameters between the two models.}\label{tab:lr}
\end{table}

\begin{figure}[H]
	\centering
	\includegraphics[width=0.48\textwidth]{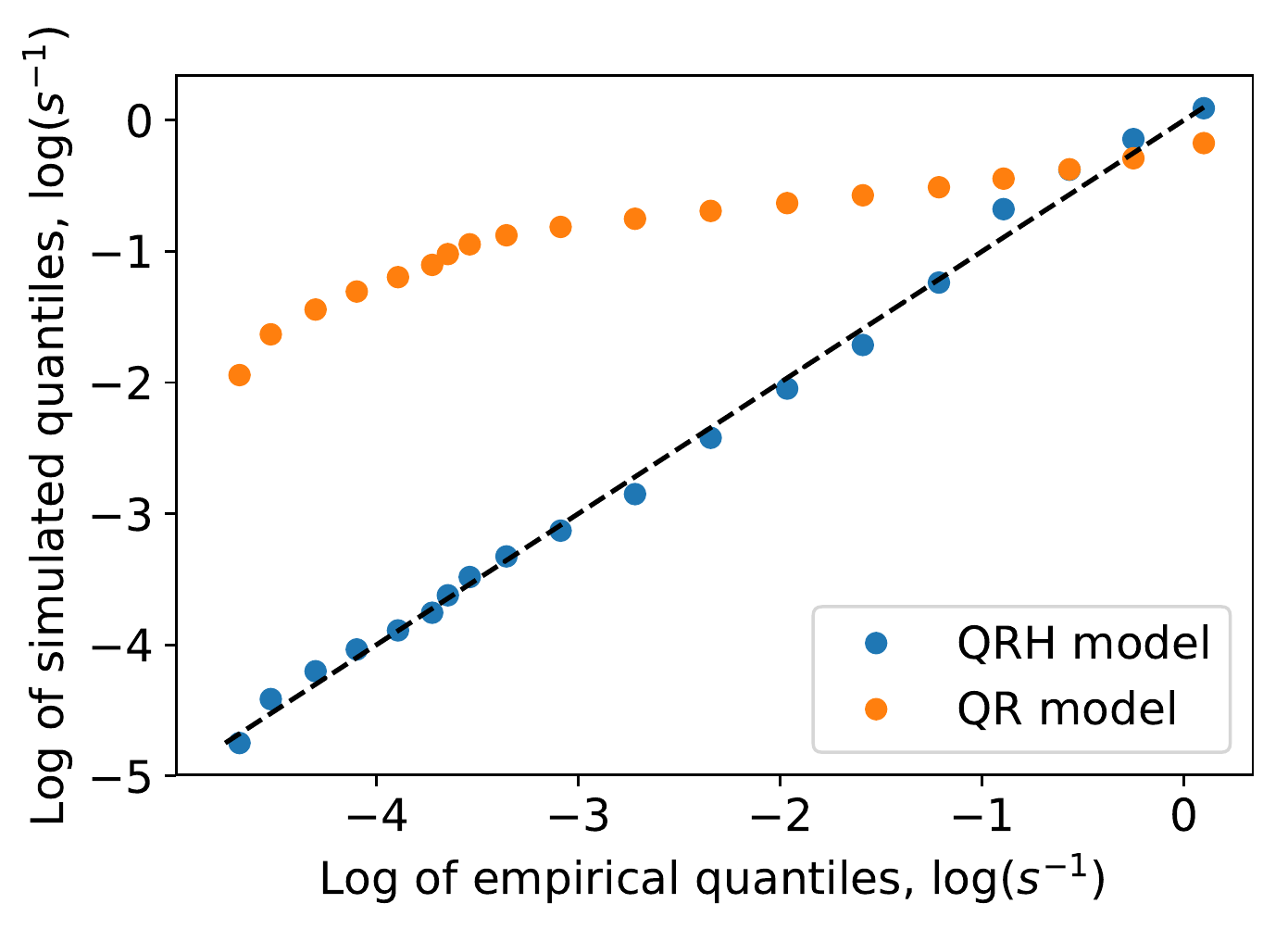}
	\includegraphics[width=0.48\textwidth]{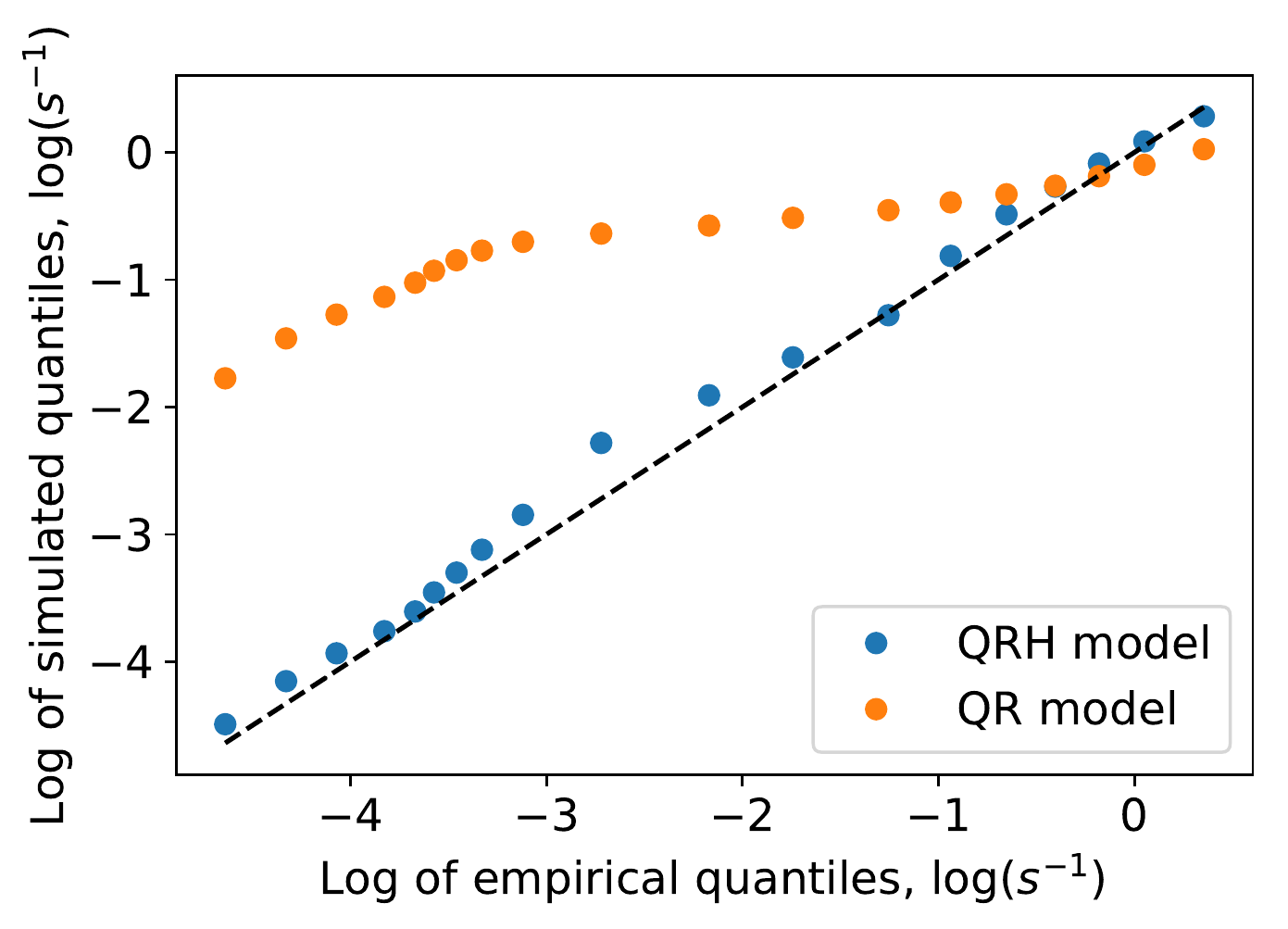}
	\caption{Log qq-plot of inter-event times. Log of quantiles of inter-events times simulated by model (horizontal) is plotted against log of empirical quantiles (vertical). Left: Bund future. Right: DAX future. }
	\label{fig:qqplot}
\end{figure}

\noindent
To complete the goodness-of-fit comparison of the models, we look at the inter-event time distribution. In particular, in Figure~\ref{fig:qqplot} we compare by means of a quantile-quantile plot the empirical inter-event times distribution with the ones produced by simulations of the calibrated QR and QRH-I models. It is clear from the figure that the QRH-I model reproduces strikingly better the empirical inter-event distribution, indicating that including the dependence on the past event is crucial in order to build a faithful model for the order flow fluctuations.

\noindent
The results presented in this section suggest that both LOB-state dependence and memory effects due to correlation in the order flow are relevant variables that need to be taken into account in order to build a faithful model for the order book dynamics. Crucially, adding a order flow dependence in the form of a Hawkes term dramatically increases the model likelihood as well as its capability of reproducing the observed inter-event time distribution.

\begin{figure}[tb]
	\centering
	\includegraphics[width=0.32\textwidth]{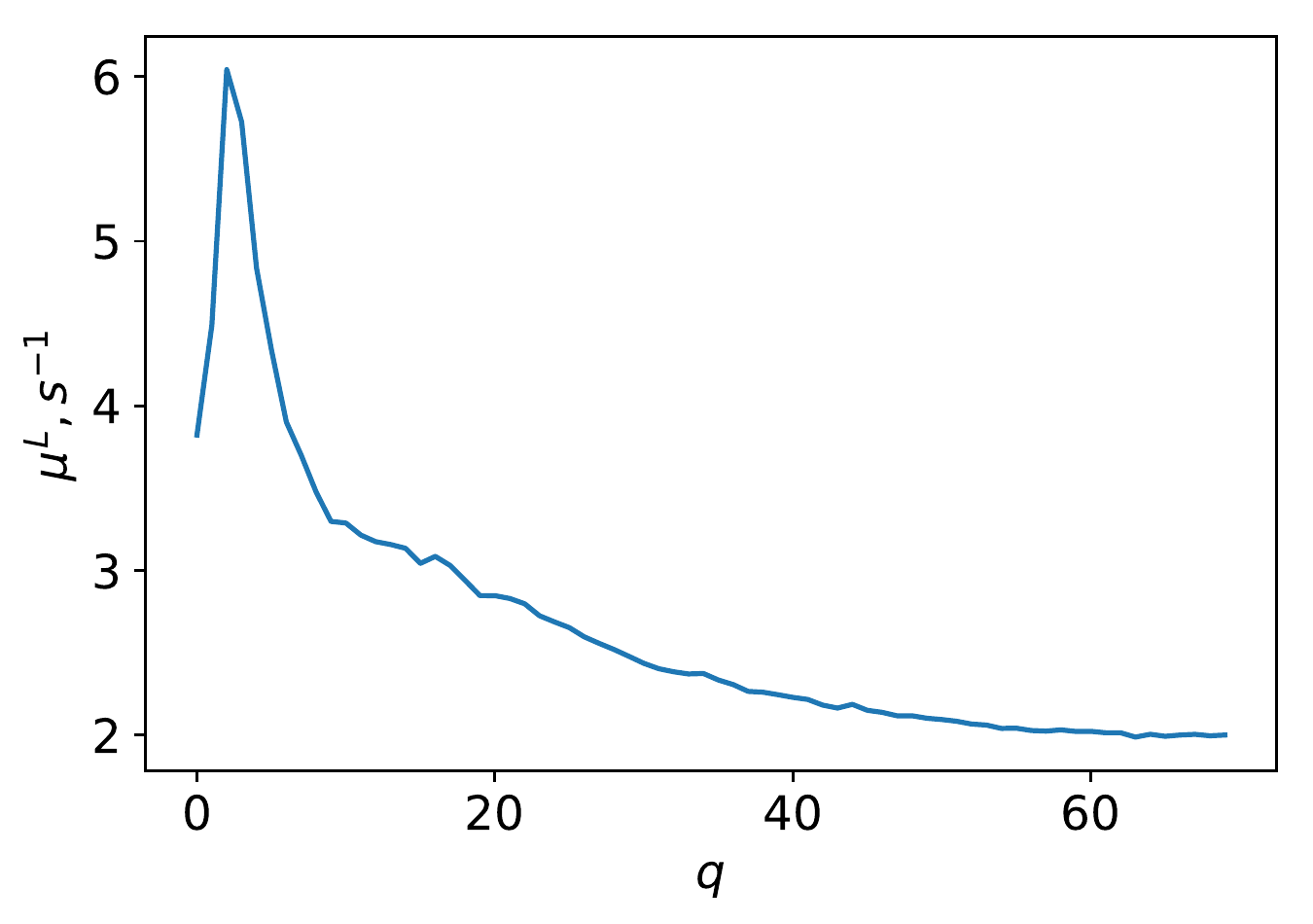}
	\includegraphics[width=0.32\textwidth]{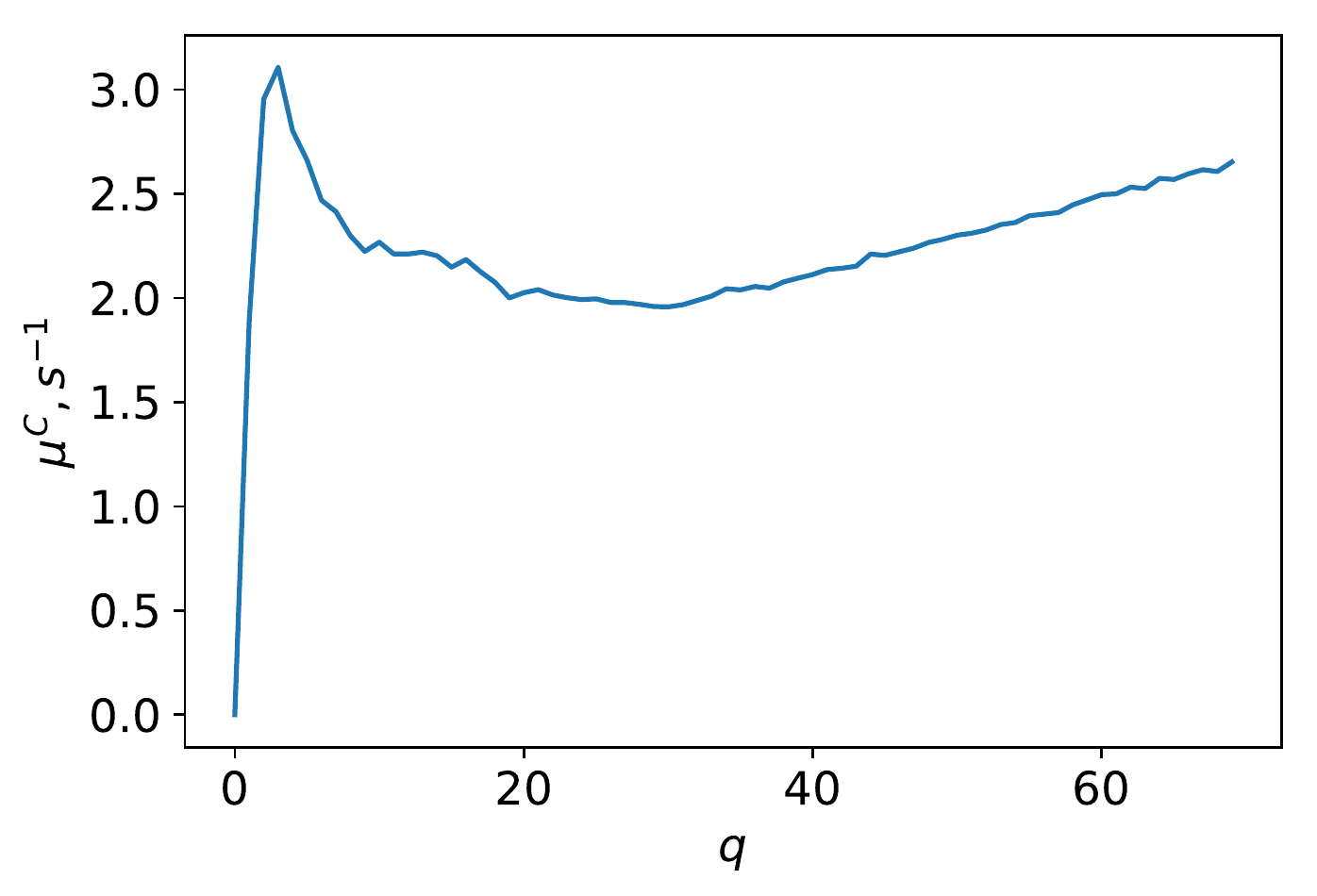}
	\includegraphics[width=0.32\textwidth]{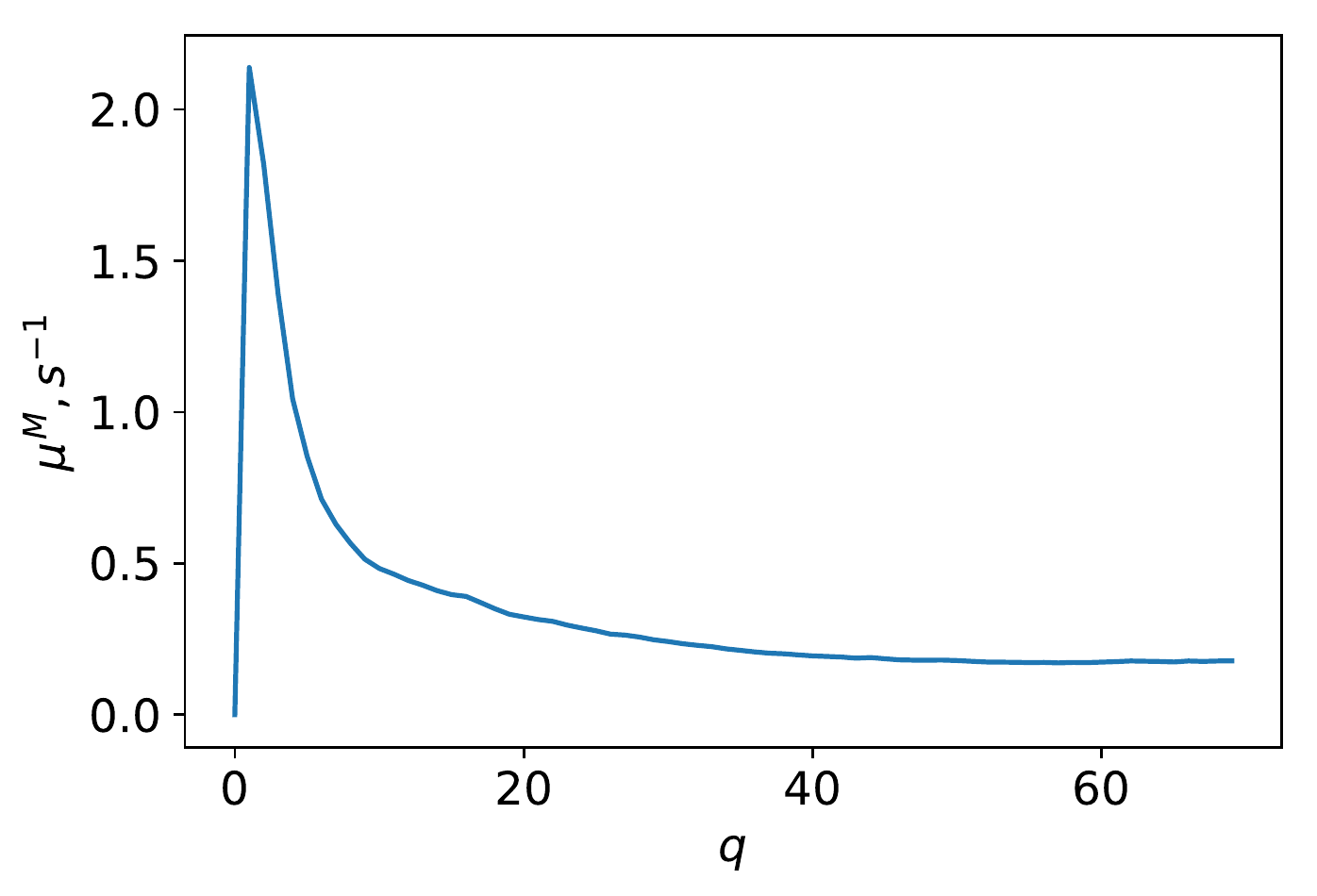}\\
	
	\includegraphics[width=0.32\textwidth]{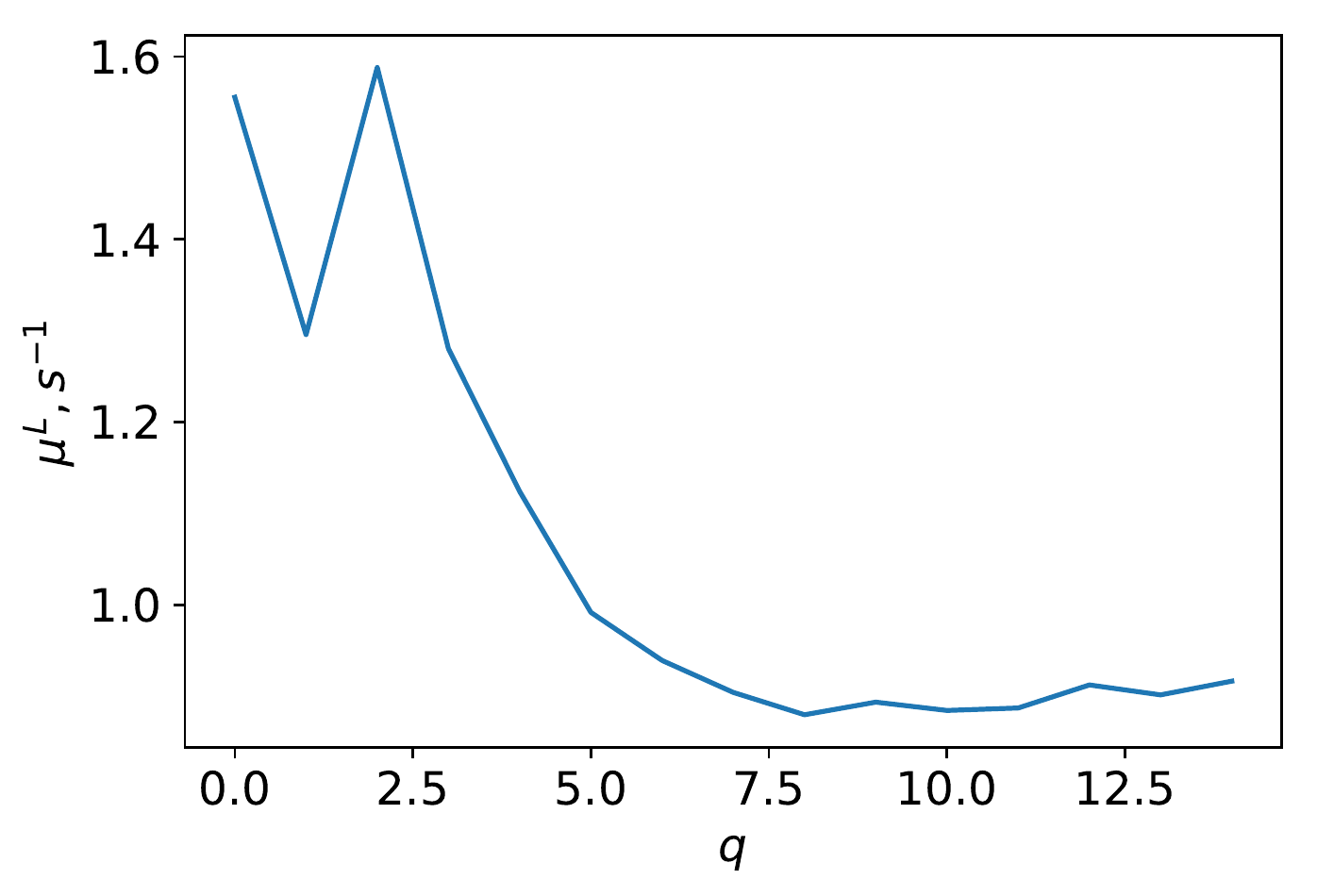}
	\includegraphics[width=0.32\textwidth]{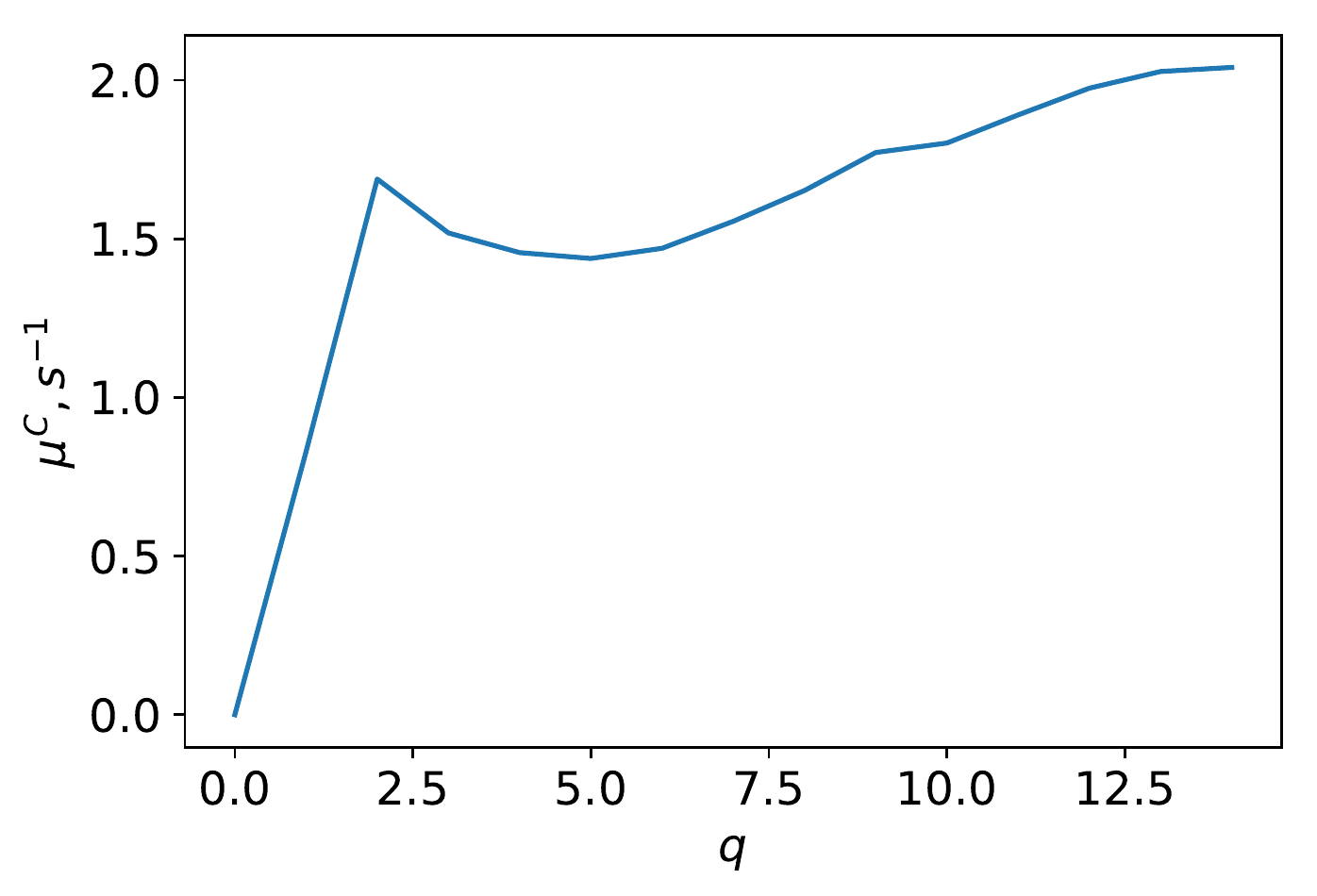}
	\includegraphics[width=0.32\textwidth]{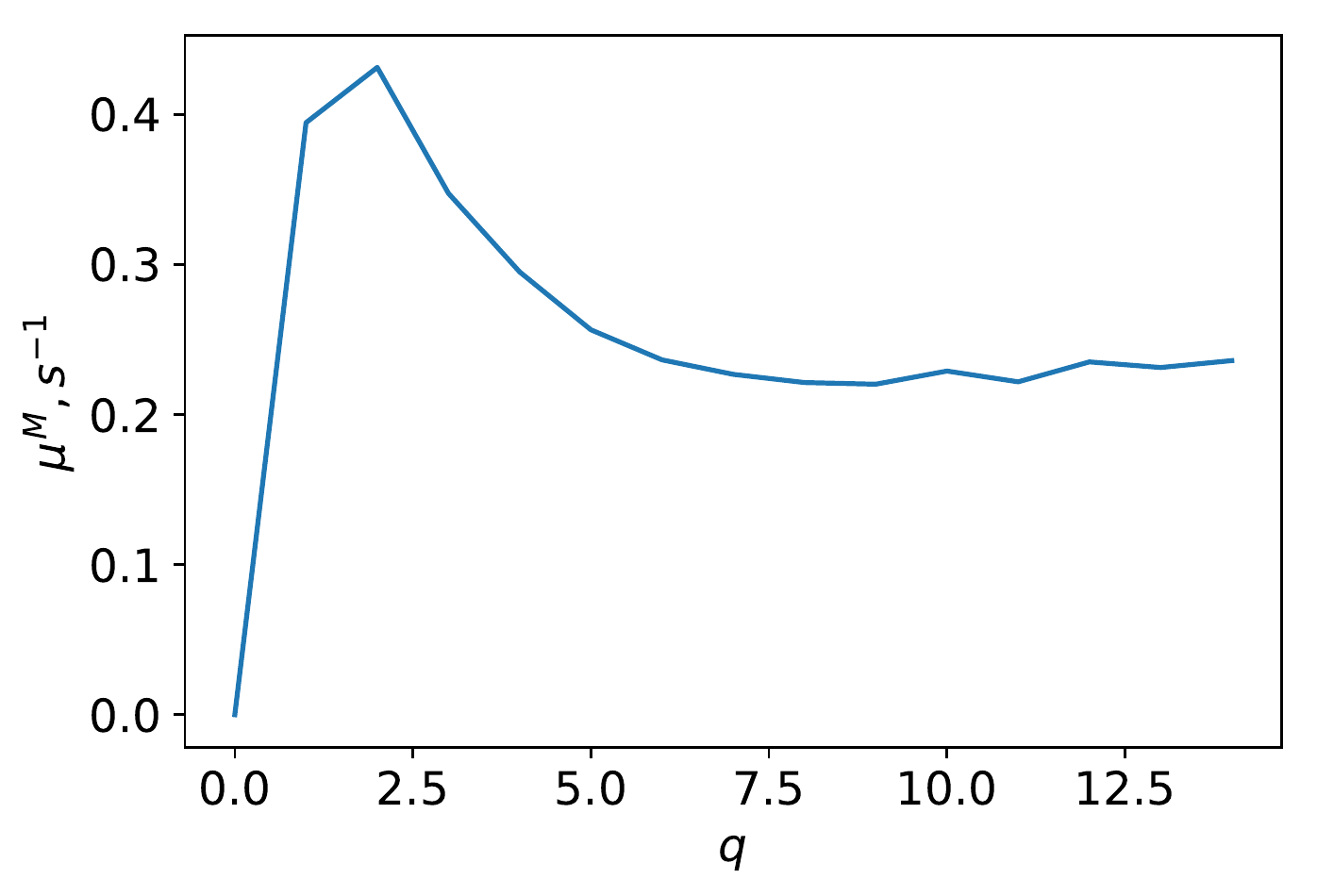}
	
	\caption{From left to right: estimated values $\mu_{q}$ for limit order insertion, limit order cancellation and market orders, QR model. Top row: Bund future. Bottom row: DAX future.}
	\label{fig:QR_mu}
\end{figure}

\begin{figure}[tb]
	\centering
	\includegraphics[width=0.32\textwidth]{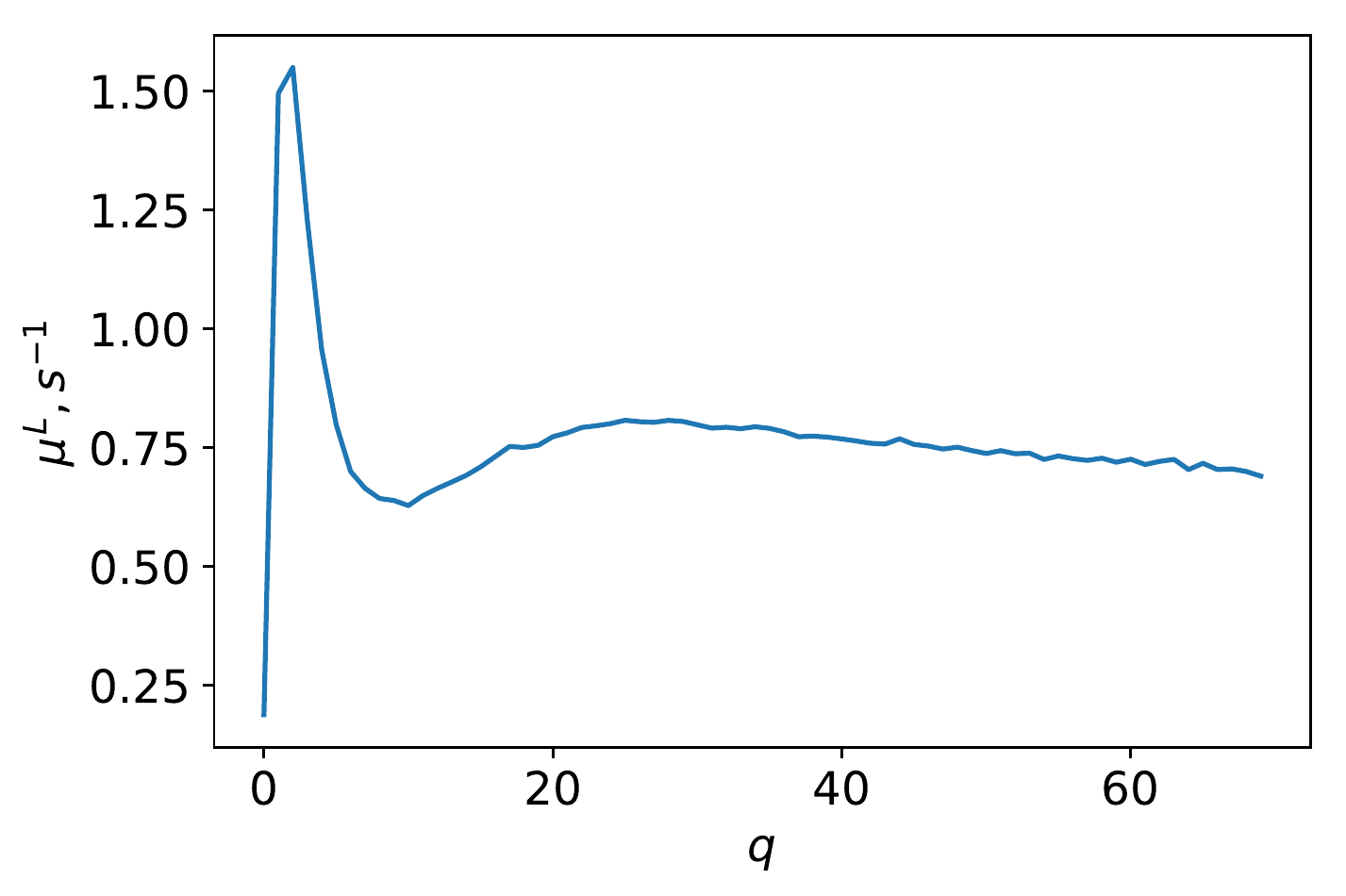}
	\includegraphics[width=0.32\textwidth]{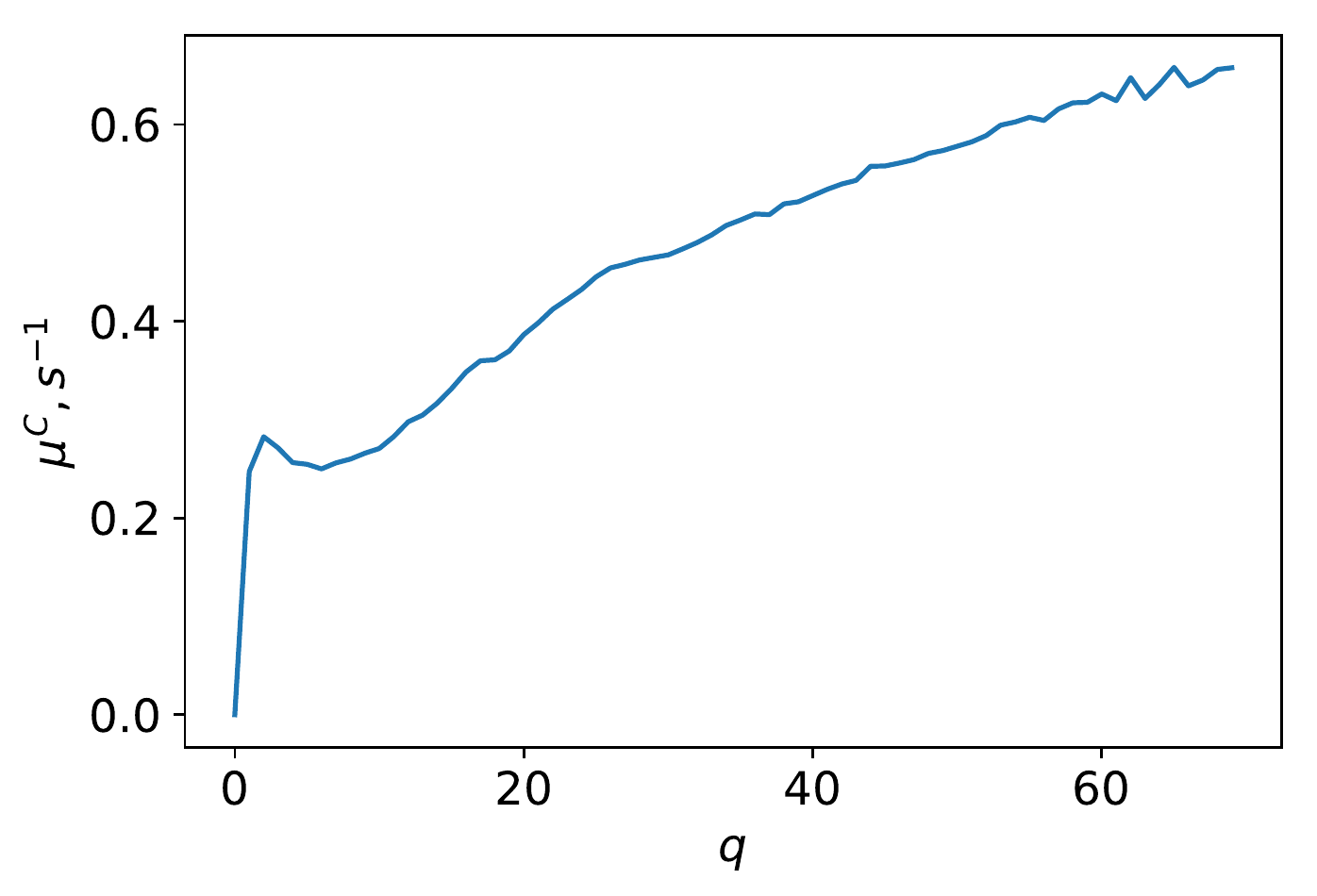}
	\includegraphics[width=0.32\textwidth]{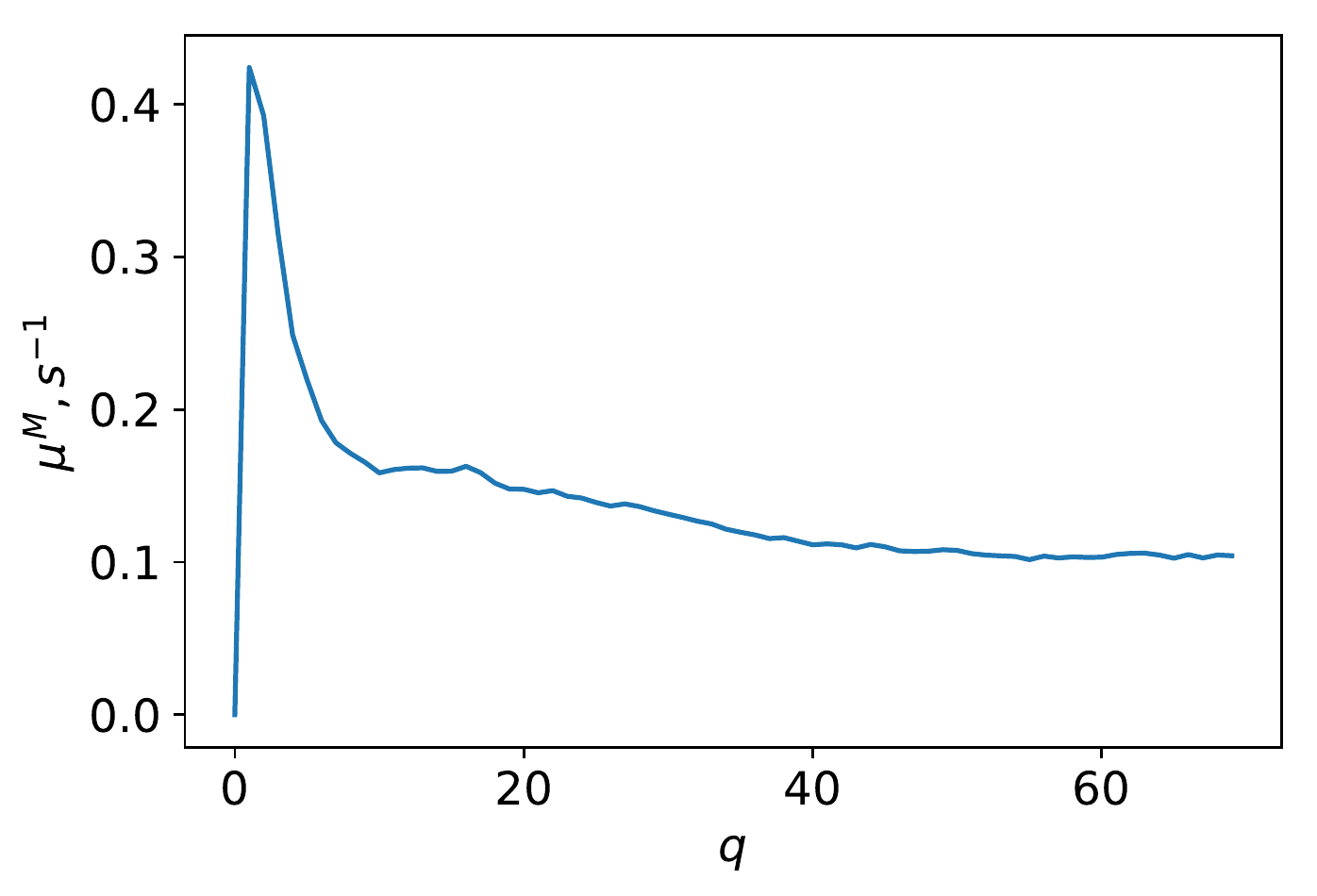}\\
	
	\includegraphics[width=0.32\textwidth]{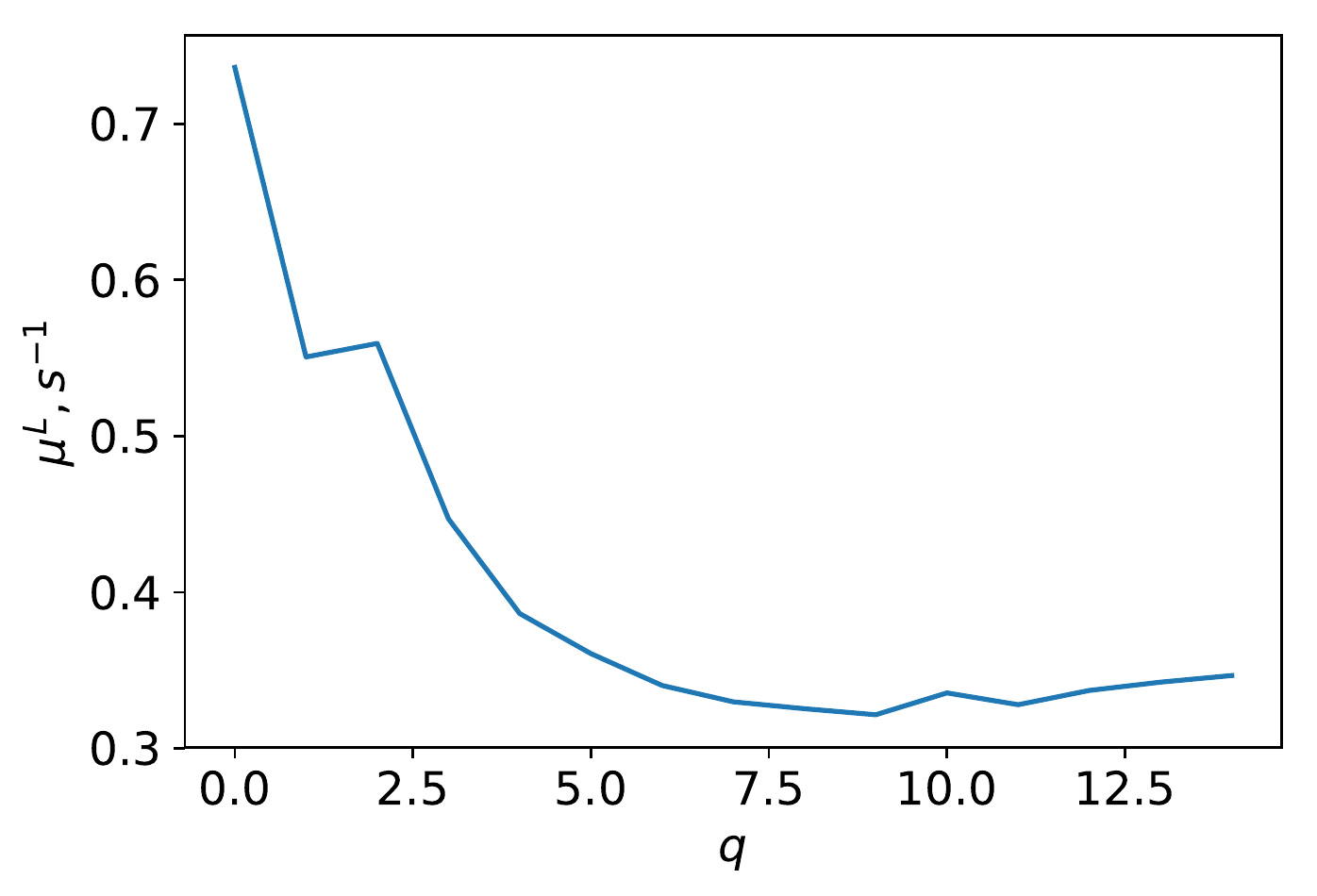}
	\includegraphics[width=0.32\textwidth]{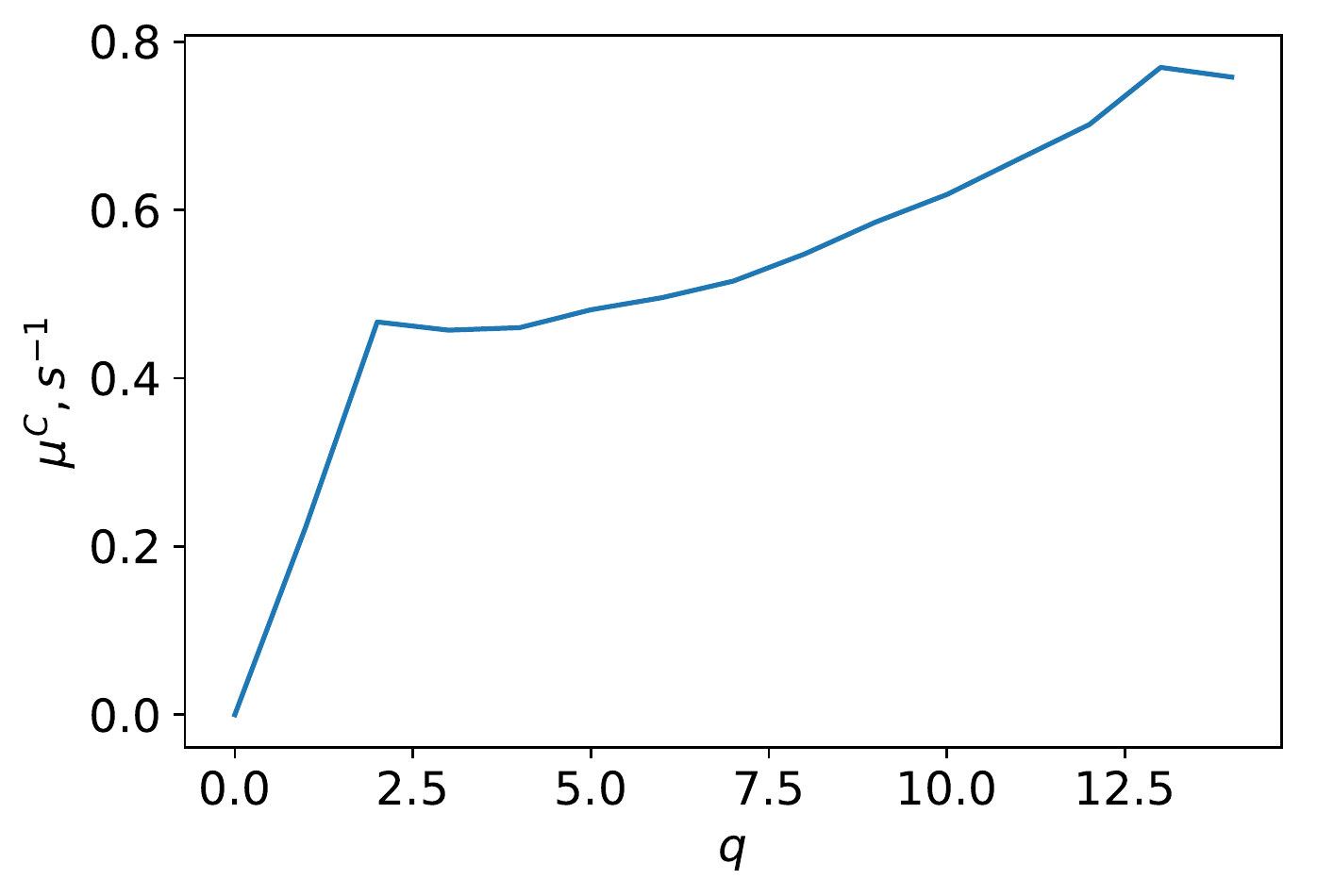}
	\includegraphics[width=0.32\textwidth]{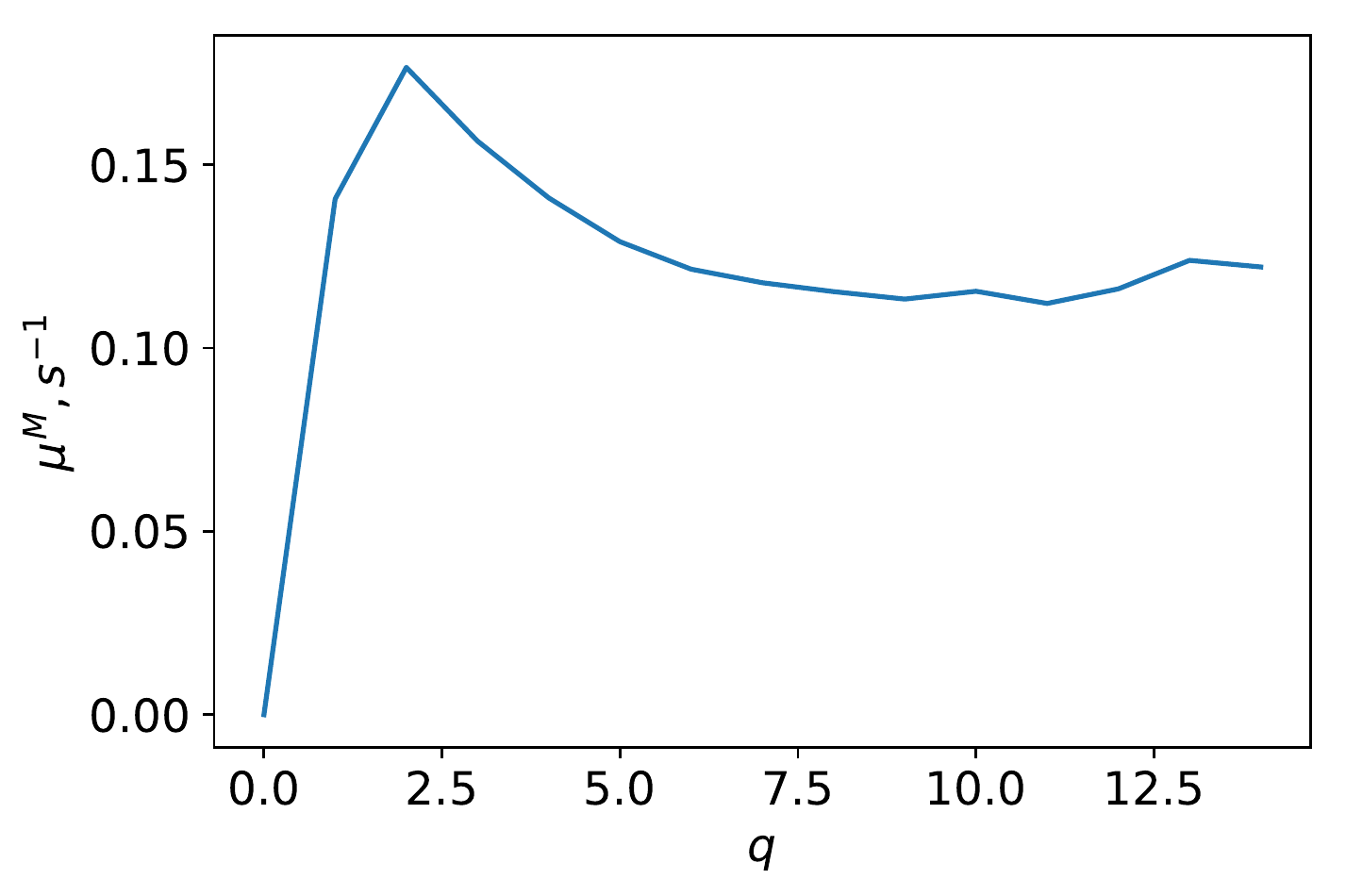}
	\caption{From left to right: estimated values $\mu_{q}$ for limit order insertion, limit order cancellation and market orders, QRH model. Top row: Bund future. Bottom row: DAX future.}
	\label{fig:QRH_mu}
\end{figure}

\paragraph{State dependency and Hawkes matrix empirical estimations}
\noindent
In Figure~\ref{fig:QR_mu} we report the estimated parameters $\mu(q)$ for the QR model, while in Figure~\ref{fig:QRH_mu} we plot the analogous quantities for the QRH-I model \eqref{eq:qrh_model}. 
We can make two general remarks while comparing these plots. First, we note that the dependence on the queue size captured by the two models are roughly concordant, in that functions $\mu^\ixa(q)$ have similar shapes in both models.
However let us remark that the values estimated within the QRH-I model are much smaller, indicating that a large part of the intensity is now explained by the self- and cross-exciting Hawkes components (see the discussion below). 
One can notice some differences between the Bund and DAX results, most likely stemming from the different order book dynamics typical of large and small tick assets respectively.
As in \cite{rsb1}, we observe a decreasing rate of Market orders arrivals 
as the queue size increases. This can be explained
by the fact that agents tend to consume liquidity faster as this liquidity becomes rare.
We also find that, when, $q(t)$ is large enough, 
the rate of cancellation is an increasing function of the queue size. This is an expected feature assumed in most former LOB models (see e.g., \cite{smith2003,cont2010}), since cancellations 
are more likely to occur when they are many active limit orders. As shown in Appendix \ref{sec:app1}, this behavior ensures the ergodicity of the queue process.
Let us finally notice, that unlike the observed behavior in \cite{rsb1} on specific stocks, 
we don't observe that the intensity of limit order insertion is almost independent of the queue size. 
It is rather a decreasing function of the queue size probably reflecting a lesser quest for priority 
when $q$ is large.

\noindent
It is also interesting to look at the quantity
\begin{equation}\label{eq:endo_fraction}
e^\ixa(q) = 1 - \frac{\mu^\ixa(q)}{\Lambda^\ixa(q)} 
\end{equation}
where 
\begin{equation}
\label{eq:Lambdal}
\Lambda^\ixa(q) = \E{\lambda^\ixa(t) |q(t^-) = q}
\end{equation}
is the average intensity in a given state $q$. $e^\ixa(q)$
corresponds to the fraction of the total average intensity explained by the endogenous self- and cross-exciting mechanism as a function of the queue size $q$,  While $\Lambda^\ixa(q)$, in the case of the QR model, is directly provided by the parameter $\mu^\ixa(q)$, for the QRH-I model it is given by the contribution of both the baseline intensity $\mu^\ixa(q)$ and the Hawkes interactions. Unlike standard multivariate Hawkes processes, the QRH-I model does not admit a closed form formula of $\Lambda^\ixa(q)$ from its parameters. Therefore, while $e^\ixa(q)$ is trivially zero for the QR model, we resort to numerical computation of $\Lambda^\ixa(q)$ in order to compute $e^\ell(q)$ for the QRH-I model.

\begin{figure}[tb]
	\centering
	\includegraphics[width=0.32\textwidth]{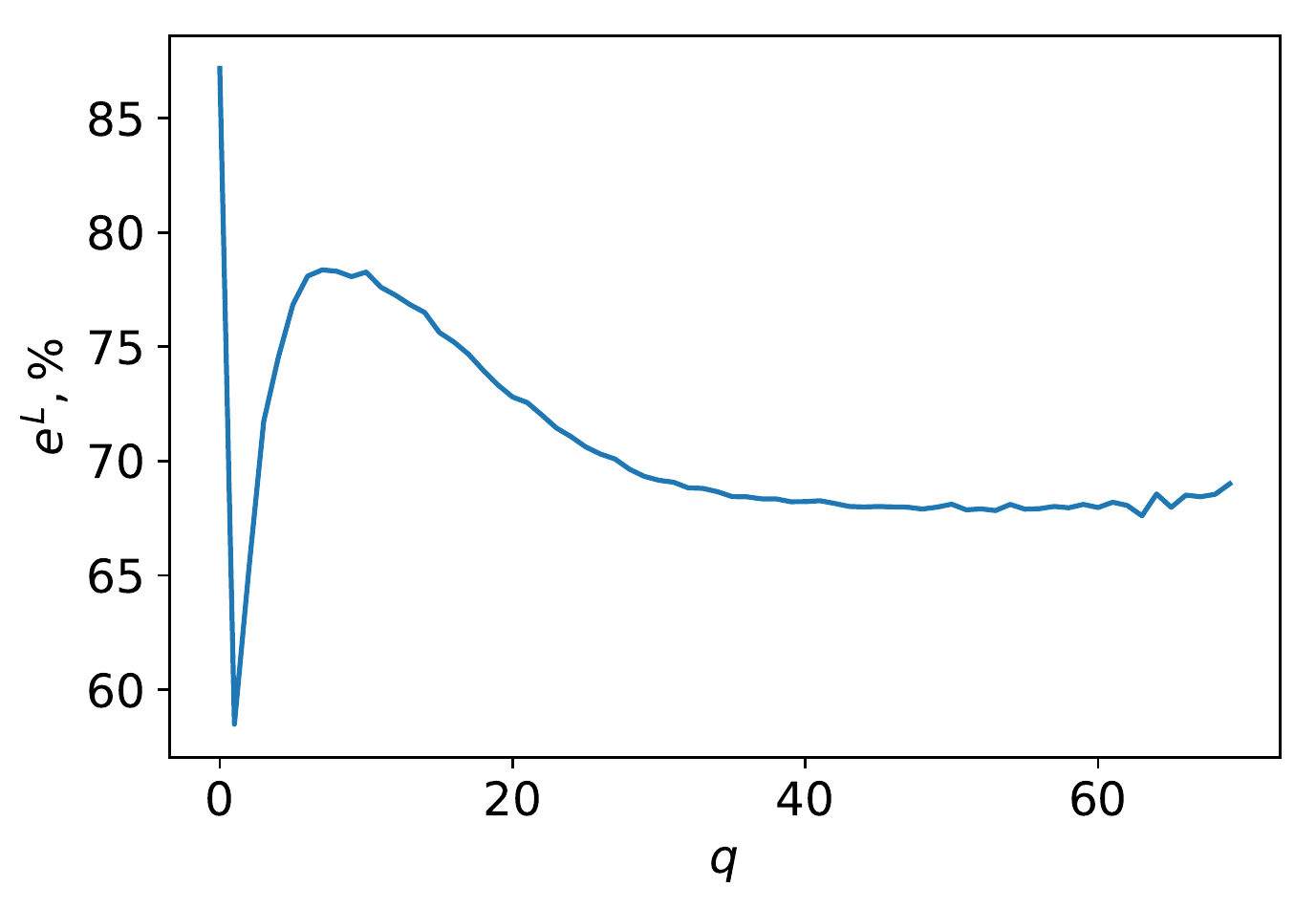}
	\includegraphics[width=0.32\textwidth]{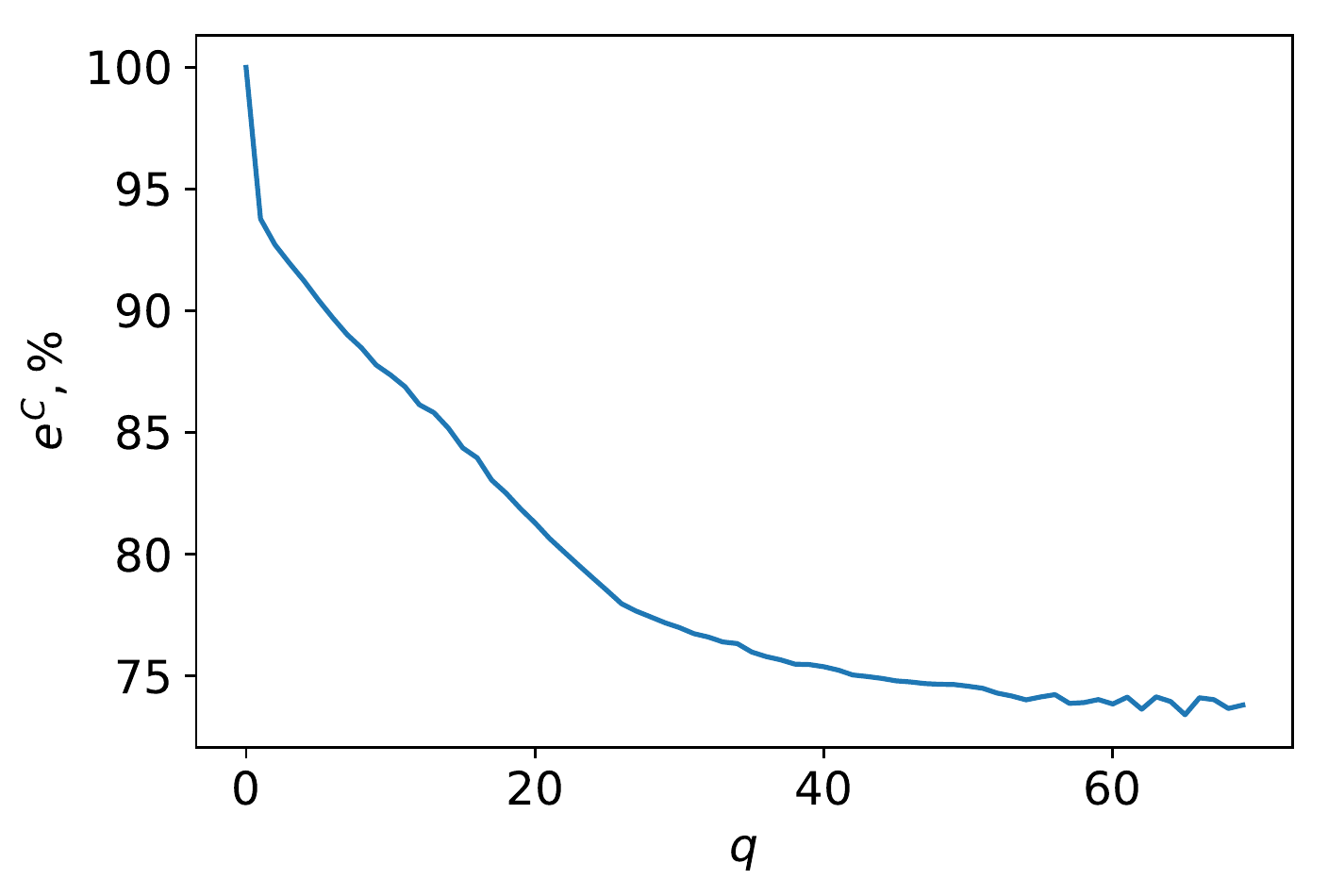}
	\includegraphics[width=0.32\textwidth]{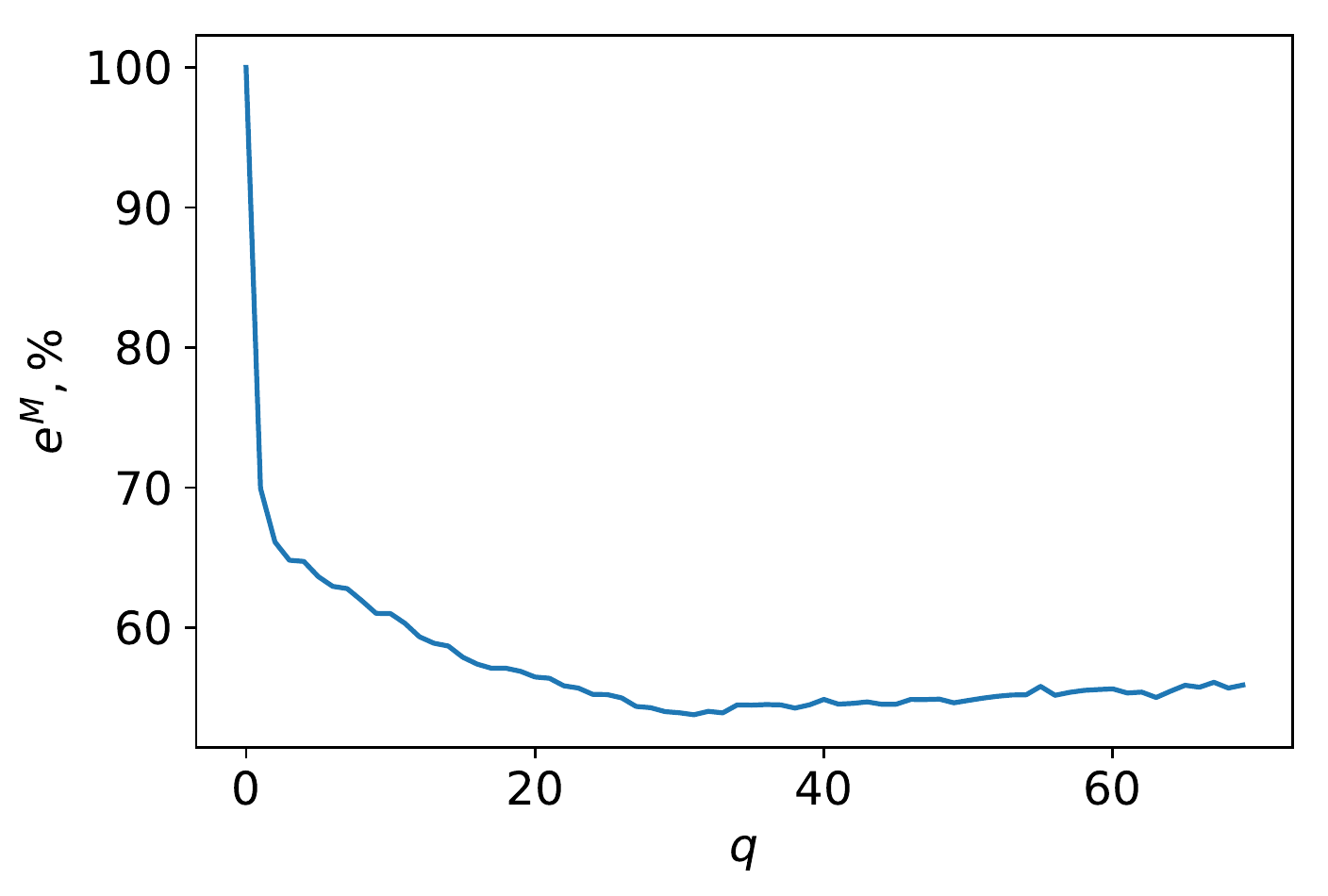}\\
	
	\includegraphics[width=0.32\textwidth]{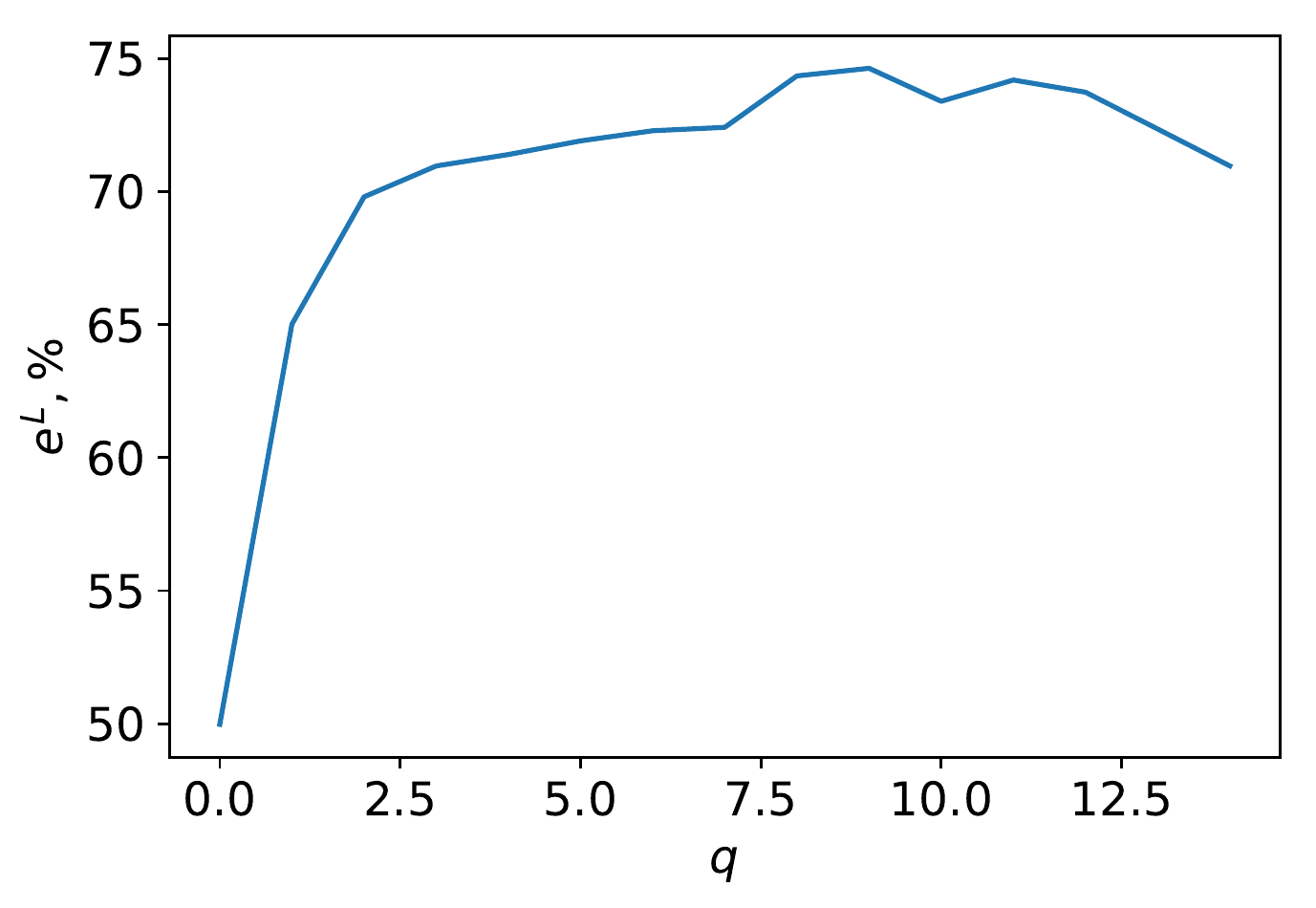}
	\includegraphics[width=0.32\textwidth]{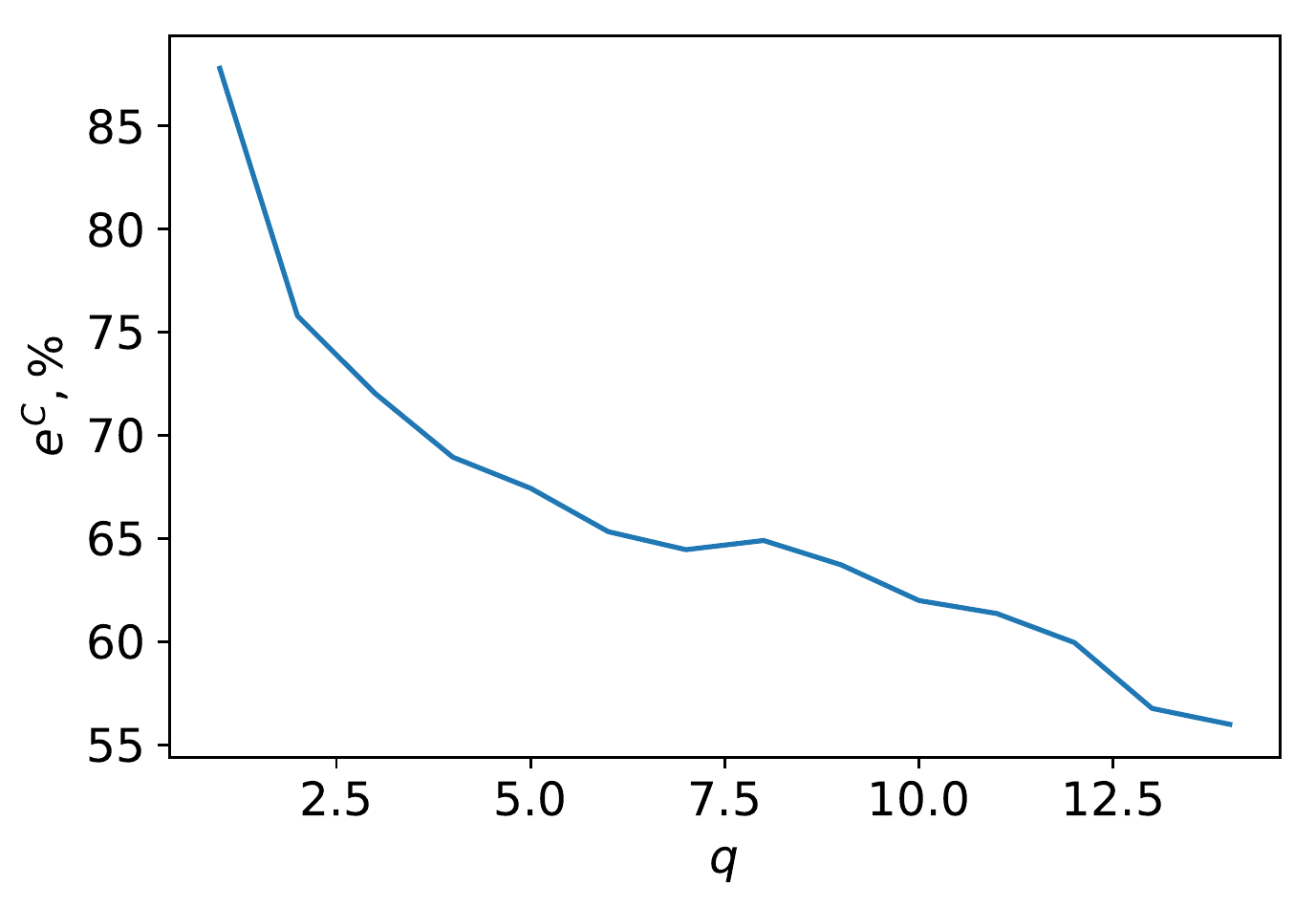}
	\includegraphics[width=0.32\textwidth]{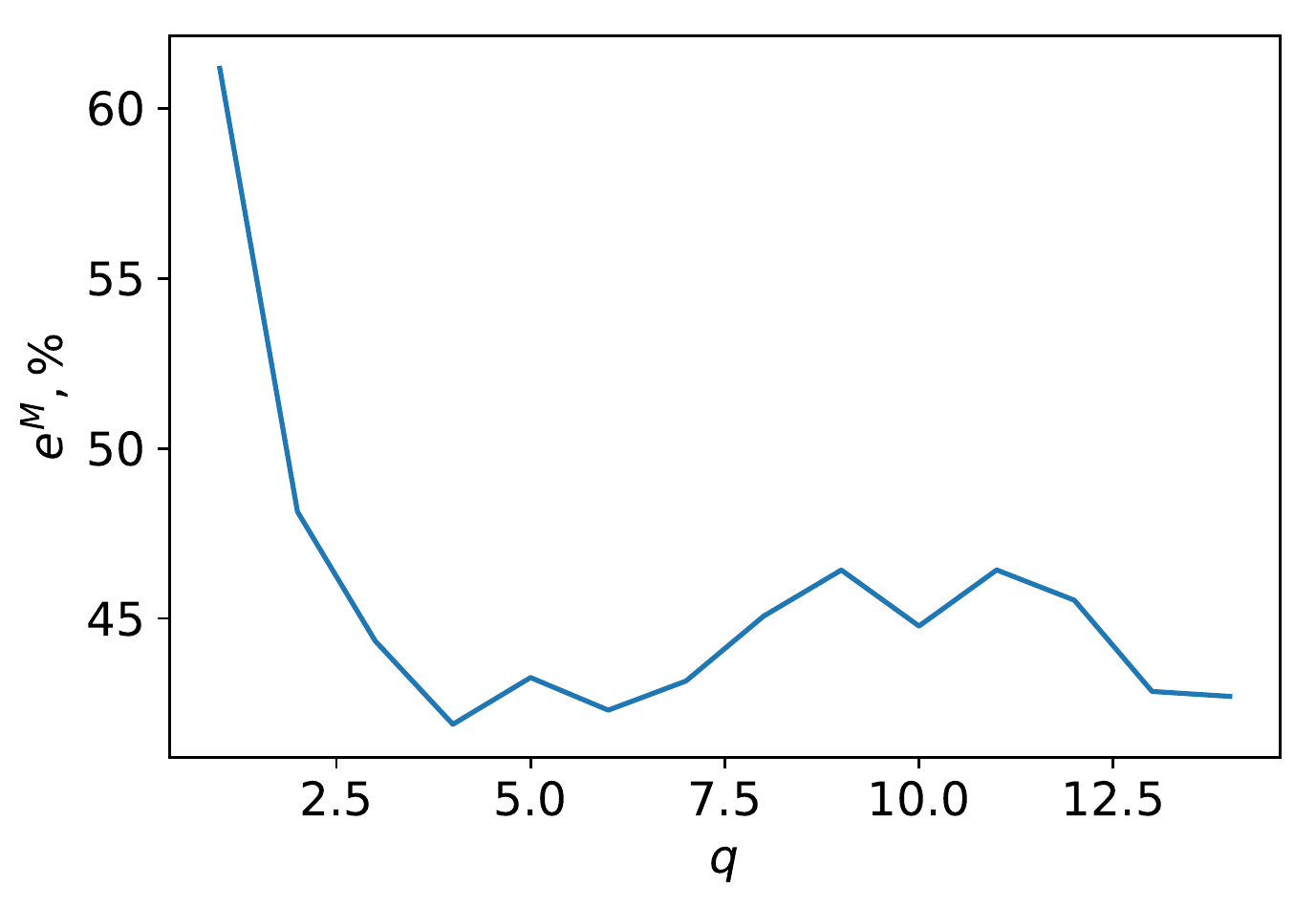}
	\caption{From left to right: Endogenous fraction for limit order insertion, $e^\ixa(q)$ as defined in Eq. \eqref{eq:endo_fraction}), for limit  ($\ixa = L$), cancellation ($\ixa = C$) and market ($\ixa = M$) orders by QRH model. Top row: Bund future. Bottom row: DAX future.}
	\label{fig:Endo_frac}
\end{figure}

\begin{figure}[tb]
	\centering
	\includegraphics[width=0.48\textwidth]{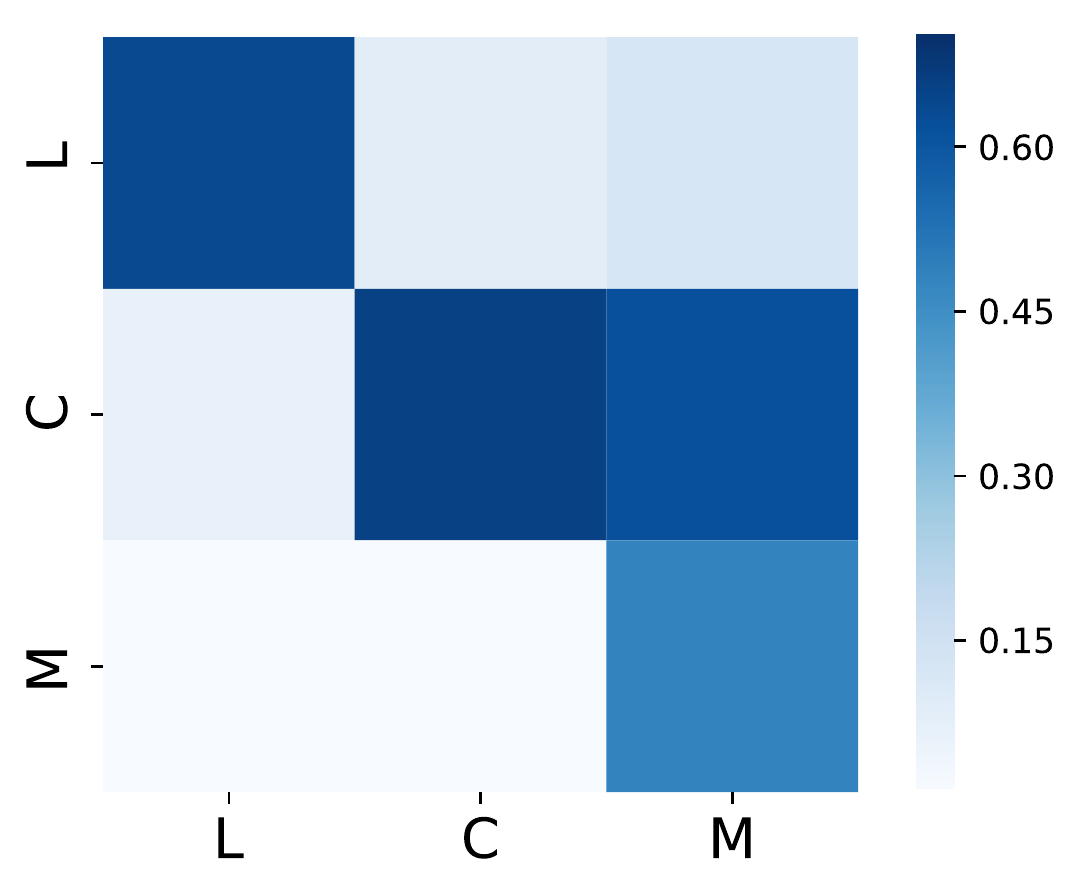}
	\includegraphics[width=0.48\textwidth]{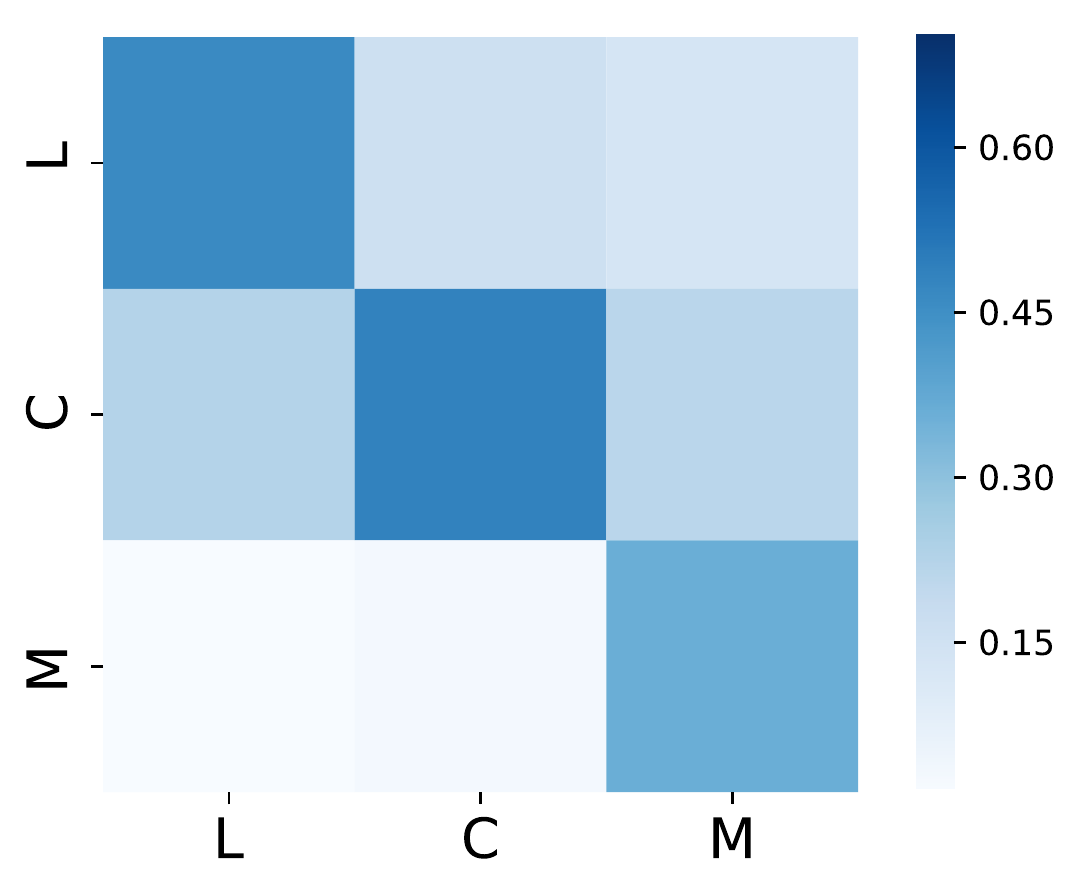}
	\caption{Kernel norm matrices for Bund future (left) and DAX future (right). }
	\label{fig:kernel_norm_plt_dim1}
\end{figure}

\noindent
The result are shown in Figure~\ref{fig:Endo_frac}, where we have plotted the estimated $e^\ixa(q)$ for all types of orders and for both the Bund (top panels) and the DAX (bottom panels) futures.
Overall we see that a large part, from $60\%$ to $80\%$, of the total average intensity is explained by the self- and cross-exciting effect. We note that for cancel and market orders, the intensity is maximally explained by the Hawkes term when the queue is small. 
This is likely to result from the persistence in the order flow and in particular of the relevance of the self-exciting term as we noted in Figure~\ref{fig:kernel_norm_plt_dim1}, which in case of market and cancel orders tends to empty the queue. This explanation is corroborated by the observation that the opposite effect is found for limit orders, namely an higher endogeneity for higher values of $q$.  

\noindent
To complete the analysis of the QRH model results, in Figure~\ref{fig:kernel_norm_plt_dim1} we plot in a color map the Hawkes kernel norms $|\phi^{\ixa \ixb}| = \int_0^\infty \phi^{\ixa \ixb} (t) dt$. Note that in our setting these quantities are simply given by $|\phi^{\ixa \ixb}| = \sum_{u=1}^U \alpha^{\ixa \ixb}_u$. As discussed in \cite{reviewHawkes}, these quantities represent the average direct effect of an event of type $\ixb$ (columns) over the intensity of type $\ixa$ (rows) events. Hawkes kernel matrices of order book events have been extensively studied in \cite{thibault1,rambaldi2017role}. Here, we note that despite the addition of the queue-dependent term, we recover many of the features already observed in previous studies, such as the strong diagonal component corresponding to self-excitation, likely the result of correlation in the order flow induced by order splitting strategies. 
We also confirm that market orders influence liquidity much more than the opposite effect. In particular, since here we look at interaction on the same side of the book, we note that market order have on average an exciting effect on cancellations. As observed in the aforementioned studies, a flow of market order at a given price signals that the ``true'' price is closer to that side and therefore liquidity adapts, with outstanding orders being canceled in order not to be adversely selected. 

\paragraph{Equilibrium and empirical queue size distributions}
In Ref.~\cite{rsb1}, the authors emphasized that the QR model provides a simple framework to account for the observed
queue size distributions in the order book. For that purpose they have shown that the model invariant distribution fits 
quite well the empirical laws notably at the first bid/ask levels. In Appendix \ref{sec:app1}, we show
that, under some conditions that appear to be empirically fulfilled, the QRH-I model is also an ergodic process and the queue size can thus be described by its invariant distribution.
Before comparing the performances of QR and QRH-I models with respect to their prediction of the 
equilibrium queue size distribution, let us emphasize that some caution is needed when addressing this issue. Indeed, this distribution, even
if reached exponentially fast, does necessarily correspond to the empirically observed queue law since when the queue
is empty, the reference price has a non-vanishing probability to change. 
This directly implies that for small values of the queue size, the invariant distribution is not supposed
to account for the observed values from snapshots of the empirical book state. Moreover,
since the initial queue size 
has no reason to be drawn with the invariant distribution, this law is pertinent
only after a short delay that has to be compared to the length of each realization, i.e., the time period between two changes of reference price. The exponential rate involved in the ergodic theorem is however hard to estimate, one can use a proxy as given for example by the exponential decay of empirical queue size autocorrelation function. That is an alternative measure of a mixing coefficient that can be, under some conditions, related to the distribution relaxation time \cite{bradley2005}.
If one assumes that the decay of autocorrelation of the queue size takes the form $\rho(t) = a \exp^{-t/\tau_c}$, we find empirically that
$\tau_c \simeq 15$ $s$ for the Bund future and $\tau_c \simeq 2$ $s$ for the DAX. 
For both assets these correlation characteristic scales have to be compared with the average realization length, namely $\tau_m \simeq 100$ s for the Bund and $\tau_m \simeq 16s$ for the DAX. 
Since in both cases we have $\tau_c \ll \tau_m$,
it is likely that the invariant distribution is pertinent to account for the queue size distribution 
as observed at randomly chosen times.

\begin{figure}[H] 
	\centering\includegraphics[width=0.48\textwidth]{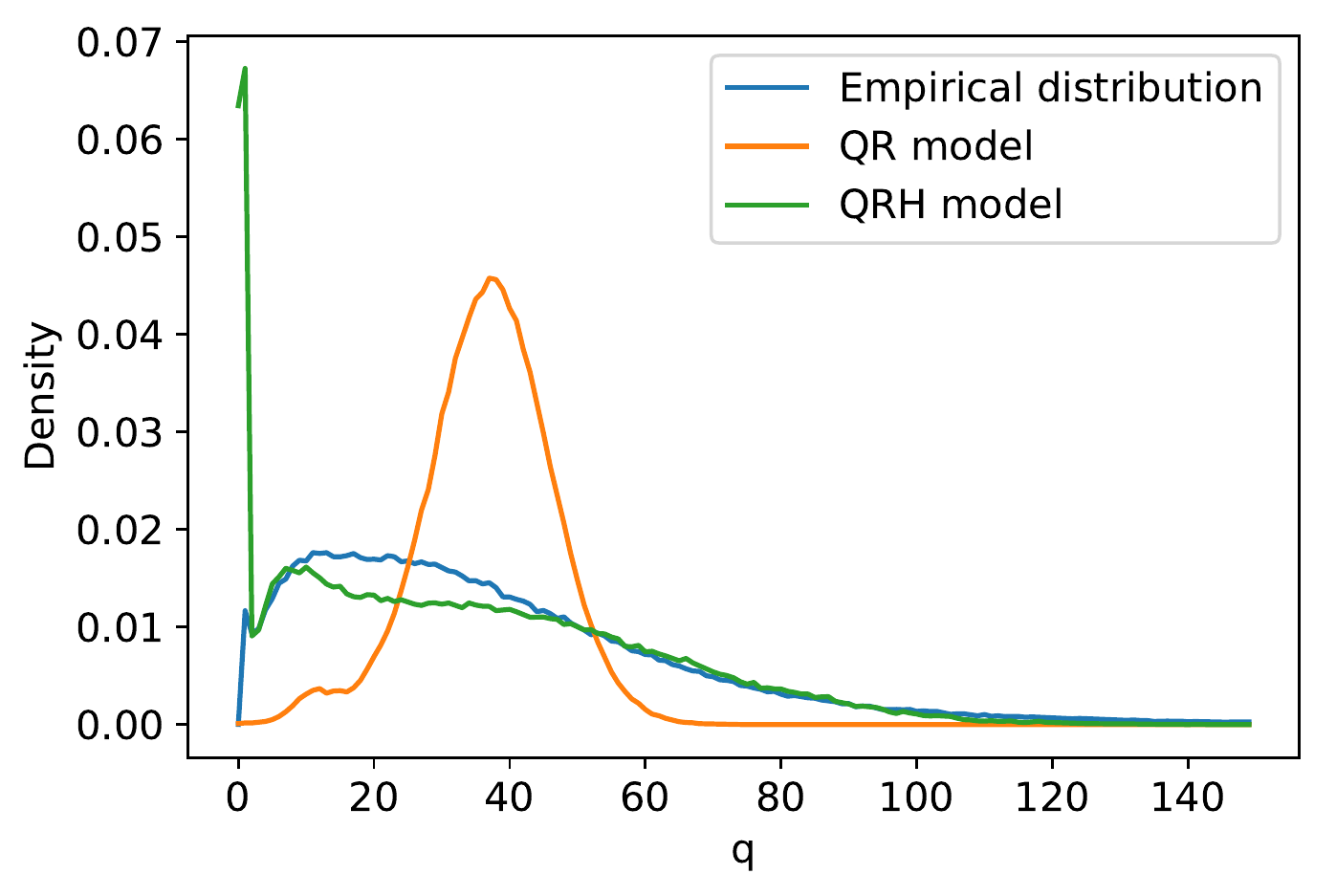} 
	\centering\includegraphics[width=0.48\textwidth]{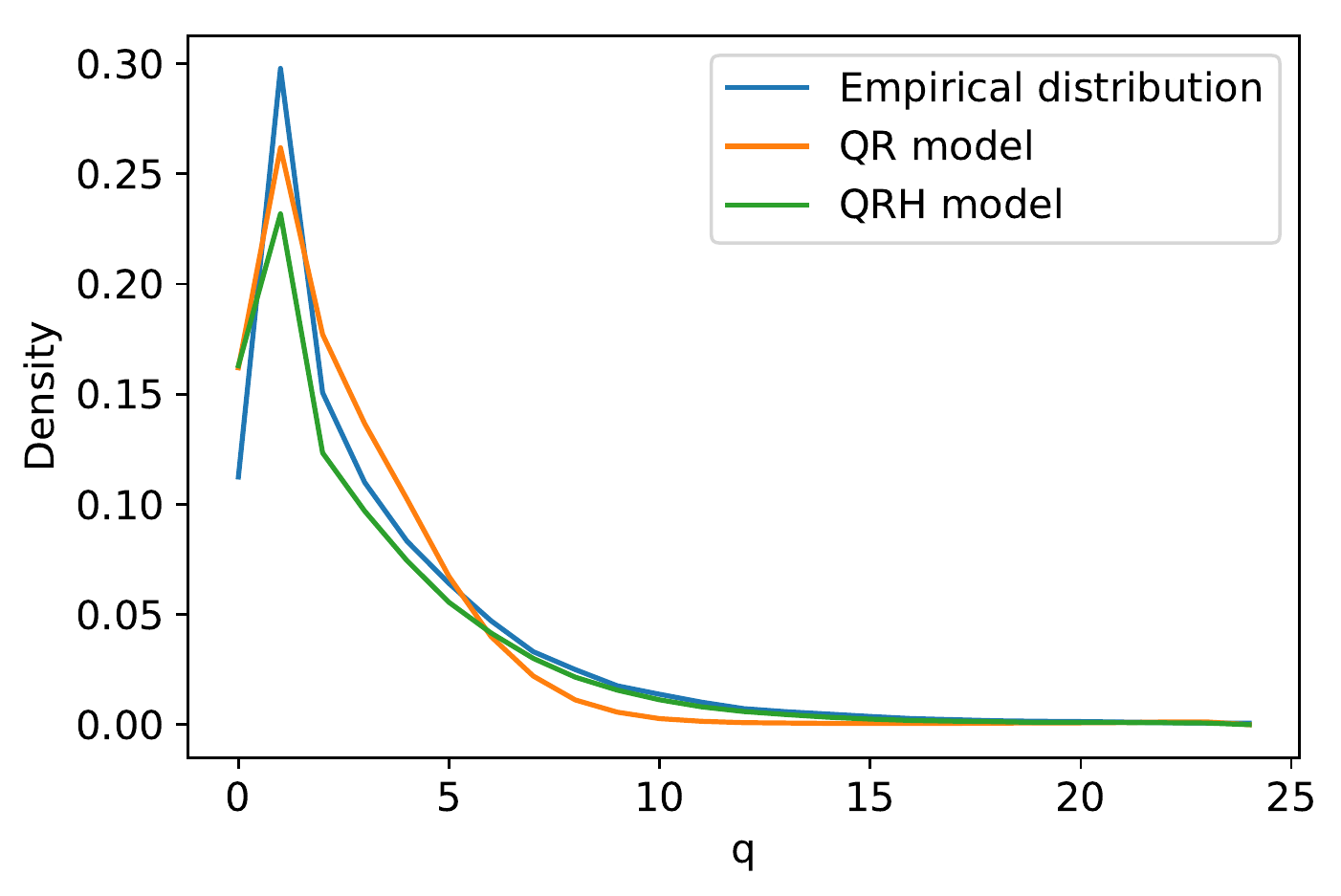} 
	\caption{Comparison of the invariant distributions of the QR and QRH models with the empirical one. Left: Bund future. Right: DAX future.}
	\label{fig:inv_dist}
\end{figure}

\noindent
With previous observations in mind, we now proceed at the comparison of the empirical queue distribution, measured by taking snapshots of the book every 30s and the invariant distributions produced by the QR and QRH-I models. 
Since we do not have any explicit formula for the QRH-I model, we 
estimate the invariant distribution of $q(t)$ by 
performing a simulation over a long time period. The invariant measure of the the QR model can be directly 
deduced from the estimations of $\mu^\ixa(q)$ in Figure \ref{fig:QR_mu} using the analytical formula in Sec. 2.3.3 of Ref. \cite{rsb1}.
The plots the both QR and QRH-I invariant measures together with the empirical queue size distribution for both Bund and DAX are reported in Figure~\ref{fig:inv_dist}. First of all, we observe that the QRH-I model provides in both cases
of better fit of empirical data, notably in the tail region, than the QR model. The latter is particularly far from
the observed distribution in the large tick case of the Bund data. Its performance for the smaller tick asset (DAX)
are closer to the results reported in \cite{rsb1} for stock data. 
Beyond the fact that this striking difference between large and small tick assets is hard to explain (though the analytical formula in \cite{rsb1} shows that the overall shape of the distribution can vary quite drastically when on changes the respective behavior of the $\mu^\ixa(q)$ functions)
our findings show that accounting for the Hawkes self-interaction within a queue reactive model 
is important not only to describe correctly the order flow dynamics but also provides a better model 
for the queue size distributions.

\noindent
We notice furthermore that the distribution of queue size simulated by QRH-I model deviates from the empirical distribution especially around states when the queue is small. As discussed previously, when
the queue is empty, the reference price has a large probability to change and therefore the corresponding empirical sample to stop. This results in a statistical bias 
with respect to the ``theoretical model" where such event type does not exist since when the 
queues empty nothing happens until a new limit order arrives. The states with a low queue values are therefore 
more likely to be visited in the model than in empirical observations.

\section{A Queue Reactive Hawkes model for the best limits of the orderbook}
\label{sec:part2}

In the previous Section, we presented a model for a single best limit with a fixed price, i.e., the model is reset when the price of the best limit changes. Thus, it does not account for the cross dynamic between the two best limits (best bid/best ask) nor it accounts for the changes of their corresponding price. This Section is devoted to a model that tackles these drawbacks. Both best limits are modeled along with mid-price changes. Hence, contrarily to the previous model (and also to the models presented in \cite{rsb1}), resetting of the model only occurs at market close time. 
This new model is built by adding a queue-dependency to the model of Bacry et al. \cite{thibault1}. The model of the previous section was referred to as QRH-I (one-side LOB modeling), this new model will be referred to as QRH-II (two-side LOB modeling).

\subsection{The QRH-II model : Adding queue-dependence to Bacry et al.'s model}
In \cite{thibault1} the authors consider eight event types at the best level of a LOB, namely
\begin{itemize}
	\item[] $P^+$ ($P^-$) for events that moves the midprice\footnote{We recall that the midprice corresponds to the average of the best ask price with the best bid price.} up (down) independently on the size of this move,
	\item[] $L^a$ ($L^b$) for limit orders at the best ask (bid) that do not change the midprice,
	\item[] $C^a$ ($C^b$) for cancellations at the best ask (bid) that do not change the midprice,
	\item[] $M^a$ ($M^b$) for market orders at the best ask (bid) that do not change the midprice,
\end{itemize}
where, for any $\ixa \in \{P^+$, $P^-$, $L^a$, $L^b$, $C^a$, $C^b$, $M^a$, $M^b \}$, the best ask (resp. bid) level refers to the first non empty level on the ask (resp. bid) side. Let
us define $N^{\ixa}_t$ as the counting
process associated with events to type $\ixa$ and $\lambda^\ixa(t)$ the associated conditional intensity.
Let us point out that that this description of the first limits of the LOB is significantly different from the approach taken in the previous section or in \cite{rsb1} in which the best limit queue size can be 0. Also, this model no longer needs to be reset regularly and does not use any notion of reference price $\refpr{}$.

\noindent
The authors in \cite{thibault1} consider a multivariate Hawkes model where each event type can influence, and be influenced by, the others so that the conditional intensities read:
\begin{equation}\label{eq:thibault}
\lambda^\ixa(t) = \mu^\ixa + \sum_{\ixb} \int_0^t \phi^{\ixa \ixb}(t-s) dN^\ixb_s,
\end{equation}
where $\ixa$ and $\ixb$ can take any value among the 8 values \{$P^+$, $P^-$, $L^a$, $L^b$, $C^a$, $C^b$, $M^a$, $M^b$\}.
In their work the kernels $\phi$ are estimated using the non parametric estimation method first described in \cite{bacry2016first}. This model allows the authors to highlight the rich influence structure between events in a limit order book, including the high-frequency midprice reversion and the persistent autocorrelation in the order flow determined by order splitting strategies but also some more refined market-maker induced dynamics (see \cite{thibault1}).

\noindent
In order to add queue-dependency to the previous Hawkes approach, we proceed in much the same way as we did in the previous section except that the order book state is no longer represented by a single queue quantity $q$ (that represented either the best ask or the best bid quantity) but by both the best bid and the best ask queue sizes $(q_a, q_b)$.
Let us point out that now the queue sizes $q_a$ and $q_b$, by definition, never reach zero, moreover each process $dN^\ixa$ encodes the full history of all the events of type $\ixa$ happening at the corresponding best side, independently of the various moves of the best ask/bid prices.
So adding queue-dependency to such a model calls for a mechanism that is able to modulate (as a function of the orderbook state) not only the exogenous intensity $\mu_\ixa$ (as in the previous model) but also the Hawkes part of the intensity. To illustrate that point, we can mention a situation where obviously the Hawkes kernel matrix should explicitely depend on the queue states:  This is for instance the case of $P^+$ events which are very unlikely to occur (for a large tick asset) when the ask queue is big and the bid queue is small. 

\noindent
We consider the simplest possibility where both the exogenous and the self-exciting part of the intensity
share the same multiplicative dependence on the states and introduce the following model, referred to in the following as QRH-II 
\begin{equation}\label{lambda_dim_2}
\lambda^\ixa(t) = f^\ixa{(q_a(t),q_b(t))} \left(\mu^\ixa + \sum_{\ixb} \int_0^t \phi^{\ixa \ixb}(t-s) dN^\ixb_s \right),
\end{equation}
where the functions $f^\ixa$ that encode the dependence on the orderbook states, modulate not only the exogenous intensity (as in Eq. \eqref{eq:qrh_model}) but also the Hawkes term. Let us notice that unlike
QRH-I model, the QRH-II model is mainly a model for the order flow, so we disregard the relationship
between the order arrivals and the queue sizes and consider these queues as (observable) 
exogenous processes.
As before, we choose a parametric form for the kernels $\phi^{\ixa \ixb}$ and in particular we adopt the same exponential-sum specification as provided in Eq. \eqref{eq:hawk_kernels}.

\subsection{From log-likelihood estimation to least square estimation}
Once the parametric form \eqref{eq:hawk_kernels} for the kernels has been specified, the model can be estimated using MLE. Although the QRH-II model slightly differs from the QRH-I model, the calculation of its log likelihood function and its gradients follows the same track presented in Appendix~\ref{sec:app2}, with only a trivial modification required. Moreover, thanks to the chosen parametrization, the log-likelihood is again a convex function of the parameters, thus guaranteeing the existence of a global optimum. 

\noindent
Since the number of configurations of the orderbook states grows quadratically in the number of states considered per-side, this can become problematic if this number is too large. Considering every possible state (i.e., every possible values for each queue) would require very long computation times. To limit the number of configurations, we thus consider the orderbook to be in state $(q_a^i,q_b^j)$ if the bid queue size is within its $i$-th quintile (i.e. the $25\cdot i$ percentile) and the ask queue is in its $j$-th quintile. So the estimated function $f^\ixa$ is actually a function of the quintiles $f^\ixa(q_a^i,q_b^j)$. Finally, we choose the normalization (note that any other normalization would be equivalent up to a rescaling of the kernels and of the $\mu^\ixa$ in Eq. \eqref{lambda_dim_2}) 
\begin{equation}
f^\ixa(q_a^1,q_b^1) = 1.
\end{equation}

\noindent
Using maximum likelihood, we calibrate the so-obtained model on the same dataset used in the previous section. 
As shown in Tables \ref{tab:likelihood_2queue} and \ref{tab:lr_2queue}, our model outperforms in terms of goodness of fit the pure-Hawkes model introduced in \cite{thibault1} when calibrated with sum of exponential kernels. In particular, a likelihood ratio test rejects the null hypothesis of a pure-Hawkes model with a $p\text{-value} < 10^{-16}$.

\begin{table}[H]
	\centering
	\begin{tabular}{lcccc}
		\toprule
		&\multicolumn{4}{c}{Bund}\\
		\midrule
		&$L$ & AIC & BIC & \# parameters\\
		\midrule
		QRH-II & {$5.348\times 10^8$} & {$-1.070\times 10^9$} & {$-1.070\times 10^9$} & 400\\
		Hawkes & {$5.200\times 10^8$} & {$-1.040\times 10^9$}& {$-1.040\times 10^9$} & 200  \\
		\midrule
		&\multicolumn{4}{c}{DAX}\\
		\midrule
		&$L$ & AIC & BIC & \# parameters\\
		\midrule
		QRH-II & {$4.626\times 10^8$} & {$-9.253\times 10^8$} & {$-9.253\times 10^8$} & 400\\
		Hawkes & {$4.488\times 10^8$} & {$-8.976\times 10^8$} & {$-8.976\times 10^8$} & 200  \\
		\bottomrule
	\end{tabular}
	\caption{Log-likelihood, AIC, and BIC values for the QRH-II model (defined by \eqref{lambda_dim_2}) and the Hawkes model (defined in \cite{thibault1}) for Bund and DAX data.}
	\label{tab:likelihood_2queue}
\end{table}

\begin{table}[H]{}
	\centering
	\begin{tabular}{lccc}
		\toprule
		\multicolumn{4}{c}{Bund}\\
		\midrule
		& LR & df & $p$-value \\
		\midrule
		$H_0 =$ Hawkes, $H_1 =$ QRH-II & $2.9\cdot 10^7$ & 200 & $< 10^{-16}$ \\
		\midrule
		\multicolumn{4}{c}{DAX}\\
		\midrule
		& LR & df & $p$-value \\
		\midrule
		$H_0 =$ Hawkes, $H_1 =$ QRH-II & $2.8\cdot 10^7$ & 200 & $< 10^{-16}$ \\
		\bottomrule
	\end{tabular}
	\caption{Likelihood ratio test statistic and $p$-values for the case where the null hypothesis is the QRH-II model (defined by \eqref{lambda_dim_2}) and for the case where the null hypothesis is the Hawkes model (defined in \cite{thibault1}).}\label{tab:lr_2queue}
\end{table}

\begin{figure}[H]
	\centering
	\includegraphics[width=0.48\textwidth]{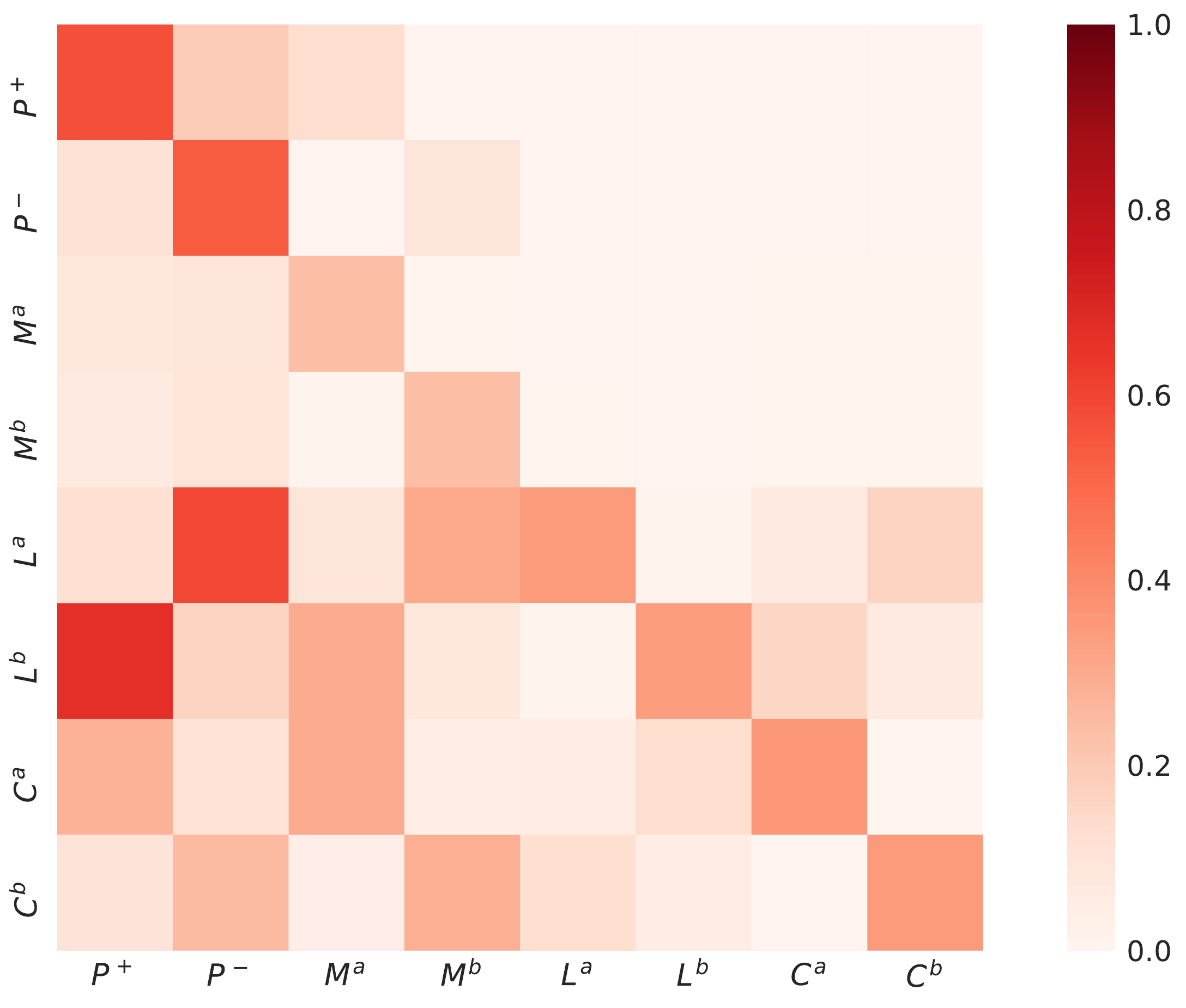}
	\includegraphics[width=0.48\textwidth]{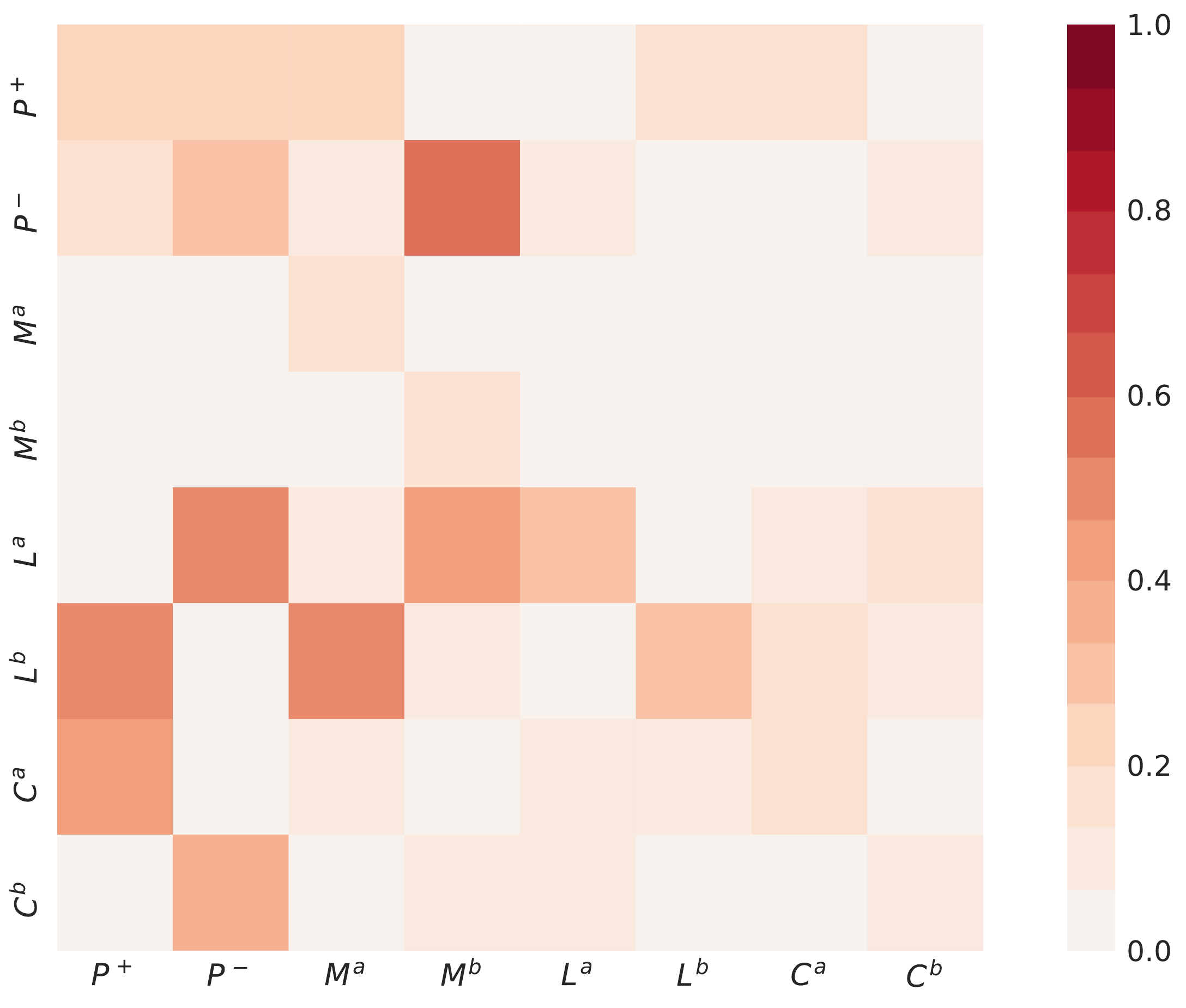}
	\caption{MLE  Kernel norm matrices $\int \phi^{lm}(t)dt$ for Bund future (left) and DAX future (right) for the QRH-II model (defined by \eqref{lambda_dim_2}). }
	\label{plot_llk_Hawkes_Kernel_Norm}
\end{figure}

\noindent
A detailed discussion on the so-obtained MLE results will be made in Section \ref{sec:comments}. For now, 
for the sake of simplicity, let us first put apart the effect of the queue dependency and just look at the kernel estimation.  

\noindent
Figure \ref{plot_llk_Hawkes_Kernel_Norm} represents the so-obtained kernel norm matrices $\{\int\phi^{\ixa \ixb}(t)dt\}_{\ixa \ixb}$ given by our model. 
This figure has to be compared with the matrix obtained in Figure~4 of \cite{thibault1} for which the same model (without queue dependency) has been used. Let us point out that in order to obtain this matrix, \cite{thibault1} performed a non-parametric estimation which allows negative values for the kernels and consequently negative values for some elements of the matrix $\{\int\phi^{\ixa \ixb}(t)dt\}_{\ixa \ixb}$. These negative values account for inhibition dynamics, i.e., decreasing the intensity of a given type of event.  
One could show that negative values for a kernel could lead at a finite time (and with a non zero probability) to some situation where the sum \eqref{lambda_dim_2} is negative leading to a negative intensity which, of course, does not make any sense. In order 
to circumvent this difficulty, it is common to replace equation \eqref{lambda_dim_2} by a non-linear equation of the form 
\begin{equation}\label{lambda_dim_2_non_linear}
\lambda^\ixa(t) = f^\ixa{(q_a(t),q_b(t))} \left(\mu^\ixa + \sum_{\ixb} \int_0^t \phi^{\ixa \ixb}(t-s) dN^\ixb_s \right)^+,
\end{equation}
where the operator $(\ldots)^+$ does not change the value of the quantity in between the parenthesis 
in the case this quantity 
is positive and is 0 if this quantity is negative.
However, in the framework of this non linear Hawkes model, 
MLE becomes intractable since it is no longer a convex problem. As explained in \cite{hansen2015lasso}, this problem is tractable in some sense when considering the least square approach (see also \cite{rambaldi2018disentangling} for example of least square estimation with negative valued kernels). 

\noindent
Thus, the least square estimations allows negative valued kernels whereas, in the MLE framework as presented above, we forced all the kernels (i.e., all the $\alpha_u^{lm}$'s in Eq. \eqref{eq:hawk_kernels}) to be positive valued. This mainly explains the differences found between Figure~4 in \cite{thibault1} and our Figure \ref{plot_llk_Hawkes_Kernel_Norm}. 
For the sake of just naming one striking difference,
in  \cite{thibault1}, the kernel integral $\int \phi^{L^bP^-}$ (resp. $\int \phi^{L^aP^+}$) is found to be strongly negative. Actually, as seen in Appendix B.6 in \cite{thibault1}, the kernel itself $\phi^{L^bP^-}$ (resp. $\phi^{L^aP^+}$) is mostly negative at all time scales. 
This fact can be seen as a natural dynamic induced by market makers: when the price goes down, the efficient price is closer to the best bid price thus less limit orders are placed on the bid size (the gain is small compared to sending an aggressive order) and more limit orders are placed on the ask side.

\noindent
Let us point out that forcing the kernels to have only positive values (i.e., forcing all the $\alpha_u^{lm}$ to be positive as assumed when performing MLE) will a priori not only lead to highly biased values for kernels with negative values but is likely also to induce high bias for kernel with only positive values since the estimation performed is a joint estimation of all the kernels involving intricate relationships between these kernels (see \cite{bacry2016first} for examples of such biases).

\noindent
Consequently, it appears that one should use least square based estimation rather than MLE.  Details about least square estimation can be found in Appendix \ref{sec:app3}. It consists in minimizing $R(\theta)$ as defined by 
\begin{equation}\label{leastsq_dim2} 
R(\theta) = \sum_{\ixa{}=1}^D R^\ixa{}(\theta), 
\quad \text{with} \quad
R^\ixa(\theta) = \int_0^T \lambda_\ixa^2(t; \theta|\mathscr{F}_t) dt\ - \sum_{k = 1}^{N^\ixa} \lambda_\ixa(t_k^\ixa; \theta|\mathscr{F}_{t_k^\ixa})
\end{equation}
where $\lambda_\ixa$ is given by Eq.~\ref{lambda_dim_2}\footnote{or alternatively by Eq~\eqref{lambda_dim_2_non_linear} in the sense given in \cite{hansen2015lasso}}.
One could easily verify that this problem is convex as a function of the $\alpha_u{lm}$ (see Eq. \eqref{eq:hawk_kernels}), thus the existence of a global optimum is guaranteed. As shown in Appendix \ref{sec:app3}, this parametrization allows a computationally efficient calculation of the squares loss function $R$ together with its gradient.

\subsection{Fitting results and comments}
\label{part3}
\label{sec:comments}
In this section we present and comment the results obtained through least square estimation of the QRH-II model as defined by Eq.~\eqref{lambda_dim_2}.

\begin{figure}[tbp]
    \centering
    \includegraphics[width=0.48\textwidth]{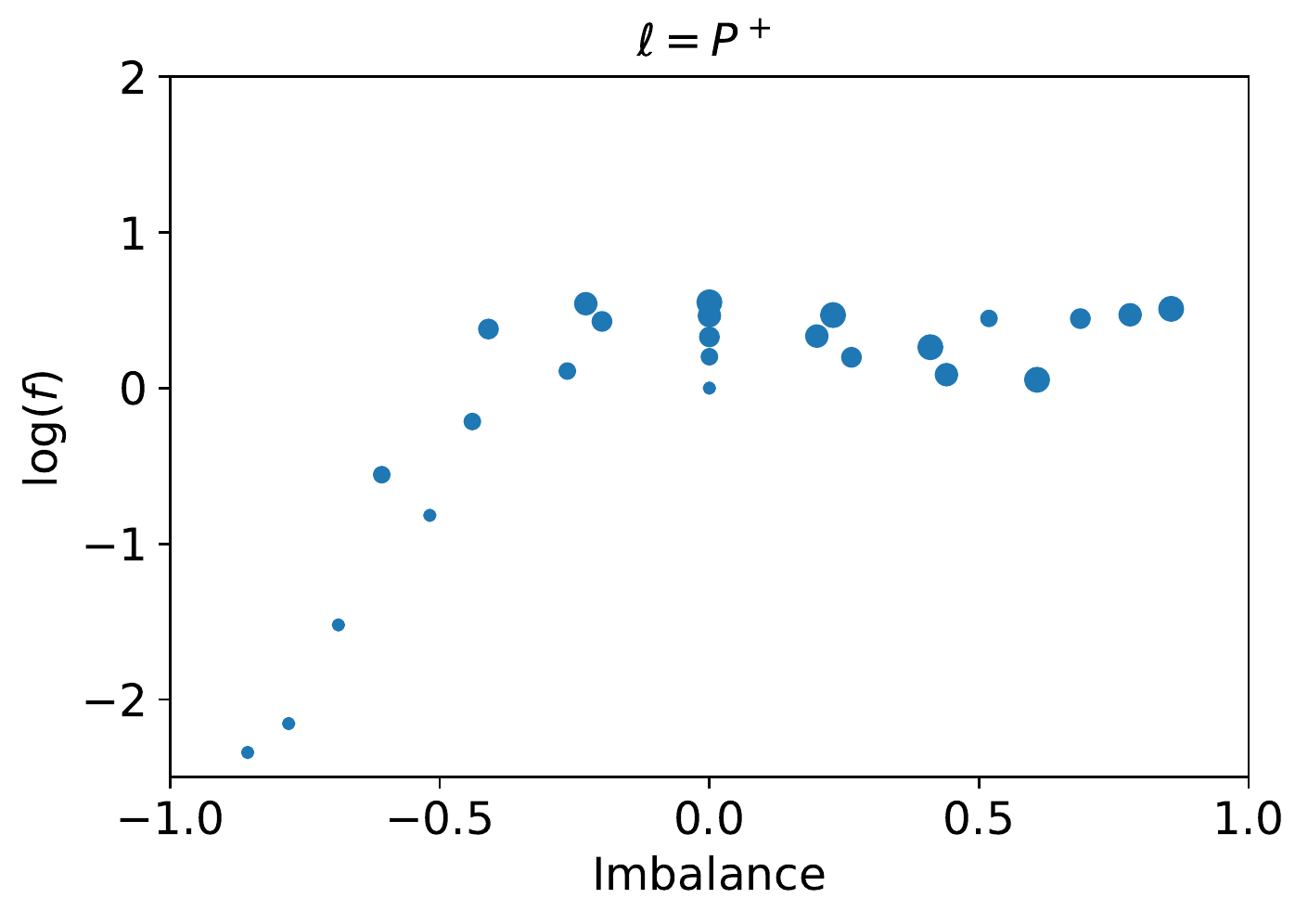}
    \includegraphics[width=0.48\textwidth]{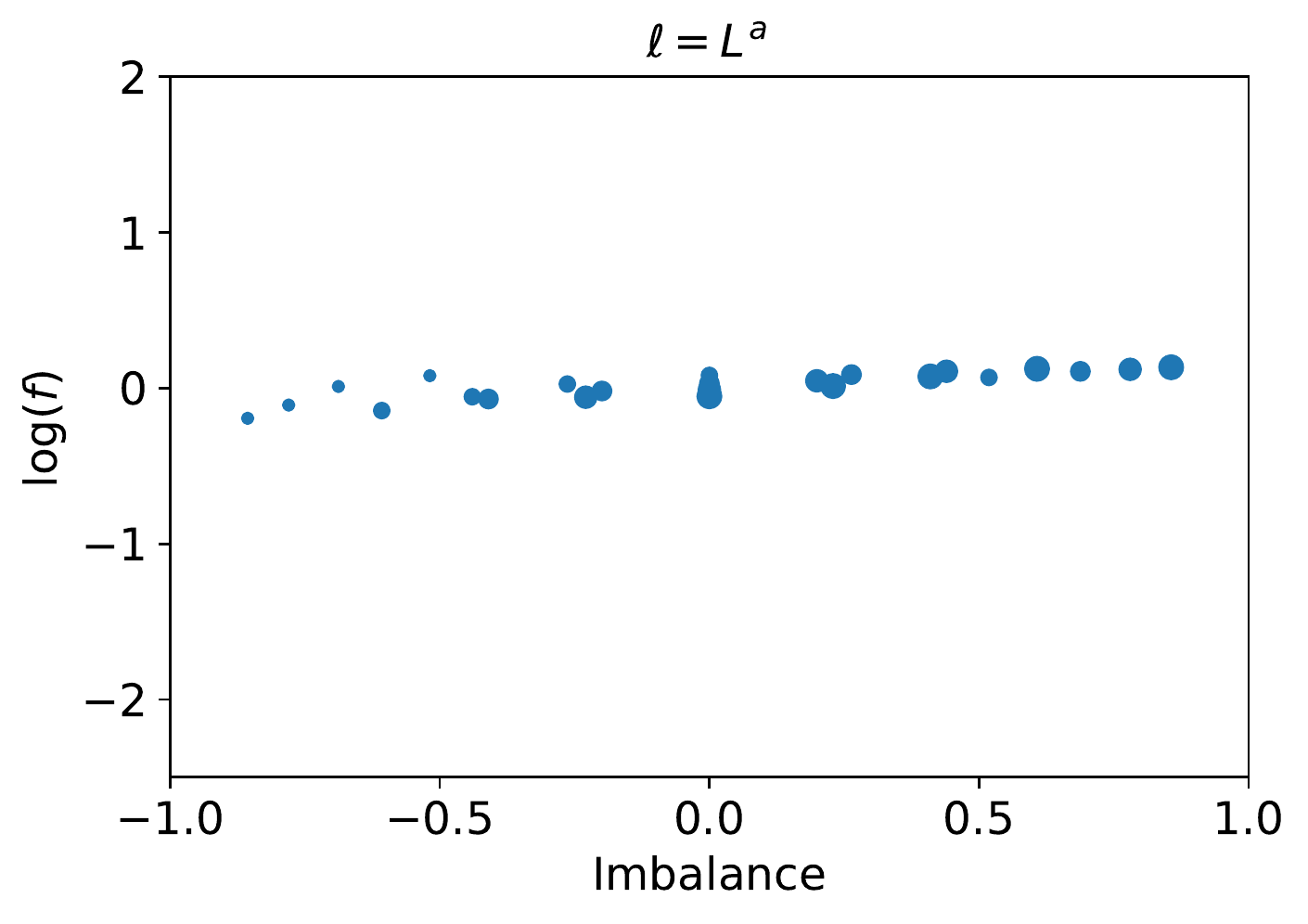}
    \includegraphics[width=0.48\textwidth]{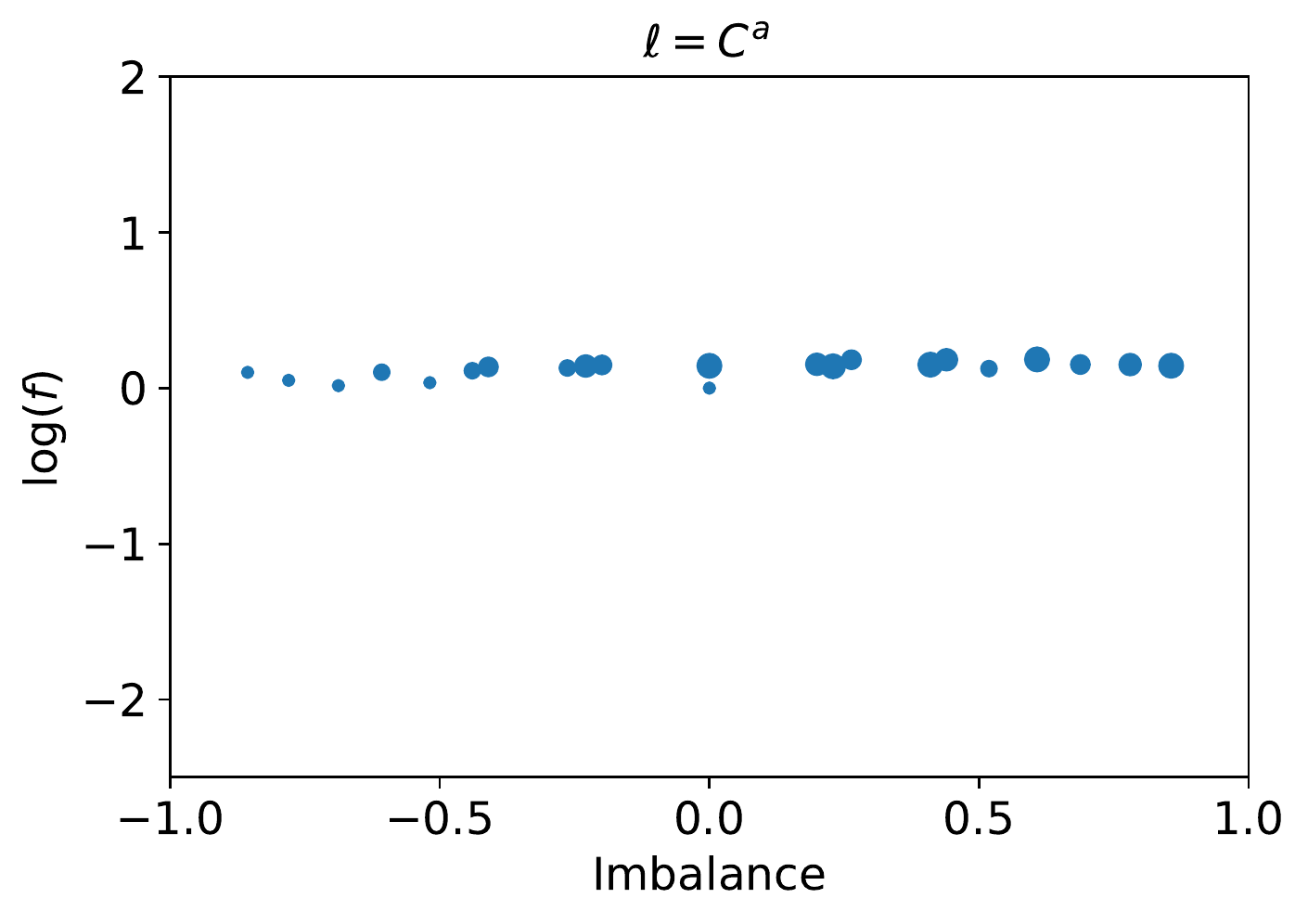}
    \includegraphics[width=0.48\textwidth]{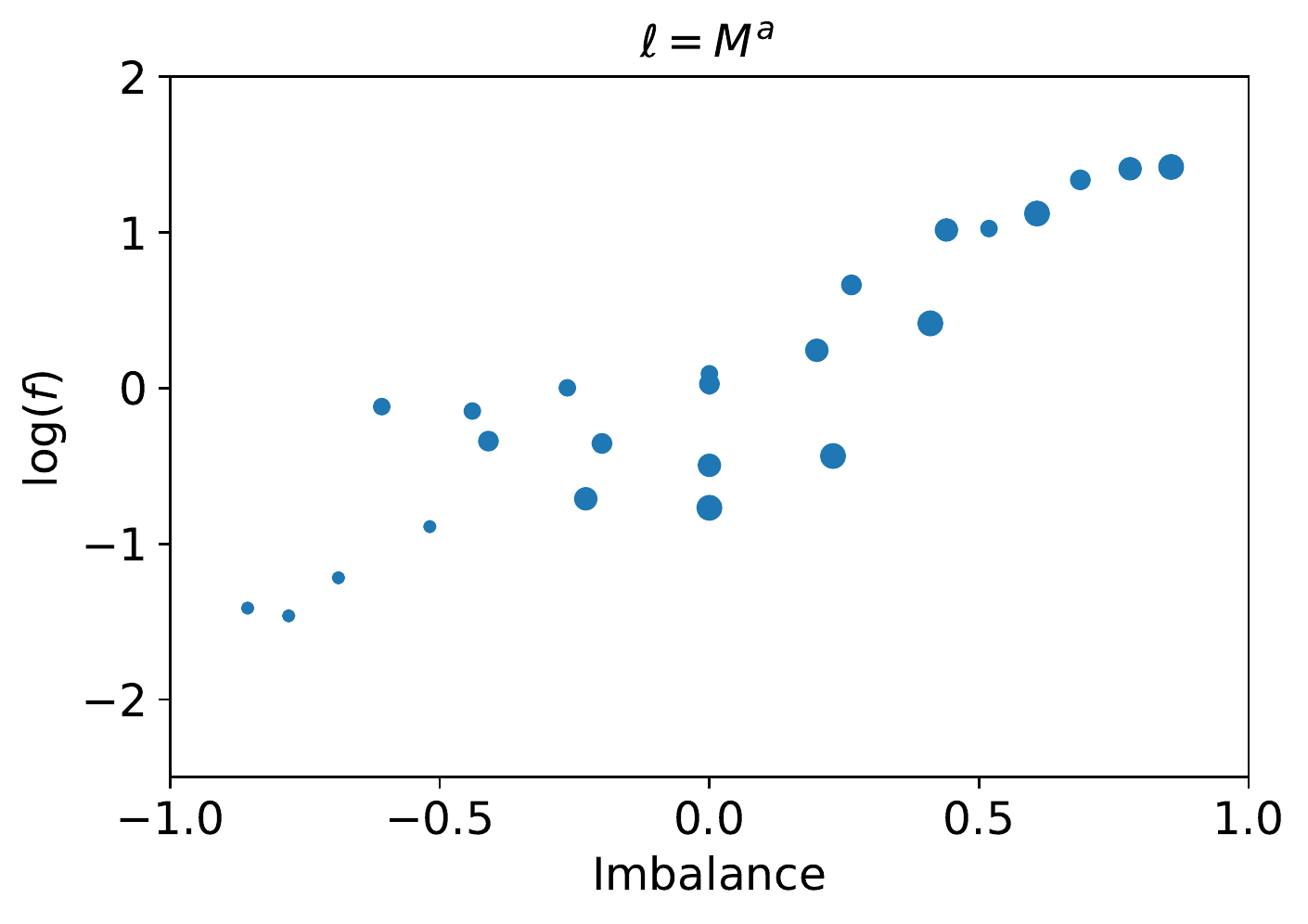}
    \caption{From left to right, upside to downside, $\log_{10}(f^l(q_a^i, q_b^j))$ for $l=P^+$, $L^a$, $C^a$, $M^a$ of QRH model as a function of the imbalance~\ref{eq:imb}, Bund future. The quantiles are the same for bid and ask sides and correspond to $q^1_a=q^1_b=]0,80]$, $q^2_a=q^2_b=]80,165]$, $q^3_a=q^3_b=]165,258]$, $q^4_a=q^4_b=]258,386]$ and $q^5_a=q^5_b=]386,+\infty[$.}
    \label{fig:Bund_logf_QRH}
\end{figure}

\begin{figure}[tbp]
    \centering
    \includegraphics[width=0.48\textwidth]{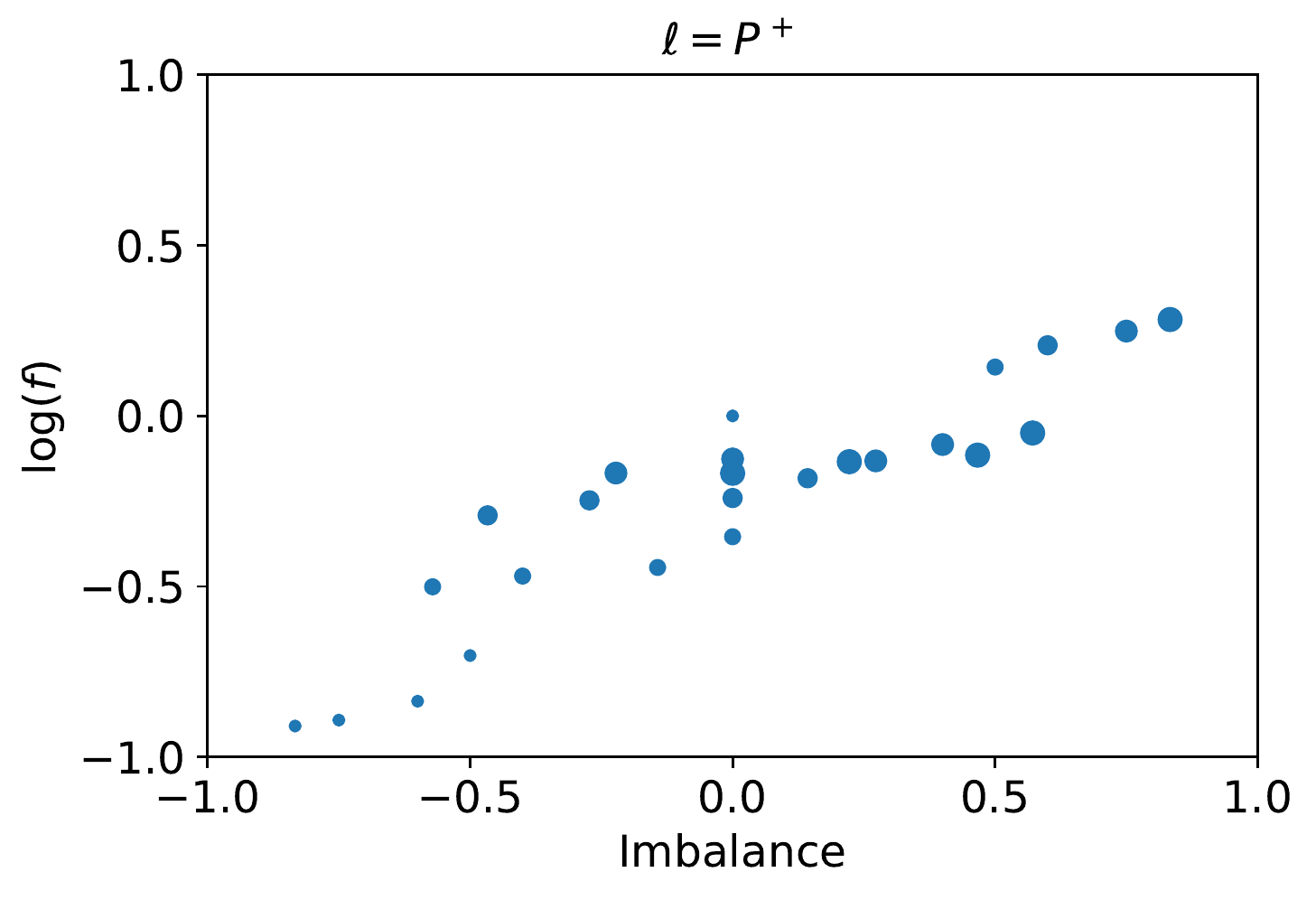}
    \includegraphics[width=0.48\textwidth]{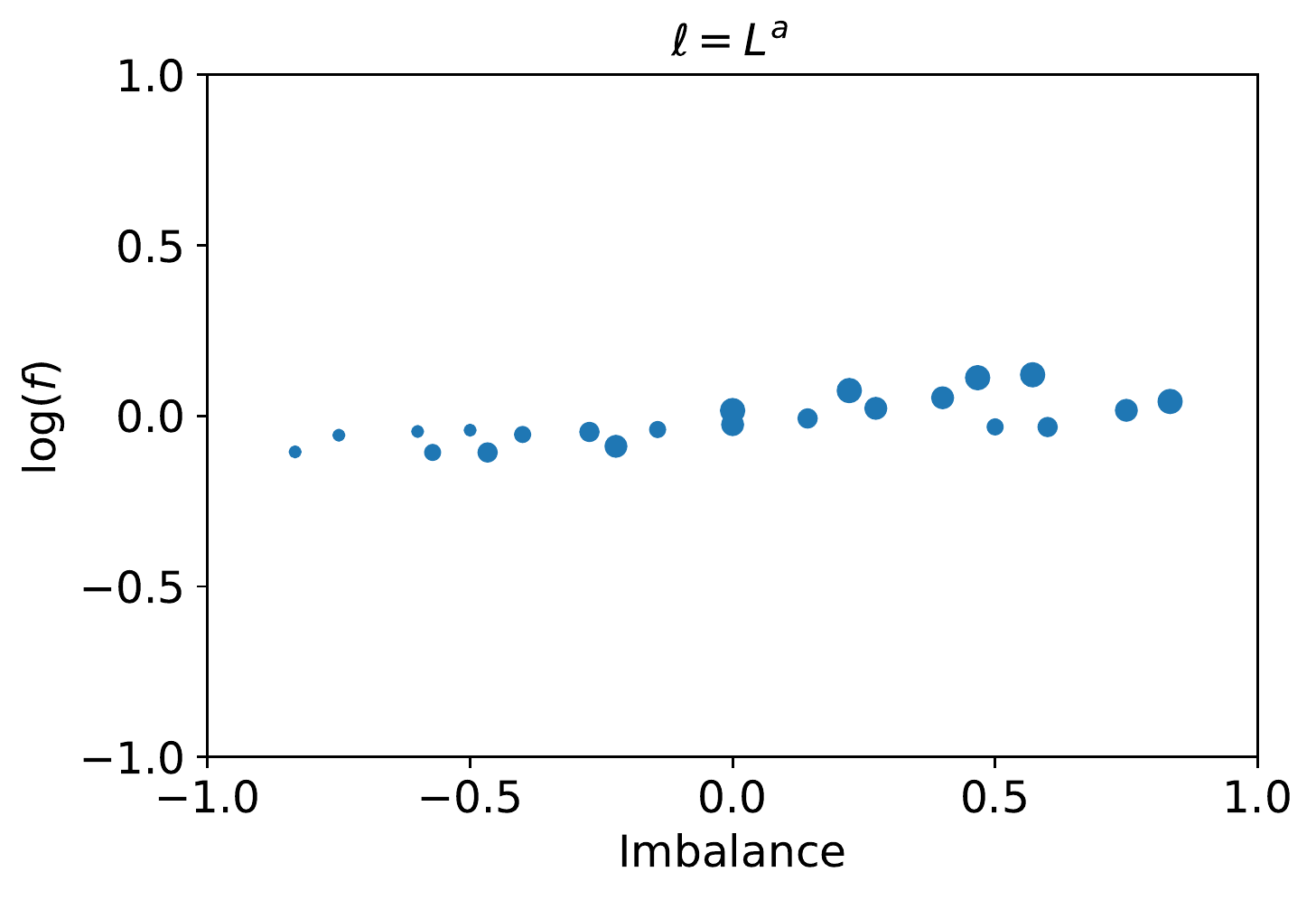}
    \includegraphics[width=0.48\textwidth]{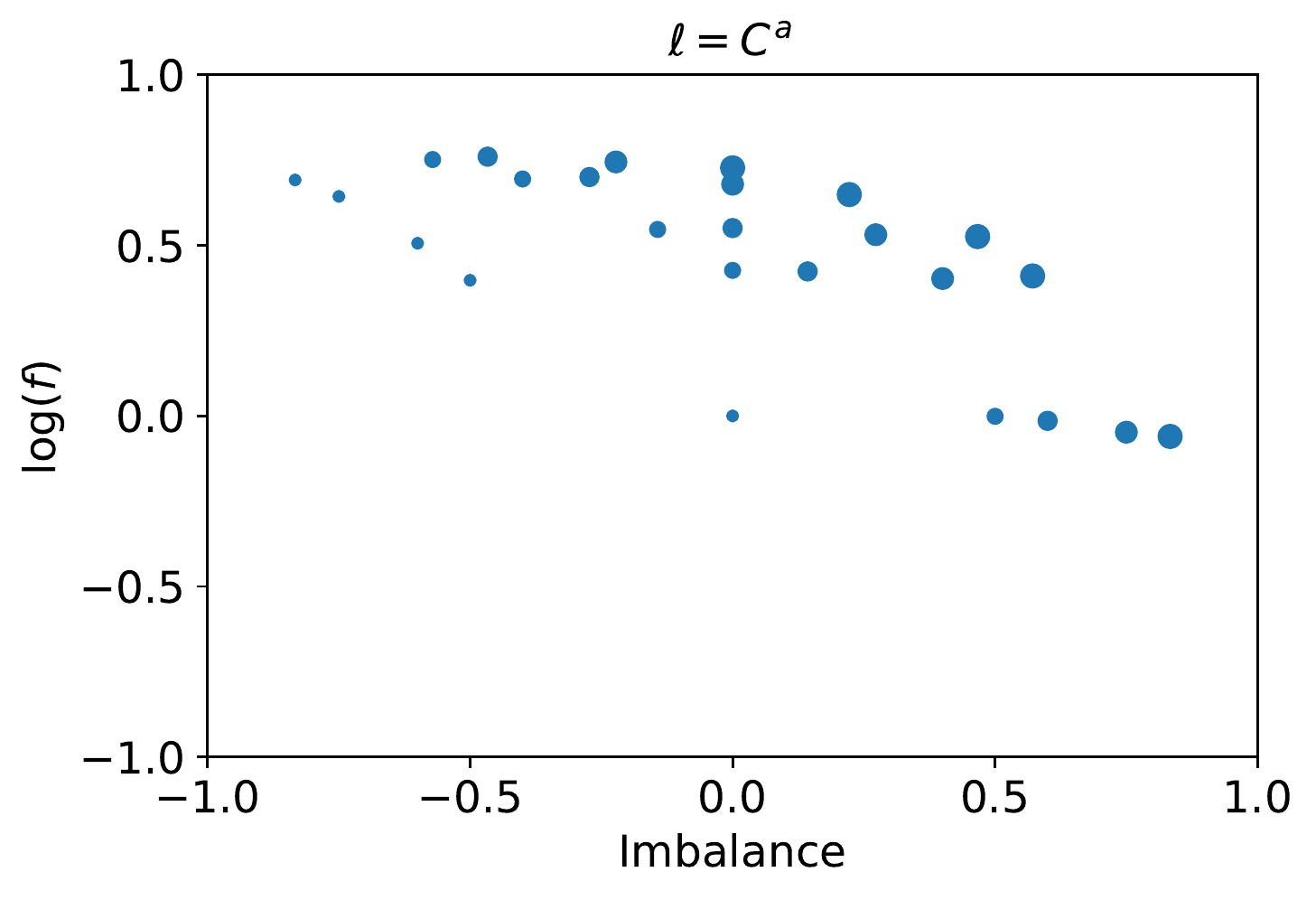}
    \includegraphics[width=0.48\textwidth]{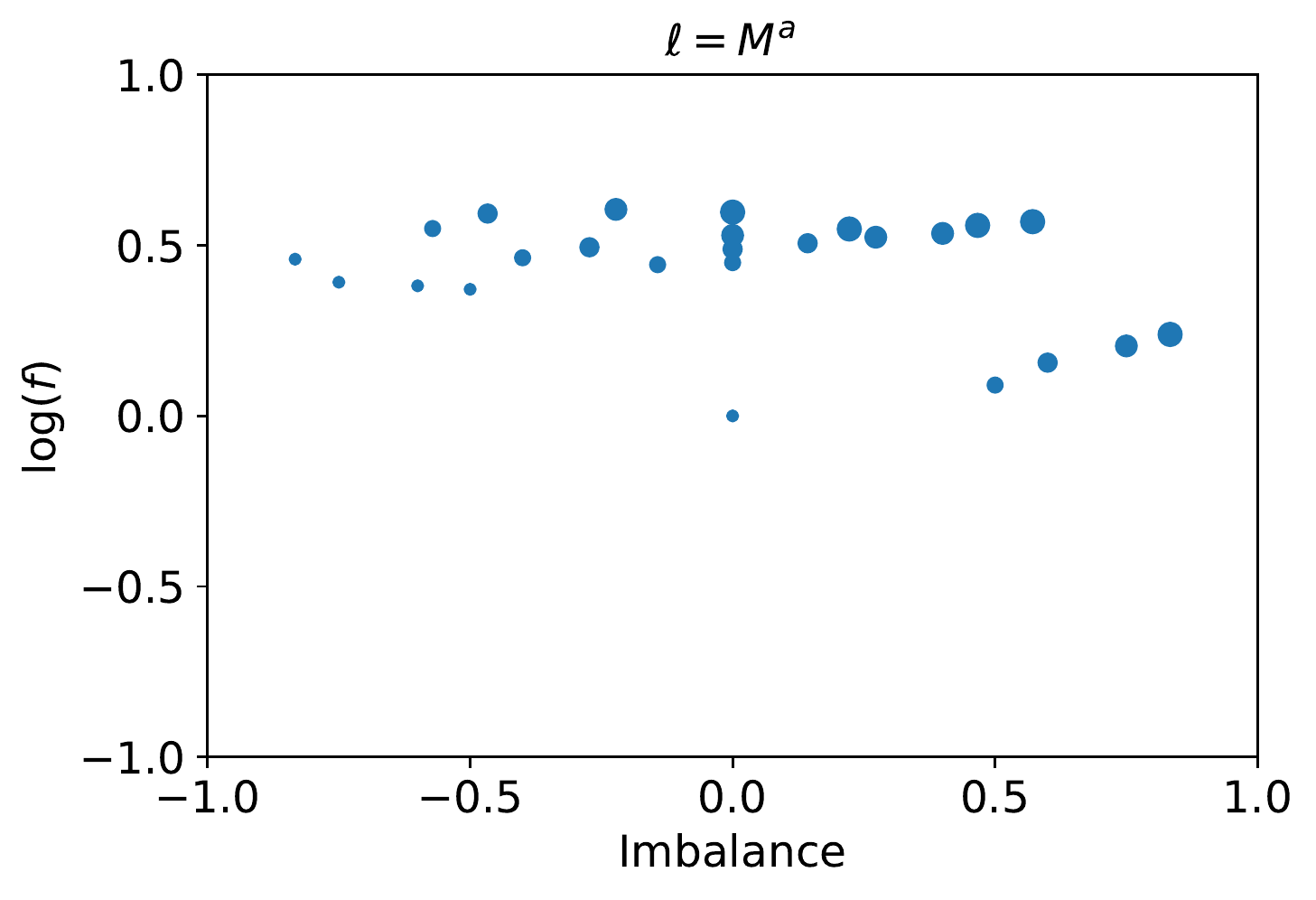}
    \caption{From left to right, upside to downside, $\log_{10}(f^l(q_a^i, q_b^j))$ for $l=P^+$, $L^a$, $C^a$, $M^a$ of QRH model as a function of the imbalance~\ref{eq:imb}, DAX index future. The quantiles are the same for bid and ask sides and correspond to $q^1_a=q^1_b=]0,2]$, $q^2_a=q^2_b=]2,3]$, $q^3_a=q^3_b=]3,5]$, $q^4_a=q^4_b=]5,8]$ and $q^5_a=q^5_b=[8,+\infty[$.}
    \label{fig:DAX_logf_QRH}
\end{figure}

\vskip .3cm
\noindent
{\bf Estimation of the $f^\ixa(q_a,q_b)$ functions.}
In our results we document a clear dependence of the order arrival rates on both $q_a^i$ and $q_b^j$, indicating that the state of the LOB has a clear influence on the order arrival rates. Let us point out that previous works (e.g., \cite{rambaldi2017role} and \cite{Lehalle2018}) suggest that this dependence is actually a dependence on the imbalance of the queue sizes as defined by 
\begin{equation}\label{eq:imb}
I(t)=\frac{v_b(t) - v_a(t)}{v_b(t) + v_a(t)} = \frac{q_b(t) - q_a(t)}{q_b(t) + q_a(t)} 
\end{equation}
where $v_{a/b}(t)$ denote the volume available at time $t$ at best ask/bid prices and we have assumed that orders have a constant volume corresponding to the AES as defined previously. The imbalance represents the simplest proxy to
account for the instantaneous buying pressure. In that respect, in Figures~\ref{fig:Bund_logf_QRH} and \ref{fig:DAX_logf_QRH} we plot the parameters $f^\ixa(q_a^i, q_b^j)$, in logarithmic scale, for each order type $\ixa$ as a function of the imbalance $I$ calculated as the median imbalance in the state interval associated with quantiles $(q_a^i, q_b^j)$.  Let us note that in these plots, the dot sizes indicate the sizes of the corresponding $q^i_b$.
By looking at Figures~\ref{fig:Bund_logf_QRH} and \ref{fig:DAX_logf_QRH} we can make the following observations: First, the variation of the imbalance on the order book captures most of the variations of the intensive parameters $f^\ixa(q_a,q_b)$, this is more evident on the large tick asset (Bund) for which $f$ can span almost three order of magnitudes (for $P$, and $M$ events) as $I$ ranges from $-1$ to $+1$. The variations of $f^\ixa(q_a,q_b)$ are smaller for the small tick asset (DAX) so the effect of the imbalance is less pronounced albeit still visible for $P$ and to a lesser extent $L$ events. 
This is in line with the observation that the imbalance is a very good predictor for midprice changes for large tick assets while its predictive power is less marked for small tick ones (see e.g. \cite{gould2016queue}). One aspect to keep in mind while analyzing these figures is that, especially for a small tick asset, there is also a considerable amount of information in the deeper levels of the book that is not taken into account here.

\noindent
In Figure~\ref{fig:Bund_logf_QRH} corresponding to the Bund results, we observe that for midprice changes ($P$ events) there seems to be a kind of threshold effect, in that for imbalance values below $I = -1/2$, upwards price increments are dramatically inhibited, while large positive imbalance only marginally increases the likelihood of upwards price changes. This is rather natural since an imbalance smaller than $-1/2$ corresponds to the case  the ask size is at least three times bigger than the bid size which makes upwards move of the midprice very unlikely.
More surprinsingly, the agents seem to strongly condition their decision to use aggressive orders ($M$ events) on the state of the order book. Maybe they rush to get last available liquidity before an upwards midprice move (indeed $f$ gets very large for $M$ events when the imbalance is large, i.e., when there is hardly anything left on the ask size). 
Limit and Cancel orders appear to be much less sensitive to the state of the book. 
For the DAX, as we already pointed out, the dependence on the imbalance is much less pronounced (see Figure~\ref{fig:DAX_logf_QRH}) and, expcept the mid-price changes, all factor $f^\ell(q_a,q_b)$ are
weaky varying with $I$. Notice hower that
for market ($M$) and cancel ($C$) orders we observe a regime switching around $I = 1/2$. These regimes 
correspond respectively to quite large and very small ask queue size. In the latter case (that turns out to occur when the spread equals one tick) the occurence of market and cancel orders at ask that do not change the midprice is very unlikely.

\noindent
Last but not least, let us remark that the intensity variations that are not captured by the imbalance are mainly located around $I=0$ where the individual queue sizes ($q_a$ or $q_b$ which are almost equal) seem to have an important impact. 
Thus, for instance, for the Bund,  $f^{P^+}(q,q)$ seems to increase with $q$. Actually, this can be explained by a more thorough analysis that shows that the average spread is increasing with $q$ and is very close to 2 when $q$ is large. The same kind of remarks could be done for market orders.
As this example illustrates other microstructural variables and notably the spread, should be taken into account for an even more complete model. This however is outside the scope of the present work.

\paragraph{Comparison of estimated and empirical intensity.} To further validate the QRH-II model, we test its ability to reproduce the empirical intensity. 
The MLE of the averaged intensity conditioned by the state $q = (q_a,q_b)$ \ref{eq:Lambdal} writes (in a non parametric framework)
\begin{equation}\label{eq:def_emp_Lambda}
\begin{aligned}
\hat{\Lambda}^\ixa(q) &= \hat{\Lambda}(q)  \frac{\sum_{t_k^\ixa} \mathds{1}_{q(t_k^{l-}) = q}}{\sum_{t_k} \mathds{1}_{q(t_k^-) = q}}\\
\text{with} \quad \hat{\Lambda}(q) &= \text{mean}(t_k - t_{k-1}|q(t_k^-) = q)^{-1},
\end{aligned}
\end{equation}
where the operator $\text{mean}(\ldots)$ corresponds to the empirical mean.
This estimation is simply based on the observations of the process $\{N^l_t\}_l$. One can compute a corresponding estimator $\hat{\Lambda}^\ixa_{\text{QRH-II}}(q)$ using QRH-II parametric form of $\hat \lambda^\ixa_{\text{QRH-II}}(t^{\ixa, -}_k | q)$, this leads to 
\begin{equation}\label{eq:def_qrh_Lambda}
\hat{\Lambda}^\ixa_{\text{QRH}}(q) = \text{mean}(\hat \lambda_{\text{QRH-II}}^\ixa(t^{\ixa, -}_k) |q(t^{\ixa, -}_k) = q)
\end{equation}
In order to synthesize the so-obtained results, we choose not to present the comparison of $\hat{\Lambda}^\ixa_{\text{QRH-II}}(q)$ with $\hat{\Lambda}^\ixa(q)$ for all types of orders and for all states $q$. Instead, for each type of order, we report the weighted relative error $\Delta^\ixa$ defined as:
\begin{equation}
\Delta^\ixa = \frac{\sum_q  \left|  \hat{\Lambda}^\ixa(q) - \hat{\Lambda}^\ixa_{\text{QRH-II}}(q)  \right|  N^\ixa(q)}{\sum_q \hat{\Lambda}^\ixa(q) N^\ixa(q)}
\end{equation}
The weighted relative error for Bund and DAX  is presented in table~\ref{tab:Compare_Lambda}. We observe that $\Delta^\ixa$ are of the order of $10\%$, which provides a satisfactory match to the empirically observed intensity. 
\begin{table}[tbh]
        \centering
    \begin{tabular}{lccccccccc}
        \toprule
        & $P^+$ & $P^-$ & $L^a$ & $L^b$ & $C^a$ & $C^b$ & $M^a$ & $M^b$ \\
        \midrule
        Bund &  $14.2\%$ & $10.5\%$ & $6.7\%$ & $6.0\%$ & $7.4\%$ & $8.1\%$ &  $4.0\%$ &  $12.0\%$ \\
        DAX  &  $8.2\%$ & $5.9\%$ & $0.5\%$ & $4.7\%$ & $7.1\%$ & $1.1\%$ &  $1.6\%$ &  $5.9\%$ \\
        \bottomrule
    \end{tabular}
\caption{Error of average intensities by order type }
\label{tab:Compare_Lambda}
\end{table}

\begin{figure}[tbh]
	\centering
	\includegraphics[width=0.48\textwidth]{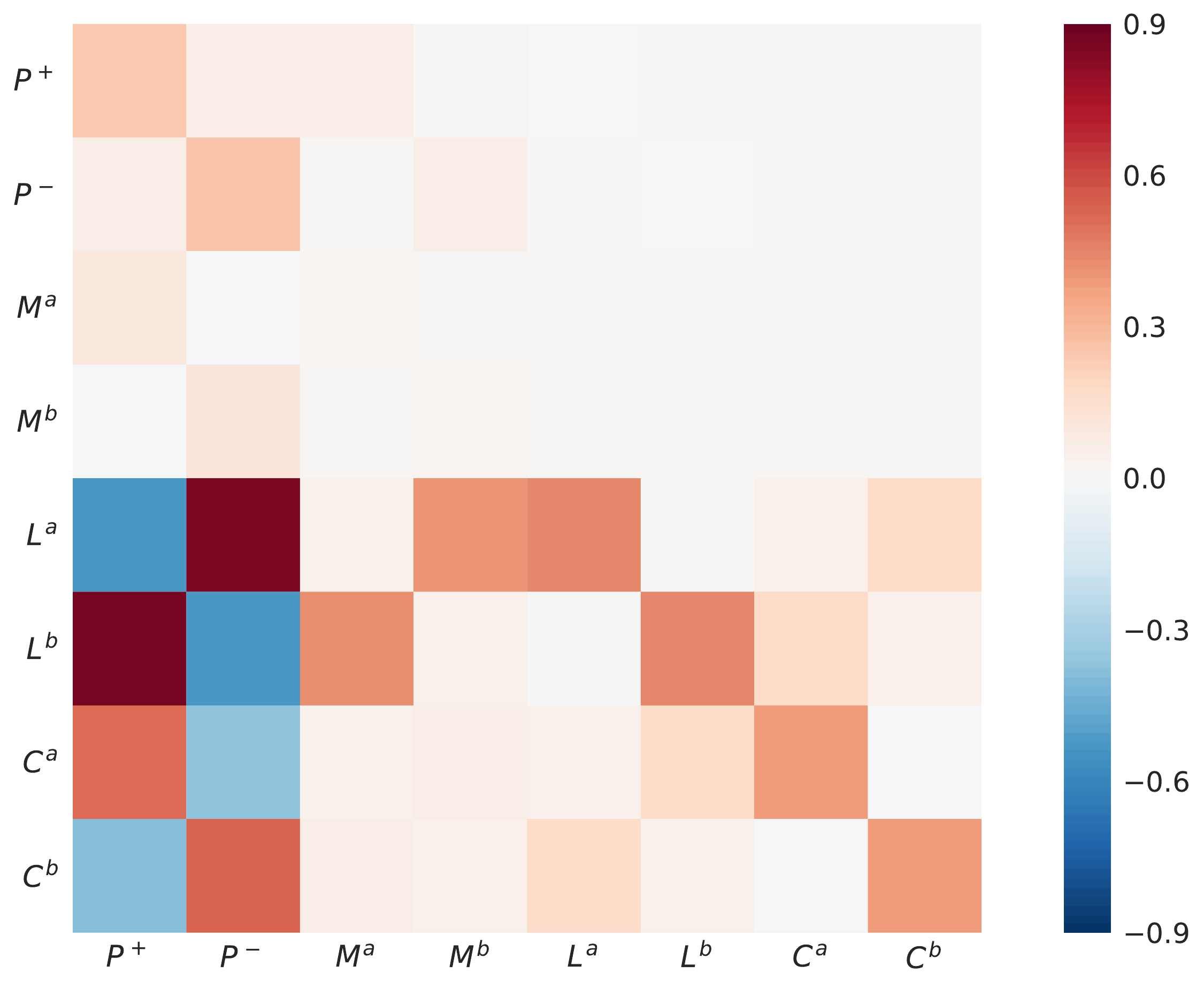}
	\includegraphics[width=0.48\textwidth]{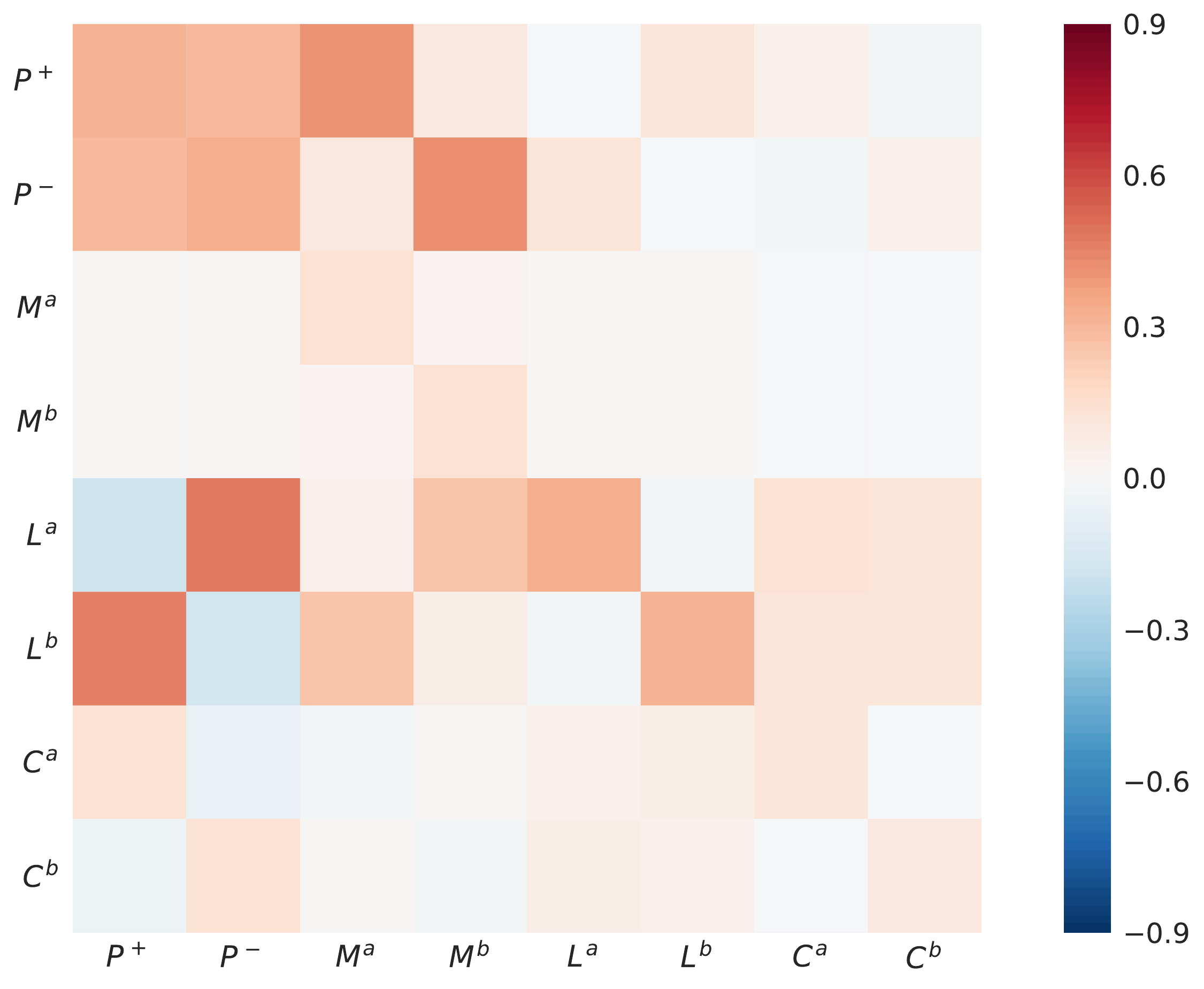}
	\caption{The estimated matrix norms $\int \phi^{lm}(t)dt$ using least square estimation QRH-II model. Bund future on the left and DAX future on the right.}
	\label{fig:LeastSq_Kernel_Norm}
\end{figure}

\paragraph{Analysis of the kernel norm matrices.} Finally, to complete the analysis of our results, in Figure~\ref{fig:LeastSq_Kernel_Norm} we display the matrices of the estimated norms $\{\int \phi^{\ixa \ixb}(t)dt\}_{\ixa \ixb}$ for both the Bund and the DAX. These matrices provide information on the average interactions between different event types when queue dependence is disregarded. As such they are the counterpart of the kernel norm matrices shown in Figure 4 of \cite{thibault1}. 
We recover a lot of the features highlighted in \cite{thibault1}, such as the strong diagonal components for limit, market and cancel orders (a signature of order splitting), as well as the fact that market orders and price movement appear to influence much more limit and cancel than the other way round. 
Notably, for the Bund, when the midprice moves up (resp. down) it decreases (resp. increases) the rate of limit orders on the ask (resp. best) side since the efficient price is close to the best ask, there is no gain to send a limit versus a market. This also explains why it increases (resp. decreases) the rate of cancel orders sent on the best ask (resp. bid) side. The same effect can be more or less seen (but attenuated a lot) on the DAX. It is attenuated because the tick is small on the DAX, so the efficient price argument is not as strong. Let us point out that the exact same argument can be used to explain the effects of  market orders on limit and cancel orders. The influence of market order over price change is mainly because that under our settings, market order consumes the liquidity at the best prices, which could eventually create a new price. Since DAX is a small tick asset and queue size at the best prices is smaller that Bund, market orders are more likely to generate new prices. So the influence of market order over price change is more visible. We can see that for DAX future, limit orders and cancellations also has stronger influence toward price changes, for the same reason.
\begin{figure}[tbh]
	\centering
	\includegraphics[width=0.48\textwidth]{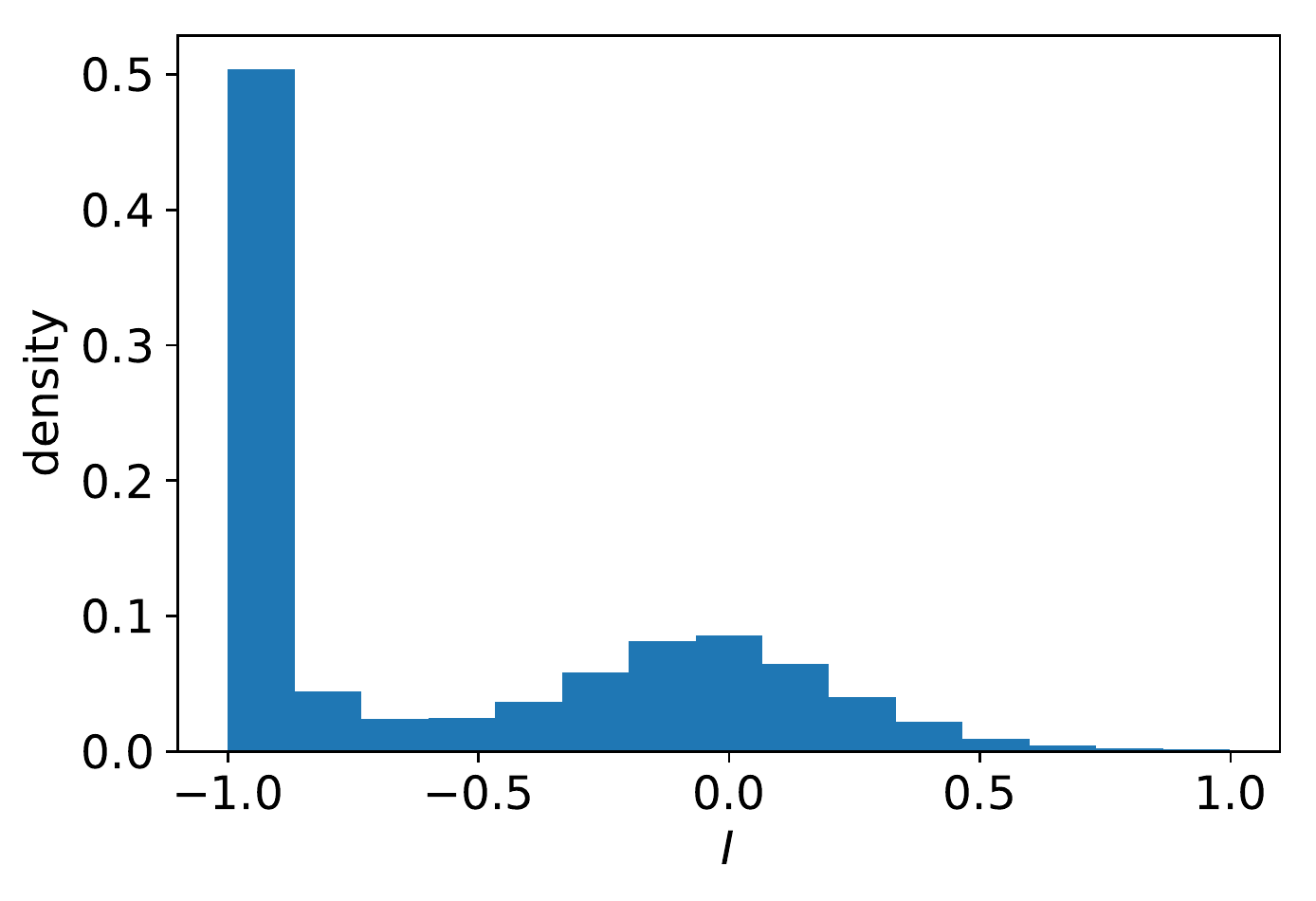}
	\includegraphics[width=0.48\textwidth]{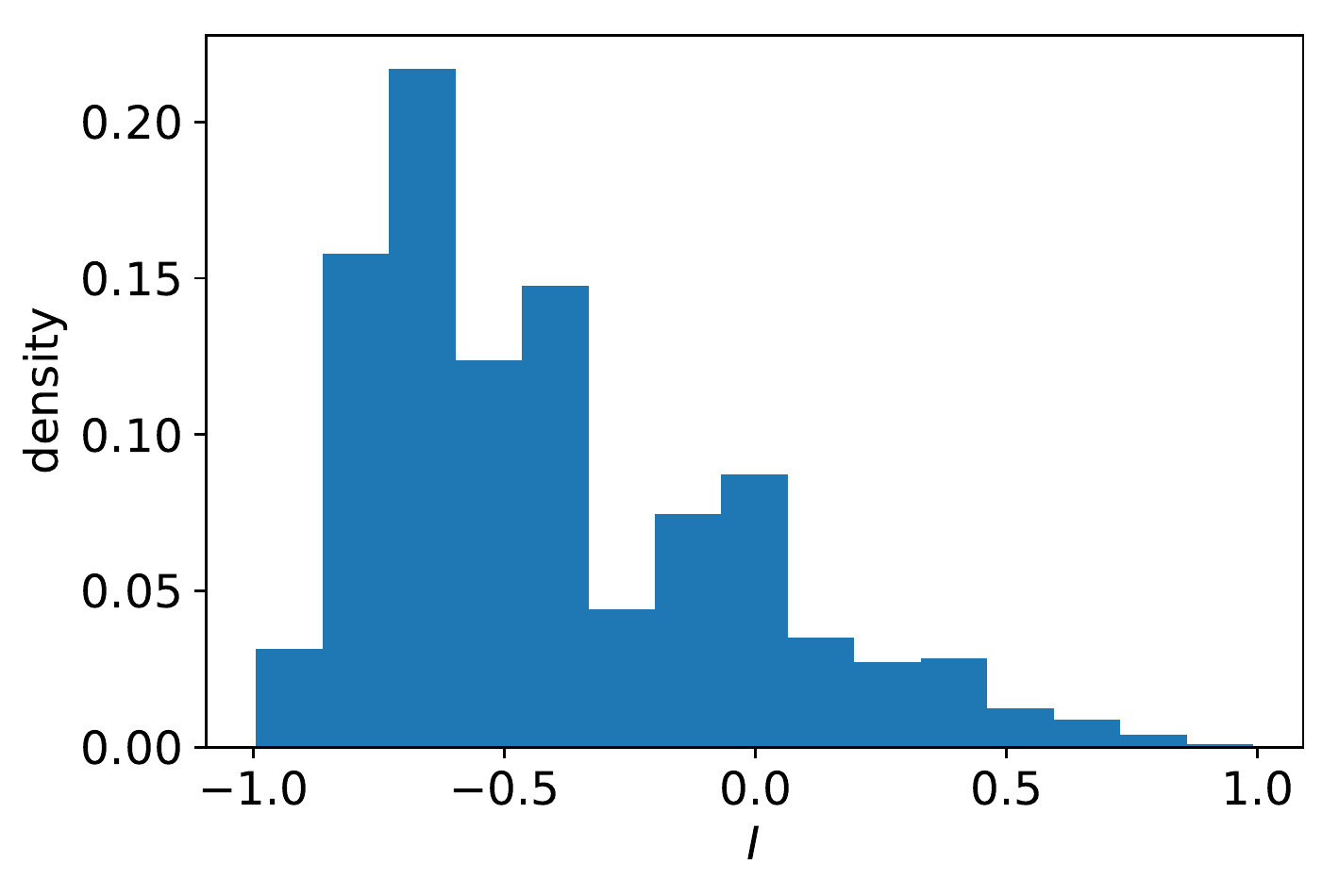}
	\caption{Empirical frequencies of observed imbalance values right after a $P^+$ event for the Bund (left panel) and the DAX (right panel). Notice the large components for negative imbalance values in both cases.}
	\label{fig:histo_imb}
\end{figure}
One can also see that, apart from  the self-exciting effect due to splitting of orders, for the Bund, limit orders on one side are coupled with cancel orders on the other side. This can be seen as a simple market maker strategy (rebalancing of the position). It does not show on the DAX certainly because, since the tick size is much smaller, the same rebalancing strategy does not affect necessary both best sides.
We refer the reader to \cite{thibault1}, Section~4 for further details on the interpretation of these features. 
\begin{table}[H]
	\centering
	\begin{tabular}{cc}
		\hline
		& Rate of mean-reversion \\ \hline
		Bund & 0.65                         \\
		DAX  & 0.56                          \\ \hline
	\end{tabular}
	\caption{Measure of mean-reversion of price: Empirical probabilities than two successive midprice change events have opposite directions.}\label{tab:stat_mean_rev}
\end{table}

\noindent
Though for most features, QRH-II model (Figure~\ref{fig:LeastSq_Kernel_Norm}) and the pure Hawkes model (Figure 4 in \cite{thibault1})) are similar, we can however observe a striking difference in the $P \rightarrow P$ and $P \rightarrow T$ submatrices. We can notice in QRH-II the absence of a strong excitation between $P^+$ to $P^-$ and $P^-$ to $P^+$ (top left $2\times 2$ submatrices  in Figure~\ref{fig:LeastSq_Kernel_Norm}) which should be the signature of the high frequency price mean reversion (as explained in \cite{thibault1}). 
For a pure Hawkes processes model, like in work \cite{thibault1}, the mean-reversion of price is reflected on the strong anti-diagonal terms of the $P \rightarrow P$ kernel norms submatrix (i.e., strong $P^+ \rightarrow P^-$ and $P^- \rightarrow P^+$ terms). This results from the fact that (in average) a $P^+$ event will generate more $P^-$ events than $P^+$ events, and vice versa. As shown in table \ref{tab:stat_mean_rev}, the actual midprice series are strongly mean-reverting, so it is likely that, within the QRH-II model this feature should be explained by the queue-reactive function $f(q_a,q_b)$. Indeed, as illustrated in Fig. \ref{fig:histo_imb}, after a midprice change the imbalance evolves in favor of a price move towards the opposite direction. For instance, an upward price jump $P^+$ leads, most of the time, to a negative imbalance either because of a refill of the best ask queue $q_a$ or simply because a single buy limit order is sent within the spread. According to results of Figs. \ref{fig:Bund_logf_QRH} and \ref{fig:DAX_logf_QRH}, in this case we have 
$f^{P^{+}}(q_a,q_b) \ll 1$ and $f^{P^{-}}(q_a,q_b) > 1$. Conversely, after a downward price change, $P^-$,
we will have $f^{P^-}(q_a,q_b) \ll 1$ and $f^{P^{+}}(q_a,q_b) > 1$.
The $2 \times 2$ submatrix $P \rightarrow P$ estimated within a pure Hawkes model is then likely to correspond to the QRH-II Hawkes submatrix multiplied  by a large factor on its anti-diagonal and 
a small factor on its diagonal. The same kind of argument based on imbalance impact can be invoked to explain the high diagonal values of the $P \rightarrow T$ submatrix while the highest values in \cite{thibault1} were rather observed on the anti-diagonal. \\

\section{Summary and prospects}\label{sec:conclusions}

In this paper, we introduced two "Queue Reactive Hawkes models" (QRH-I and QRH-II) with the ambition to improve respectively the approach of 
Huang et al. \cite{rsb1} on the queue reactive nature of the LOB dynamics and the model of Bacry et al. \cite{thibault1}.
We show that shuch models can be easily calibrated within a parametric approach. Our empirical findings
on two different future asset from Eurex, namely Bund and DAX order book data, suggest
that both queue reactive and past order flow dependencies are relevant to account for the occurrence likelihood
of future order book events.
Indeed, both models outperforms in terms of goodness of fit a pure Hawkes 
model as well as a pure queue-reactive one.
As far as QRH-I model is concerned, our framework allows one to remain within the framework
of Markov processes that has ergodic properties so, along the same line as in Huang et al. approach,
we can define and estimate the queue size distribution associated with the invariant measure of the model.
The QRH-II model lead us to refine Bacry et al. findings \cite{thibault1} by accounting for the states 
of best bid and best ask queues. It turns out that the volume imbalance allows us to explain most of the queue 
dependence.

\noindent
The QRH-II model could be improved by accouting for the interactions between the queue sizes (and thus the imbalance) and the order flow, as in the QRH-I model and also to include an explicit dependence on the spread for small tick assets. Some substantial simplifications we made could also be removed in order to have a even more realistic model such as, for instance, dropping the assumption of unitary order sizes by adopting a similar approach to \cite{Lu2018}.
Besides considering various applications of these models to design and optimize high frequency trading strategies,
from a numerical point of view, it remains to define approaches that allow one handle the full QR-model
as defined in Eq. \eqref{eq:general_qrh} where not only the queue dependencies of the exogenous and
Hawkes kernels are arbitrary but that simultaneously account for all the order book queues up to a given depth. Maybe some recent approaches and techniques developed in statistical learning would be helpful to handle such high dimensional problems. From a mathematical point of view, it remains to develop a deeper understanding of the stability and stationarity conditions for queue dependent Hawkes models. More fundamentally, a clear understanding to the observed shapes of the exogenous intensities and the imbalance dependence of the order flow arrival rates in term of the (rational) behavior of various market participants 
remain open questions.

\appendix
\section{Appendix}

\subsection{Proof of the existence of invariant distribution}
\label{sec:app1}

We are motivated to prove the existence of invariant distribution of QRH-I model presented in Part I, for the reason that if such invariant distribution exists, we could approximate the empirical distribution of queue size by simulating the QRH-I model for a long time. The proof is made with the help of Lyapunov function. First let's define:
\begin{equation}
o_{\ixa \ixb u}(t) = \int_0^{t} \alpha^{\ixa \ixb}_u e^{-\beta_u(t - s)} dN_s^\ixb
\end{equation}
By adopting defintion \eqref{eq:def_qrh_Lambda} and \eqref{eq:hawk_kernels}, the intensity function now takes the form:
\begin{equation}
\lambda^\ixa(t) = \mu^\ixa(q(t^-)) + \sum_{m=1}^D \sum_{u = 1}^U o_{\ixa \ixb u}(t) \; ,
\end{equation}
up to an overall factor $\mathbbm{1}_{q(t) \neq 0}$ that has to be considered for orders that 
consume the queue size.
We note $J$ the set of order types that increase the queue size and $I$ the set of order types that decrease the queue size. We further assume that $\sum_{\ixa \in I}\mu^\ixa(q) \geq c_\ixa q$ and $\mu^\ixb(q) \leq c^* ,\forall \ixb \in J$. The queue size is determined by the sum of order flows:
\begin{equation}
q(t) = \sum_{\ixb \in J} N_\ixb - \sum_{\ixa \in I} N_\ixb
\end{equation}
Let $\vec{o}(t)$ be the vector obtained as a vertical stacking of the components $o_{\ixa \ixb u}(t)$ for all
$(\ixa, \ixb, u) \in \{1,\ldots, D\}^2 \times \{1,\ldots U\}$.
It is not difficult to verify that the vector process $(q, \vec{o})^T$ is Markovian. We then aim at constructing a Lyapunov function for $(q, \vec{o})^T$. To start with, we construct Lyapunov functions for $q$ and $\vec{o}$ seperately, then we combine them together.

\subsubsection{Lyapunov function for $\vec{o}$}
Let us first write the differential form of $\vec{o}(t)$:
\begin{equation}
\mathrm{d} o_{\ixa \ixb u}(t) = - \beta_u o_{\ixa \ixb u}(t)  \mathrm{d}t + \alpha^{\ixa \ixb}_u \mathrm{d}N^\ixb_t
\end{equation}
Then for any arbitrary suitable function $F$ who maps $\mathbb{R}^{2D+U}$ to $\mathbb{R}$, the infinitesimal generator takes the form:
\begin{equation}\label{infinite_generator_1}
\mathcal{L} F(\vec{o}) = \sum_{\ixb} \lambda_\ixb (F(\vec{o} + \Delta_\ixb(\vec{o})) - F(\vec{o})) - \sum_{\ixa,\ixb,u} \beta_{u} o_{\ixa \ixb u} \frac{\partial F}{\partial o_{\ixa \ixb u}}
\end{equation}
Where $\lambda_\ixb$ is the probability for an event to arrive in dimension $\ixb$ of the point process $\vec{N}$, and $\Delta_\ixb(\vec{o})$ is jump of $\vec{o}$ caused by this event. We then define the matrix $A$ as:
\begin{equation}
A_{\ixa \ixb} = \sum_u \frac{\alpha^{\ixa \ixb}_u}{\beta_{u}}
\end{equation}
We assume that the dynamics of order flows described by the Hawkes part of the QRH-I model corresponds to a stable Hawkes process. According to Perron-Frobenius theorem, the maximal eigenvalue $\kappa$ of $A$ satisfies $0< \kappa < 1$, and $\vec{\epsilon}$ the associated eigenvector of $\kappa$ satisfies $ \forall l, \epsilon_l > 0$. We then note
\begin{equation}\delta_{\ixa \ixb u} :=\delta_{\ixa u} = \frac{\epsilon_\ixa}{\beta_u}
\end{equation}
We choose function $V_1$ of $\vec{o}$ as:
\begin{equation}V_1(\vec{o}) = \sum_{\ixa,\ixb,u} \delta_{\ixa \ixb u} o_{\ixa \ixb u}
\end{equation}
With notations defined above, we could then verify that
\begin{equation}
\begin{aligned}
\mathcal{L} V_1(\vec{o}) &=\sum_\ixb(\mu_\ixb + \sum_{p,q} o_{\ixb pq})\mathbbm{1}_{\ixa = 0 \vee q(t)>0} \sum_{\ixa,u}\delta_{\ixa \ixb u}\alpha^{\ixa \ixb}_u - \sum_{\ixa,\ixb,u} \beta_u o_{\ixa \ixb u}\delta_{\ixa \ixb u}\\
&\leq \sum_\ixb(\mu_\ixb + \sum_{p,q} o_{mpq}) \sum_{\ixa,u}\delta_{\ixa \ixb u}\alpha^{\ixa \ixb}_u - \sum_{\ixa,\ixb,u} \beta_uo_{\ixa \ixb u}\delta_{\ixa \ixb u}\\
&= \sum_{\ixa,\ixb,u}\mu_\ixb \delta_{\ixa \ixb u}\alpha^{\ixa \ixb}_u + \sum_\ixb (\sum_{p,q} o_{mpq})(\sum_{\ixa, u} \frac{\epsilon_\ixa}{\beta_u} \alpha^{\ixa \ixb}_u) - \sum_{\ixa,\ixb,u} \beta_u o_{\ixa \ixb u}\delta_{\ixa \ixb u}\\
&= C_1 + \sum_\ixb (\sum_{p,q} o_{\ixb pq})(\sum_\ixa \epsilon_\ixa \sum_u \frac{\alpha^{\ixa \ixb}_u}{\beta_u}) - \sum_{\ixa,\ixb,u} \beta_u o_{\ixa \ixb u}\delta_{\ixa \ixb u}\\
&= C_1 + \sum_\ixb (\sum_{p,q} o_{\ixb pq}) \epsilon_\ixb \kappa - \sum_{\ixa,\ixb,u} \epsilon_\ixa o_{\ixa \ixb u}\\
&= C_1 - (1 - \kappa) \sum_{\ixa,\ixb,u} \epsilon_\ixa o_{\ixa \ixb u} \\
&= C_1 - (1 - \kappa) \sum_{\ixa,\ixb,u} \beta_u \delta_{\ixa \ixb u} o_{\ixa \ixb u} \\
&\leq -\rho_1 V + C_1
\end{aligned}
\end{equation}
Where the constant $\rho_1$ is chosen as
\begin{equation}
\rho_1 = (1-\kappa) \inf \beta_u
\end{equation}

\subsubsection{Lyapunov function for $q$}
Similarly, let us first write the differential form of $q(t)$:
\begin{equation}
\mathrm{d} q(t) = \sum_{\ixb \in J} \mathrm{d} N^\ixb_t - \sum_{\ixa \in I} \mathrm{d} N^\ixa_t
\end{equation}
\noindent
We keep using the same notation of $\Delta$ as in Eq~\ref{infinite_generator_1}. For any arbitrary suitable function $F$ who maps $\mathbb{R}^+$ to $\mathbb{R}$, the infinitesimal generator takes the form:
\begin{equation}
\mathcal{L} F(q) = \sum_{\ixb} \lambda_\ixb (F(q + \Delta_\ixb(q)) - F(q)) 
\end{equation}
Next, we choose $V_2$ as the identity function of $q$:
\begin{equation}
\begin{aligned}
V_2(q) &:= q
\end{aligned}
\end{equation}
Then it could be easily verified that $V_2(q)$ is a Lyapunov function for $q$:
\begin{equation}
\begin{aligned}
\mathcal{L} V_2(q) &= \sum_{\ixb \in J} \lambda_\ixb(q) - \sum_{\ixa \in I} \lambda_\ixa(q)\\
&\leq \sum_{\ixb \in J} \mu_\ixb(q) - \sum_{\ixa \in I} \mu_\ixa(q) + \sum_{\ixa,\ixb,u} o_{\ixa \ixb u}\\
&\leq \sum_{\ixb \in J} c^* -\sum_{\ixa \in I} c_\ixa q +  \sum_{\ixa,\ixb,u} o_{\ixa \ixb u}\\
&\leq  -\sum_{\ixa \in I} c_\ixa q + C_2\\
&\leq -\rho_2 V_2 + C_2
\end{aligned}
\end{equation}
Remind that constants $c^*$, $c^\ixa$ are defined before. The constant $\rho_2 $ is simply chosen as $\rho_2  = \max_\ixa c_\ixa$. Also, it must be noted that here
\begin{equation}
C_2 := \sum_{\ixb \in J} c^* + \sum_{\ixa,\ixb,u} o_{\ixa \ixb u}
\end{equation}
which depends on $\vec{o}$. Another remark is since $\ixb$ and $c^*$ are fixed, there must exist $\rho^*$ who satisfy
\begin{equation}
C_2 < \rho^* \sum_{\ixa,\ixb,u} o_{\ixa \ixb u}
\end{equation}

\subsubsection{Lyapunov function for $(q,\vec{o})$}
The final step is to build a function $V$ for both $q$ and $\vec{o}$:
\begin{equation}
V(q, \vec{o}) = V_2(q) + \frac{1}{\eta} V_1( \vec{o})
\end{equation}
We then apply previous conclusions made for $V_1$ and $V_2$. Direct calculation shows:
\begin{equation}\label{L_for_both}
\begin{aligned}
\mathcal{L} V(q, \vec{o}) &= \mathcal{L} V_2(q) + \frac{1}{\eta}\mathcal{L} V_1( \vec{o})\\
&= -\rho_2 V_2 + C_2 - \frac{\rho_1}{\eta}V_1 + \frac{C_1}{\eta}\\
&\leq -\rho_2 V_2 + \rho^* \sum_{\ixa,\ixb,u} o_{\ixa \ixb u} - \frac{\rho_1}{\eta}V_1 + C^{'}\\
\end{aligned}
\end{equation}
Notice that since both the coefficient $\delta_{\ixa \ixb u}$ in $V_1$ and $o_{\ixa \ixb u}$ are positive, there must exist an $\eta$ who satisfies
\begin{equation}
\frac{\rho_1}{2\eta}V_1(\vec{o}) > \rho^* \sum_{\ixa,\ixb,u} o_{\ixa \ixb u}
\end{equation}
By substituting this inequality into Eq~\ref{L_for_both}, we finally prove that $V$ is a Lyapunov function for $(q, \vec{o})$:
\begin{equation}
\begin{aligned}
\mathcal{L} V(q, \vec{o}) & \leq -\rho_2 V_2 - \frac{\rho_1}{2\eta}V_1 + C^{'}\\
& \leq -\min(\rho_2,\frac{\rho_1}{2\eta}) V +C
\end{aligned}
\end{equation}
Given the existence of the Lyapunov function and the geometric drift condition above, under the assumption that the spectral radius of $A$ is smaller than one, Theorem 7.1 in \cite{abergelref22} guarantees that the process $q$ is ergodic. Also, it converges exponentially fast towards its unique stationary distribution.

\subsection{Calculation of the log-likelihood function of QRH model and its gradient}\label{sec:app2}
For QRH model, the log-likelihood is a function of $\mu$ and $\alpha$. Let's note $t^\ixa_k$ the timestamp of the $k$th event of type $\ixa$, and $N^\ixa$ the total number of event of type $\ixa$. With such notation,
\begin{equation}
\begin{aligned}
L(\vec{\alpha},\vec{\mu}) &= 
\sum_{\ixa=1}^D \sum_{k=1}^{N^l} \log \Big( \mu^\ixa(q(t_k^{\ixa,-}) + \sum_{\ixb=1}^D \sum_{u=1}^U \alpha^{\ixa\ixb}_u \beta_u \int_0^{t_k} e^{-\beta_u(t - s)} dN_s^\ixb \Big)\\
&- \sum_{\ixa=1}^D \int_0^{T} \Big( \mu^\ixa(q(t)) + \sum_{\ixb=1}^D \sum_{u=1}^U \alpha^{\ixa\ixb}_u \beta_u \int_0^{s} e^{-\beta_u(s - v)} dN_v^\ixb \Big) ds \\
\end{aligned}
\end{equation}
We first define $g$ and $G$, whose value doesn't depend on the parameters $\vec{\mu}$ and $\vec{\alpha}$. The interest is that $g$ and $G$ only need to be calculated once, then they could be reused to accelerate the calculation of log-likelihood function and its gradient.
\begin{equation}
g^{\ixb}_u(t) = \sum_{t_k^\ixb < t} \beta_u e^{-\beta_u (t - t_k^\ixb)}
\end{equation}
And
\begin{equation}
G^\ixb_u(t) = \int_0^t g^\ixb_u(s) ds = \int_0^t \int_0^{s} \beta_u e^{-\beta_u(s - v)} dN_v^\ixb ds
\end{equation}
Both $g$ and $G$ could be calculated using the following recurrence relation:

\paragraph{Computation of $g$}

\begin{align}
g^{m}_u(t_k) &= \sum_{t^m_{k'} < t_k} \beta_u e^{-\beta_u(t_k - t^m_{k'})} \\
&= \sum_{t^m_{k'} < t_{k-1}} \beta_u e^{-\beta_u(t_{k-1} - t^m_{k'})} e^{-\beta_u(t_k - t_{k-1})} +
\sum_{t_{k-1} \leq t^m_{k'} < t_{k}} \beta_u e^{-\beta_u(t_k - t^m_{k'})} \\
&= e^{-\beta_u(t_k - t_{k-1})} g^{m}(t_{k-1}) + \beta_u e^{-\beta_u(t_k - t_{k-1})} \mathds{1}_{type(t_{k-1}^+) = m}
\end{align}

\paragraph{Computation of $G$}
\begin{align}
G^{m}_u(t_k) - G^{m}_u(t_{k-1})
&= \int_{t_{k-1}}^{t_k} g^{m}_u(s) ds \\
&= \frac{1- e^{-\beta_u(t_k - t_{k-1})}}{\beta_u} g^{m}_u(t_{k-1}) + \Big(1 - e^{-\beta_u(t_k - t_{k-1})} \Big) \mathds{1}_{type(t_{k-1}^+) = m}
\end{align}
We will see later that we only need the value of $g$ and $G$ at for every $t_k$.

\subsubsection{Log-likelihood function}
By exploiting the definition of $g$ and $G$, the log-likelihood function could be rewritten in the following way:
\begin{equation}
\begin{aligned}
L(\alpha,\mu) 
&= \sum_{\ixa=1}^D \sum_{k=1}^{N^\ixa} \log \Big( \mu^\ixa(q(t_k^{\ixa,-}))
 + \sum_{\ixb=1}^D \sum_{u=1}^U \alpha^{\ixa \ixb}_u g^{\ixb}_u(t_k) \Big)  \\
&  - \sum_{\ixa=1}^D \sum_{k=1}^N \mu^\ixa(q(t_k^-)) (t_k - t_{k-1})- \sum_{\ixa=1}^D \mu^\ixa_{q(t_N^+)} (T-t_N)\\
&  - \sum_{\ixa=1}^D \sum_{k=1}^{N} \sum_{\ixb=1}^D \sum_{u=1}^U \alpha^{\ixa \ixb}_u  ( G^{\ixb}_u(t_k) - G^{\ixb}_u(t_{k-1}) ) - \sum_{\ixa=1}^{D} \sum_{\ixb=1}^D \sum_{u=1}^U \alpha^{\ixa \ixb}_u ( G^{\ixb}_u(T) - G^{\ixb}_u(t_N) )
\end{aligned}
\end{equation}

\subsubsection{Gradients}
With a little abuse use of the symbol $q$ and $q(t)$, direct calculation shows that:
\begin{equation}
\frac{\partial L}{\partial \mu^\ixa(q)} = \sum_{k=1}^{N^\ixa} \frac{\mathds{1}_{q(t_k^{\ixa,-}) = q}}{\mu^\ixa(q(t_k^{\ixa,-})) + \sum_{\ixb=1}^D \sum_{u=1}^U \alpha^{\ixa \ixb}_u g^{\ixb}_u(t_k^\ixa)}
- \sum_{k=1}^N \mathds{1}_{q(t_k^-) = q} (t_k - t_{k-1})
- \mathds{1}_{q(t_N^+) = q} (T-t_N)
\end{equation}

\begin{equation}
\frac{\partial L}{\partial \alpha^{\ixa \ixb}_u} = \sum_{k=1}^{N^\ixa} \frac{ g^\ixb_u(t_k^\ixa)}{\mu^\ixa(q(t_k^{\ixa,-}))  + \sum_{\ixb=1}^D \sum_{u=1}^U \alpha^{\ixa \ixb}_u g^{\ixb}_u(t_k^\ixa)}
- \sum_{k=1}^{N}( G^{\ixb}_u(t_k) - G^{\ixb}_u(t_{k-1}) )
- ( G^{\ixb}_u(T) - G^{\ixb}_u(t_N) )
\end{equation}

\subsection{Calculating the least squares function of QRH model and its gradient}\label{sec:app3}
\subsubsection{Least squares function}
For the QRH model described in the Part II.  Let's note $q$ the state of the LOB by combining the state of both the ask side and the bide side.
\begin{equation}
q=q_a \times q_b \quad f(q(t)) := f(q_a(t^-), q_b(t^-))
\end{equation}
Then at time $t$, the intensity function of dimension $\ixa$ is:
\begin{equation}
\lambda^\ixa_t = f^\ixa(q(t)) \Big( \mu^\ixa + \sum_{\ixb} \sum_u \alpha^{\ixa\ixb}_u \int_0^t \beta_u e^{-\beta_u(t-s)} d N_s^m\Big)
\end{equation}
Similar to the calculation of log-likelihood funtion and its gradients, we first define some intermediate vairables whose value doesn't depend $\mu$, $\alpha$ or $f$. They only need to be calculated once in the pre-processing stage. Then they could be reused during the calculation of least squares function and its gradient.

\begin{equation}
g^\ixa_u(t) = \sum_{t_k^\ixa < t} \beta_u e^{-\beta_u (t - t_k^\ixa)}
\end{equation}

\begin{equation}
G^{\ixb}_u(q) = \int_0^T g^\ixb_u(s)  \mathds{1}_{type(q(t)) = q} dt
\end{equation}

\begin{equation}
H^{\ixb \ixb^\prime}_{uu^\prime} (q) = \int_{0}^{T} \mathds{1}_{type(q(t)) = q} g^\ixb_u(t) g^{\ixb\prime}_{u^\prime}(t) d t
\end{equation}

\begin{equation}
D(q) = \int_{0}^{T} \mathds{1}_{type(q(t)) = q} d t
\end{equation}

\begin{equation}
C^\ixb(q) =  \sum_{k = 1}^{N^\ixb} \mathds{1}_{type(t_k^{\ixb,-}) = q}
\end{equation}

\paragraph{Computation of $g$}

\begin{align}
g^\ixb_u(t_k) &= \sum_{t^\ixb_{k'} < t_k} \beta_u e^{-\beta_u(t_k - t^\ixb_{k'})} \\
&= \sum_{t^\ixb_{k'} < t_{k-1}} \beta_u e^{-\beta_u(t_{k-1} - t^\ixb_{k'})} e^{-\beta_u(t_k - t_{k-1})} +
\sum_{t_{k-1} \leq t^\ixb_{k'} < t_{k}} \beta_u e^{-\beta_u(t_k - t^\ixb_{k'})} \\
&= e^{-\beta_u(t_k - t_{k-1})} g^\ixb(t_{k-1}) + \beta_u e^{-\beta_u(t_k - t_{k-1})} \mathds{1}_{type(t_k^-) = m}
\end{align}

\paragraph{Computation of $G$}
\begin{align}
G^{m}_u(t_k) - G^{m}_u(t_{k-1})
&= \int_{t_{k-1}}^{t_k} g^{m}_u(s) ds \\
&= \frac{1- e^{-\beta_u(t_k - t_{k-1})}}{\beta_u} g^{m}_u(t_{k-1}) + \Big(1 - e^{-\beta_u(t_k - t_{k-1})} \Big) \mathds{1}_{type(t_{k-1}^+) = m}
\end{align}

\paragraph{Computation of $H$}
\begin{align}
H^{\ixb \ixb^\prime}_{uu^\prime} (q) &= \sum_{k} \int_{t_{k-1}}^{t_k} \mathds{1}_{type(q(t)) = q} g^\ixb_u(t) g^{\ixb\prime}_{u^\prime}(t) d t\\
&= \sum_{k} \mathds{1}_{type(q(t_k^-)) = q} g^\ixb_u(t_{k-1}) g^{\ixb\prime}_{u^\prime}(t_{k-1}) e^{-(\beta_u + \beta_{u^\prime})(t_k - t_{k-1})}
\end{align}

\subsubsection{Least squares function}
With the intermidiate variables defined in the previous part, we can rewrite the least squares function defined in \ref{leastsq_dim2} in a more efficient form for calculation. In dimension $\ixa$, 
\begin{equation}
\begin{aligned}
R^\ixa(\theta) &= \int_0^T \lambda_\ixa^2(t; \theta|\mathscr{F}_t) dt\ - \sum_{k = 1}^{N^\ixa} \lambda_\ixa(t_k^\ixa; \theta|\mathscr{F}_{t_k^\ixa})\\
&= \int_0^T {f^\ixa(q(t))}^2 \Big( \mu^\ixa + \sum_{\ixb} \sum_u \alpha^{\ixa\ixb}_u \int_0^t \beta_u e^{-\beta_u(t-s)} d N_s^m\Big)^2 dt\ \\
&- \sum_{k = 1}^{N^\ixa} f^\ixa(q(t_k^\ixa)) \Big( \mu^\ixa + \sum_{\ixb} \sum_u \alpha^{\ixa\ixb}_u \int_0^{t_k^\ixa} \beta_u e^{-\beta_u(t_k^\ixb-s)} d N_s^m\Big)\\
\end{aligned}
\end{equation}
Let's name the first item and the second item in the formula above \textit{term I} and \textit{term II} seperately. We could further decomposite \textit{term I} into new items \textit{term I.1}, \textit{term I.2} and etc: 
\begin{equation}
\begin{aligned}
\textit{term I} &=  \int_0^T {f^\ixa(q(t))}^2  \Big( \mu^\ixa + \sum_{\ixb} \alpha^{\ixa\ixb}_u \sum_u \int_0^t \beta_u e^{-\beta_u(t-s)} d N_s^m\Big)^2 dt\\
&=  \int_0^T {f^\ixa(q(t))}^2  {\mu^\ixa}^2 dt\\
&+ \int_0^T 2 {f^\ixa(q(t))}^2  \mu^\ixa \Big( \sum_{\ixb} \sum_u \alpha^{\ixa\ixb}_u \int_0^t \beta_u e^{-\beta_u(t-s)} d N_s^m \Big) dt\\
&+ \int_0^T 2{f^\ixa(q(t))}^2  \Big( \sum_{\ixb} \sum_u \alpha^{\ixa\ixb}_uu \int_0^t \beta_u e^{-\beta_u(t-s)} d N_s^m \Big)^2 dt\\
\end{aligned}
\end{equation}
For \textit{term I.1}, it could be calculated from the intermediate variables defined above:
\begin{equation}
\begin{aligned}
\int_0^T {f^\ixa(q(t))}^2  {\mu^\ixa}^2 dt = {\mu^\ixa}^2 \sum_{q} D(q) f^\ixa(q)^2
\end{aligned}
\end{equation}
For \textit{term I.2}, 
\begin{equation}
\begin{aligned}
&\int_0^T 2{f^\ixa(q(t))}^2 \mu^\ixa \Big( \sum_{\ixb}  \sum_u \alpha^{\ixa\ixb}_u \int_0^t \beta_u e^{-\beta_u(t-s)} d N_s^m \Big) dt\\
&= \int_0^T 2{f^\ixa(q(t))}^2 \mu^\ixa \Big( \sum_{\ixb}  \sum_u \alpha^{\ixa\ixb}_u g^\ixb_u(t) \Big) dt\\
&= 2\mu^\ixa \sum_{q} f^\ixa(q)^2 \Big( \sum_{\ixb}  \sum_u \alpha^{\ixa\ixb}_u \sum_u G^\ixb_u(q) \Big) 
\end{aligned}
\end{equation}
And for \textit{term I.3}, 
\begin{equation}
\begin{aligned}
&\int_0^T 2{f^\ixa(q(t))}^2  \Big( \sum_{\ixb} \sum_u \alpha^{\ixa\ixb}_u \int_0^t \beta_u e^{-\beta_u(t-s)} d N_s^m \Big)^2 dt\\
&= \int_0^T 2{f^\ixa(q(t))}^2  \Big( \sum_{\ixb} \sum_{\ixb^\prime} \sum_u\sum_{u^\prime} \alpha^{\ixa\ixb}_u \alpha^{\ixa\ixb^\prime}_{u^\prime} g^\ixb_u(t) g^{\ixb^\prime}_{u^\prime}(t)  \Big) dt\\
&= \sum_{\ixb} \sum_{\ixb^\prime} \sum_u\sum_{u^\prime} \alpha^{\ixa\ixb}_u \alpha^{\ixa\ixb^\prime}_{u^\prime} \Big( \sum_{q} f^\ixa(q)^2 H^{\ixb \ixb^\prime}_{uu^\prime} (q)  \Big) 
\end{aligned}
\end{equation}
For the convinience of notation, we flip the sign of \textit{term II}. Then we decomposite it into \textit{term II.1} and \textit{term II.2}:
\begin{equation}
\begin{aligned}
II &=\sum_{k = 1}^{N^\ixa} f^\ixa(q(t_k^\ixa)) \Big( \mu^\ixa + \sum_{\ixb} \sum_u \alpha^{\ixa\ixb}_u \int_0^{t_k^\ixa} \beta_u e^{-\beta_u(t_k^\ixb-s)} d N_s^m\Big)\\
&= \sum_{k = 1}^{N^\ixa} f^\ixa(q(t_k^\ixa)) \mu^\ixa + \sum_{k = 1}^{N^\ixa} f^\ixa(q(t_k^\ixa)) \Big(\sum_{\ixb} \sum_u \alpha^{\ixa\ixb}_u \int_0^{t_k^\ixb} \beta_u e^{-\beta_u(t_k^\ixb-s)} d N_s^m\Big)\\
\end{aligned}
\end{equation}
For \textit{term II.1}, 
\begin{equation}
\begin{aligned}
\sum_{k = 1}^{N^\ixa} f^\ixa(q(t_k^\ixa)) \mu^\ixa = \mu^\ixa \sum_{q} f(q) C^\ixa(q)\\
\end{aligned}
\end{equation}
For \textit{term II.2}, 
\begin{equation}
\begin{aligned}
&\sum_{k = 1}^{N^\ixa} f^\ixa(q(t_k^\ixa)) \Big(\sum_{\ixb} \sum_u \alpha^{\ixa\ixb}_u \int_0^{t_k^\ixb} \beta_u e^{-\beta_u(t_k^\ixb-s)} d N_s^m\Big)
=\sum_{k = 1}^{N^\ixa} f^\ixa(q(t_k^\ixa)) \Big(\sum_{\ixb} \sum_u \alpha^{\ixa\ixb}_u g^\ixb_u(t_k^\ixa) \Big)\\
\end{aligned}
\end{equation}
The least square function defined in \ref{leastsq_dim2} could be calculated by summing up all these terms.

\subsubsection{Gradients}
Using the intermediate variables and results presented above, direct calculation shows that:
\begin{equation}
\begin{aligned}
\frac{\partial R}{\partial \mu^\ixa} &= 2{\mu^\ixa} \sum_{q} D(q) f^\ixa(q)^2\\
&+2 \sum_{q} f^\ixa(q)^2 \Big( \sum_{\ixb} \alpha^{\ixa\ixb}_u \sum_u G^\ixb_u(q) \Big) \\
&-2 \sum_{q} f(q) C^\ixb(q)\\
\end{aligned}
\end{equation}

\begin{equation}
\begin{aligned}
\frac{\partial R}{\partial \alpha^{\ixa\ixb}_u} &= 2\mu^\ixa \sum_{q} f^\ixa(q)^2 G^\ixb_u(q) \\
&+2\sum_{\ixb^\prime} \sum_{u^\prime} \alpha^{\ixa\ixb^\prime}_{u^\prime} \Big( \sum_{q} f^\ixa(q)^2 H^{\ixb \ixb^\prime}_{uu^\prime} (q)  \Big) \\
&-2\sum_{k = 1}^{N^\ixb} f^\ixa(q(t_k^\ixb)) g^\ixb_u({t_k^\ixb}^-) \\
\end{aligned}
\end{equation}

\begin{equation}
\begin{aligned}
\frac{\partial R}{\partial f^\ixa(q)} &= \sum_q 2{f^\ixa(q)}  {\mu^\ixa}^2 D(q)
+ 4\mu^\ixa f^\ixa(q) \Big( \sum_{\ixb} \alpha^{\ixa\ixb}_u \sum_u G^\ixb_u(q) \Big) \\
&+ 2 \sum_{\ixb} \sum_{\ixb^\prime} \sum_u\sum_{u^\prime} \alpha^{\ixa\ixb}_u \alpha^{\ixa\ixb^\prime}_{u^\prime} \Big( \sum_{q} f^\ixa(q) H^{\ixb \ixb^\prime}_{uu^\prime} (q)  \Big) \\
&- \mu^\ixa C^\ixa(q)
- 2\sum_{k = 1}^{N^\ixa} \mathds{1}_{type({t_k^\ixa}^-) = q} \sum_{\ixb} \sum_u \alpha^{\ixa\ixb}_u g^\ixb_u(t_k^\ixa) \\
\end{aligned}
\end{equation}

\bibliographystyle{plain}

\end{document}